\documentclass{cernrep} 
\usepackage{texnames}
\usepackage[T1]{fontenc}
\usepackage[bookmarks, colorlinks=true, linktoc=page, pdftex, linkcolor=red, citecolor=red, urlcolor=red]{hyperref}
\newcommand{\eqn}[1]{(\ref{#1})}
\newcommand{\be}{\begin{equation}}
\newcommand{\ee}{\end{equation}}
\newcommand{\no}{\nonumber}
\newcommand{\bel}[1]{\be\label{#1}}
\newcommand{\ba}{\begin{array}{c}}
\newcommand{\bat}{\begin{array}{cc}}
\newcommand{\bath}{\begin{array}{ccc}}
\newcommand{\ea}{\end{array}}
\newcommand{\beqn}{\begin{eqnarray}}
\newcommand{\eeqn}{\end{eqnarray}}

\newcommand{\bi}{\begin{itemize}}
\newcommand{\ei}{\end{itemize}}

\newcommand{\rms}{\rm\scriptstyle}

\newcommand{\toLow}{\stackrel{q^2 \ll M_W^2}{\,\longrightarrow\,}}

\def\gap{\;\lower3pt\hbox{$\buildrel > \over \sim$}\;}
\def\lap{\;\lower3pt\hbox{$\buildrel < \over \sim$}\;}
\newcommand{\cL}{{\cal L}}

\newcommand{\cM}{{\cal M}}
\newcommand{\cN}{{\cal N}}
\newcommand{\cO}{{\cal O}}
\newcommand{\cP}{{\cal P}}

\newcommand{\cG}{{\cal G}}
\newcommand{\cH}{{\cal H}}
\newcommand{\cY}{{\cal Y}}
\newcommand{\cA}{{\cal A}}
\newcommand{\cI}{{\cal I}}

\newcommand{\cJ}{{\cal J}}
\newcommand{\cC}{{\cal C}}

\newcommand{\cCP}{{\cal CP}}
\newcommand{\cT}{{\cal T}}
\newcommand{\cCPT}{{\cal CPT}}
\def\bI{\mathbf{I}}
\def\bV{\mathbf{V}}
\def\bM{\mathbf{M}}
\def\bU{\mathbf{U}}
\def\bS{\mathbf{S}}
\def\bH{\mathbf{H}}
\def\bmd{\mathbf{d}}
\def\bmu{\mathbf{u}}
\def\bml{\boldsymbol{\ell}}
\def\bmf{\mathbf{f}}

\newcommand{\e}{\mbox{\rm e}}

\def\dop{d\hskip .7pt'}
\def\up{u\hskip .6pt'}
\def\lp{\ell\hskip .7pt'}
\def\nup{\nu\hskip .7pt'}
\def\sp{s\hskip .7pt'}
\def\bp{b\hskip .6pt'}

\def\barup{\bar u\hskip .6pt'}
\def\bardop{\bar d\hskip .7pt'}
\def\barnup{\bar\nu\hskip .7pt'}
\def\barlp{\bar\ell\hskip .7pt'}
\begin{document}
\title{Flavour Dynamics and Violations of the CP Symmetry}

\author{Antonio Pich}

\institute{ %%%Departament de F\'{\i}sica Te\`orica,
IFIC, University of Val\`encia -- CSIC,
%Edifici d'Instituts d'Investigaci\'o, Apt. Correus 22085, 
E-46071
Val\`encia, Spain}

\begin{abstract}
%%%These lectures present 
An overview of flavour physics and\ $\cCP$-violating phenomena is presented. The Standard Model quark-mixing mechanism is discussed in detail and its many successful experimental tests are summarized. Flavour-changing transitions put very stringent constraints on new-physics scenarios beyond the Standard Model framework. Special attention is given to the empirical evidences of $\cCP$ violation and their important role in our understanding of flavour dynamics.
The current status of the so-called flavour anomalies is also reviewed.
\end{abstract}


\maketitle % this produces the title block

%%%%%%%%%% FLAVOUR DYNAMICS %%%%%%%%%%

\section{Fermion families}
\label{sec:families}

We have learnt experimentally that there are six different quark
flavours; three of them, $u\,$, $c\,$, $t\,$, with electric charge $Q=+\frac{2}{3}$  (up-type), and the other three, $d\,$, $s\,$, $b\,$, 
with $Q=-\frac{1}{3}$ (down-type). There are also
%%%  ($u\,$, $d\,$, $s\,$, $c\,$, $b\,$, $t\,$), 
three different charged leptons, $e\,$, $\mu\,$, $\tau$, with $Q=-1$ and their corresponding neutrinos, $\nu_e\,$, $\nu_\mu\,$, $\nu_\tau\,$, with $Q=0$. We can include all these particles into the
$SU(3)_C \otimes SU(2)_L \otimes U(1)_Y$ Standard Model (SM) framework
\cite{GL:61,WE:67,SA:69}, by organizing them into three families of quarks and leptons:
\bel{eq:families}
\left[\bat \nu_e & u \\  e^- & \dop \ea \right]\, , \qquad\quad
\left[\bat \nu_\mu & c \\  \mu^- & \sp \ea \right]\, , \qquad\quad
\left[\bat \nu_\tau & t \\  \tau^- & \bp \ea \right]\, ,
\ee
where (each quark appears in three different colours)
\bel{eq:structure}
\left[\bat \nu^{}_i & u_i \\  \ell^-_i & \dop_i \ea \right] \quad\equiv\quad
\left(\ba \nu^{}_i \\ \ell^-_i \ea \right)_L\, , \quad
\left(\ba u_i \\ \dop_i \ea \right)_L\, , \quad \ell^-_{iR}\; , \quad
u^{\phantom{j}}_{iR}\; , \quad \dop_{iR}\; ,
\ee
plus the corresponding antiparticles. Thus, the left-handed fields
are $SU(2)_L$ doublets, while their right-handed partners transform
as $SU(2)_L$ singlets. The three fermionic families
appear to have identical properties (gauge
interactions); they differ only by their mass and their flavour
quantum numbers.

The fermionic couplings of the photon and the $Z$ boson are flavour
conserving, \ie the neutral gauge bosons couple to a fermion and its
corresponding antifermion. In contrast, the $W^\pm$ bosons couple
any up-type quark with all down-type quarks because the weak doublet
partner of $u_i$ turns out to be a quantum superposition of
down-type mass eigenstates:
$\dop_{i} = \sum_j \bV_{ij}\, d^{\phantom{j}}_{j}$.
This flavour mixing generates a rich variety of observable phenomena,
including $\cCP$-violation effects, which can be described in a very
successful way within the SM \cite{SM:11,Pich:2011nh}.

In spite of its enormous phenomenological success, The SM
does not provide
any real understanding of flavour. We do not know yet why fermions are replicated in three (and only three) nearly identical copies. Why the pattern of masses and mixings is what it is? Are the masses the only difference among the three families? What is the origin of the SM flavour structure? Which dynamics is responsible for the observed $\cCP$ violation?
The fermionic flavour is the main source of arbitrary free parameters in the SM:
9 fermion masses, 3 mixing angles and 1 complex phase, for massless neutrinos.
7 (9) additional parameters arise with non-zero Dirac (Majorana) neutrino masses:
3 masses, 3 mixing angles and 1 (3) phases.
The problem of fermion mass generation is deeply related with the
mechanism responsible for the electroweak Spontaneous Symmetry Breaking (SSB).
Thus, the origin of these parameters lies in the most obscure part of the SM Lagrangian: the scalar sector. Clearly, the dynamics of flavour appears to be ``terra incognita'' which deserves a careful investigation.

The following sections contain a short overview of the quark flavour sector
and its present phenomenological status. The most relevant experimental tests are briefly described. A more pedagogic introduction to the SM can be found in
Ref.~\cite{SM:11}.

\section{Flavour structure of the Standard Model}
\label{sec:flavour}

%%%%%%%%%%%%%%%  FIGURE %%%%%%%%%%%%%%%%%%%%%%%%%
\begin{figure}[tb]\centering
\begin{minipage}[c]{.45\linewidth}\centering
\includegraphics[width=4cm]{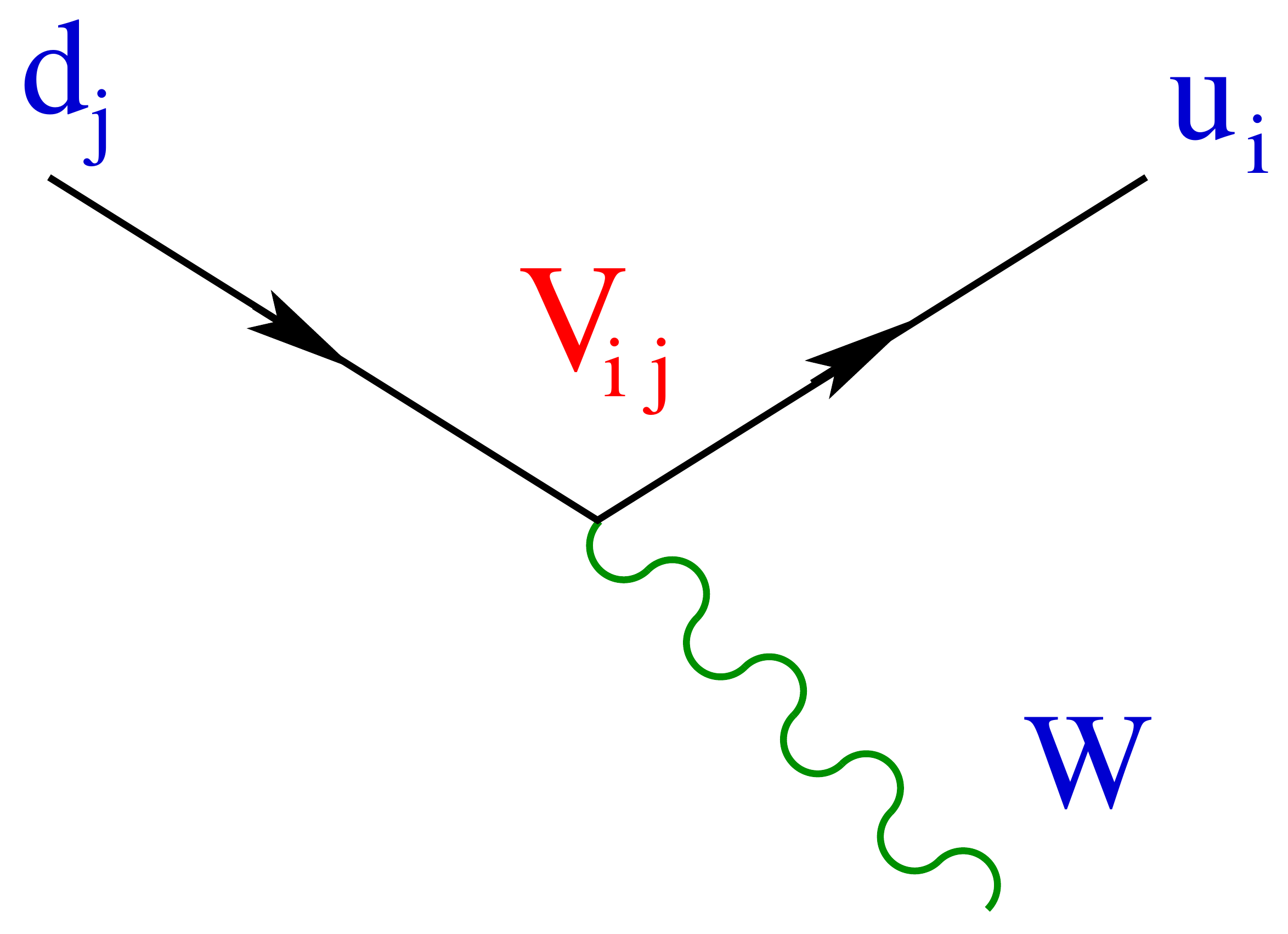}
\end{minipage}
\hskip 1cm
\begin{minipage}[c]{.45\linewidth}\centering
\includegraphics[width=4cm]{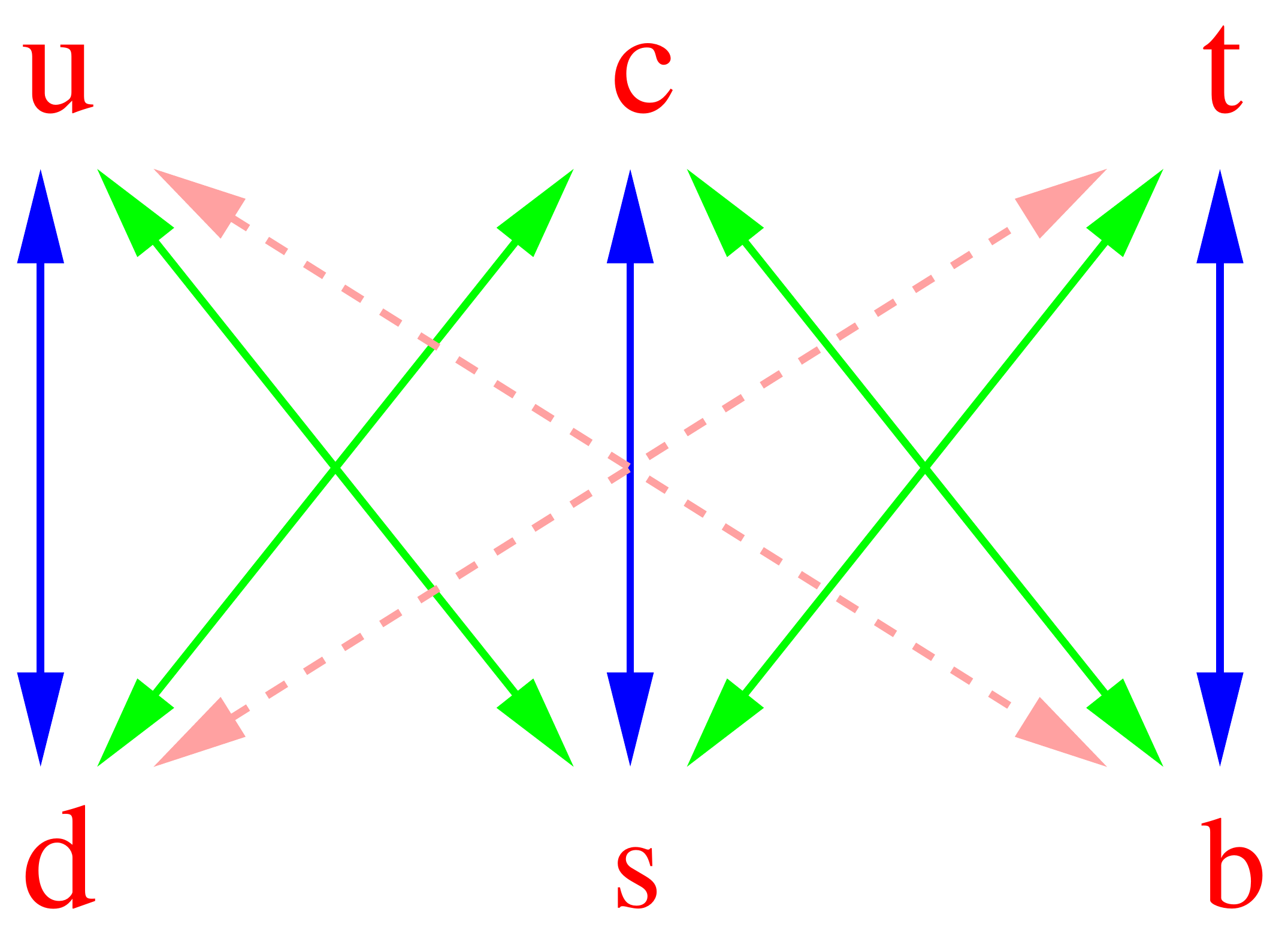}
\end{minipage}
\caption{Flavour-changing transitions through the charged-current
couplings of the $W^\pm$ bosons.}
\label{fig:CKM}
\end{figure}
%%%%%%%%%%%%%%%%%%%%%%%%%%%%%%%%%%%%%%%%%%%%%%%%%

In the SM flavour-changing transitions occur only in the charged-current sector
(Fig.~\ref{fig:CKM}):
\bel{eq:cc_mixing}
\cL_{\rms CC}\, = \, - {g\over 2\sqrt{2}}\,\left\{
W^\dagger_\mu\,\left[\,\sum_{ij}\;
\bar u_i\,\gamma^\mu(1-\gamma_5) \,\bV_{\! ij}\, d_j
\; +\;\sum_\ell\; \bar\nu_\ell\,\gamma^\mu(1-\gamma_5)\, \ell
\,\right]\, + \, \mathrm{h.c.}\right\}\, .
\ee
The so-called Cabibbo--Kobayashi--Maskawa (CKM) matrix $\bV$
\cite{cabibbo,KM:73} is generated by the same Yukawa couplings giving
rise to the quark masses.
Before SSB, there is no mixing among the different quarks, \ie $\bV = \bI$.
In order to understand the origin of the matrix $\bV$ ,
let us consider the general case of $N_G$ generations of fermions,
and denote $\nup_j$, $\lp_j$, $\up_j$, $\dop_j$ the members of the
weak family $j$ \ ($j=1,\ldots,N_G$), with definite transformation
properties under the gauge group. Owing to the fermion replication,
a large variety of fermion-scalar couplings are allowed by the gauge
symmetry. The most general Yukawa Lagrangian has the form
\beqn\label{eq:N_Yukawa}
\cL_Y &=&-\,\sum_{jk}\;\left\{
\left(\barup_j , \bardop_j\right)_L \left[\, c^{(d)}_{jk}\,
\left(\ba \phi^{(+)}\\ \phi^{(0)}\ea\right)\, \dop_{kR} \; +\;
c^{(u)}_{jk}\,
\left(\ba \phi^{(0)*}\\ -\phi^{(-)}\ea\right)\, \up_{kR}\,
\right]
\right.\no\\ && \qquad\quad\left. +\;\;
\left(\barnup_j , \barlp_j\right)_L\, c^{(\ell)}_{jk}\,
\left(\ba \phi^{(+)}\\ \phi^{(0)}\ea\right)\, \lp_{kR}
\,\right\}
\; +\; \mathrm{h.c.},
\eeqn
where
$\phi^T(x)\,\equiv\,\left( \phi^{(+)}, \phi^{(0)}\right)$
is the SM scalar doublet and
$c^{(d)}_{jk}$, $c^{(u)}_{jk}$ and $c^{(\ell)}_{jk}$
are arbitrary coupling constants.
The second term involves the charge-conjugate
%%% $\cC$-conjugate 
scalar field
$\phi^c(x) = i\,\sigma_2\,\phi^*(x)$.

In the unitary gauge $\phi^T(x)\,\equiv\,\frac{1}{\sqrt{2}}\,\left( 0, v+H(x)\right)$, where $v$ is the electroweak vacuum expectation value and
$H(x)$ the Higgs field. The Yukawa Lagrangian can then be written as
\bel{eq:N_Yuka}
\cL_Y\, =\, - \left(1 + {H\over v}\right)\,\left\{\,
\overline{\bmd}\hskip .6pt'_L \,\bM_d'\,
\bmd\hskip .6pt'_R \; + \;
\overline{\bmu}\hskip .6pt'_L \,\bM_u'\,
\bmu\hskip .6pt'_R
\; + \;
\overline{\bml}\hskip .2pt'_L \,\bM'_\ell\,
\bml\hskip .2pt'_R \; +\;
\mathrm{h.c.}\right\} .
\ee
Here, $\bmd\hskip .6pt'$, $\bmu\hskip .6pt'$
and $\bml\hskip .2pt'$ denote vectors in
the $N_G$-dimensional flavour space,
with components  $\dop_j$, $\up_j$ and $\lp_j$, respectively,
and the corresponding mass matrices are given by
\bel{eq:M_c_relation}
(\bM'_d)^{}_{ij}\,\equiv\, c^{(d)}_{ij}\, {v\over\sqrt{2}}\, ,\qquad
(\bM'_u)^{}_{ij}\,\equiv\, c^{(u)}_{ij}\, {v\over\sqrt{2}}\, ,\qquad
(\bM'_\ell)^{}_{ij}\,\equiv\, c^{(\ell)}_{ij}\, {v\over\sqrt{2}}\, .
\ee
The diagonalization of these mass matrices determines the mass
eigenstates $d_j$, $u_j$ and $\ell_j$,
which are linear combinations of the corresponding weak eigenstates
$\dop_j$, $\up_j$ and $\lp_j$, respectively.

The matrix $\bM_d'$ can be decomposed as\footnote{
%%%%%%%%%%%%%%%%%
The condition $\det{\bM'_f}\not=0$ \ ($f=d,u,\ell$)
guarantees that the decomposition
$\bM'_f=\bH_f\bU_f$ is unique:
$\bU_f\equiv\bH_f^{-1}\bM_f'$.
The matrices $\bS_f$ are completely determined (up to phases)
only if all diagonal elements of $\bM_f$ are different.
If there is some degeneracy, the arbitrariness of $\bS_f$
reflects the freedom to define the physical fields.
When $\det{\bM'_f}=0$, the matrices $\bU_f$ and $\bS_f$ are not
uniquely determined, unless their unitarity is explicitly imposed.}
%%%%%%%%%%%%%%%%%
$\bM_d'=\bH^{}_d\,\bU^{}_d=\bS_d^\dagger\, \mathbf{\cM}^{}_d
\,\bS^{}_d\,\bU^{}_d$, where \ $\bH^{}_d\equiv
\sqrt{\bM_d'\bM_d'^{\dagger}}$ is an Hermitian positive-definite
matrix, while $\bU^{}_d$ is unitary. $\bH^{}_d$ can be diagonalized
by a unitary matrix $\bS^{}_d$; the resulting matrix
$\mathbf{\cM}^{}_d$ is diagonal, Hermitian and positive definite.
Similarly, one has \ $\bM_u'= \bH^{}_u\,\bU^{}_u= \bS_u^\dagger\,
\mathbf{\cM}^{}_u\, \bS^{}_u\,\bU^{}_u$ \ and \ $\bM_\ell'=
\bH^{}_\ell\,\bU^{}_\ell= \bS_\ell^\dagger\, \mathbf{\cM}^{}_\ell
\,\bS^{}_\ell\,\bU^{}_\ell$. In terms of the diagonal mass matrices
\bel{eq:Mdiagonal}
\mathbf{\cM}^{}_d =\mathrm{diag}(m_d,m_s,m_b,\ldots)\, ,\quad\;
\mathbf{\cM}^{}_u =\mathrm{diag}(m_u,m_c,m_t,\ldots)\, ,\quad\;
\mathbf{\cM}^{}_\ell=\mathrm{diag}(m_e,m_\mu,m_\tau,\ldots)\, ,
\ee
the Yukawa Lagrangian takes the simpler form
\bel{eq:N_Yuk_diag}
\cL_Y\, =\, - \left(1 + {H\over v}\right)\,\left\{\,
\overline{\bmd}\,\mathbf{\cM}^{}_d\,\bmd \; + \;
\overline{\bmu}\, \mathbf{\cM}^{}_u\,\bmu \; + \;
\overline{\bml}\,\mathbf{\cM}^{}_\ell\,\bml \,\right\}\, ,
\ee
where the mass eigenstates are defined by
\beqn\label{eq:S_matrices}
\bmd^{}_L &\!\!\equiv &\!\!
\bS^{}_d\, \bmd\hskip .6pt'_L \, ,
\qquad\qquad  \hskip 15.4pt
\bmu^{}_L \:\equiv\; \bS^{}_u \,\bmu\hskip .6pt'_L \, ,
\qquad\qquad\,\,\,\,\,\,\,\,\,
\bml^{}_L \:\equiv\; \bS^{}_\ell \,\bml\hskip .2pt'_L \, ,
\no\\
\bmd^{}_R &\!\!\equiv &\!\!
\bS^{}_d \bU^{}_d\,\bmd\hskip .6pt'_R \, , \qquad\qquad
\bmu^{}_R \:\equiv\; \bS^{}_u\bU^{}_u\, \bmu\hskip .6pt'_R \, , \qquad\qquad
\bml^{}_R \:\equiv\; \bS^{}_\ell\bU^{}_\ell \, \bml\hskip .2pt'_R \, .
\eeqn
Note, that the Higgs couplings are flavour-conserving and proportional to the
corresponding fermion masses.

\begin{figure}[t]\centering
\begin{minipage}[t]{.19\linewidth}\centering
\includegraphics[height=1.75cm,clip]{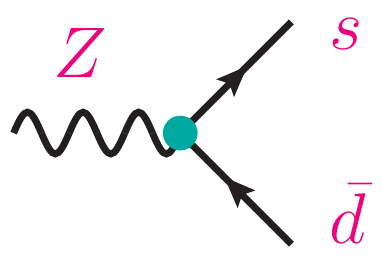}
\end{minipage}
\hskip 1cm
\begin{minipage}[t]{.3\linewidth}\centering
\includegraphics[height=1.75cm,clip]{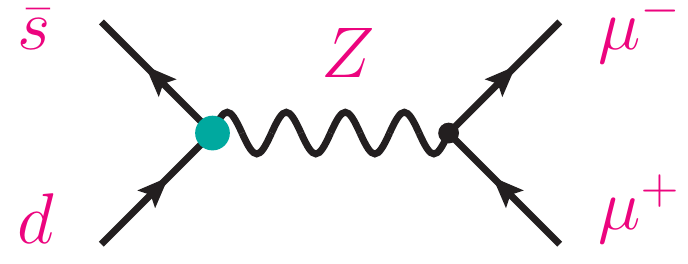}
\end{minipage}
\hskip 1cm
\begin{minipage}[t]{.3\linewidth}\centering
\includegraphics[height=1.75cm,clip]{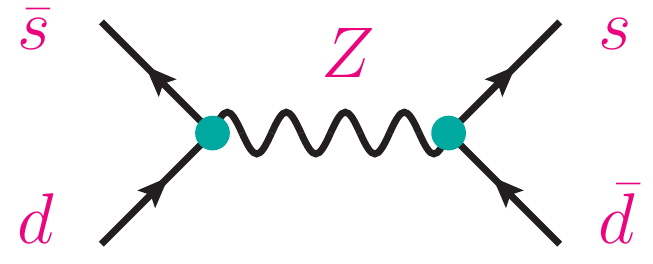}
\end{minipage}
\caption{Tree-level FCNC couplings (green solid vertices) are absent in the SM. Therefore, very suppressed (experimentally) transitions such as \ $K^0\to\mu^+\mu^-$ or \ $K^0$--$\bar K^0$ \ mixing cannot occur through tree-level exchange.}
\label{fig:No-Zsd-vertex}
\end{figure}

Since, \ $\overline{\bmf}\hskip .7pt'_L\, \bmf\hskip .7pt'_L =
\overline{\bmf}^{}_L \,\bmf^{}_L$ \ and \
$\overline{\bmf}\hskip .7pt'_R \,\bmf\hskip .7pt'_R =
\overline{\bmf}^{}_R \,\bmf^{}_R$ \
($f=d,u,\ell$), the form of the neutral-current part of the
$SU(3)_C\otimes SU(2)_L\otimes U(1)_Y$ Lagrangian does not change when expressed in terms of mass eigenstates. Therefore, there are no flavour-changing neutral currents (FCNCs) in the SM. This is a consequence of treating all equal-charge fermions on the same footing (GIM mechanism \cite{GIM:70}), and guarantees that weak transitions such as $B_{s,d}^0\to\ell^+\ell^-$, $K^0\to\mu^+\mu^-$ or  $K^0$--$\bar K^0$ mixing  (Fig.~\ref{fig:No-Zsd-vertex}), which are known experimentally to be very suppressed, cannot happen at tree level. The absence of FCNCs is crucial for the phenomenological success of the SM. 
However, $\overline{\bmu}\hskip .7pt'_L \,\bmd\hskip .7pt'_L =
\overline{\bmu}^{}_L \,\bS^{}_u\,\bS_d^\dagger\,\bmd^{}_L\equiv
\overline{\bmu}^{}_L \bV\,\bmd^{}_L$. In general, $\bS^{}_u\not=
\bS^{}_d\,$; thus, if one writes the weak eigenstates in terms of
mass eigenstates, a $N_G\times N_G$ unitary mixing matrix $\bV$
appears in the quark charged-current sector as indicated in
Eq.~(\ref{eq:cc_mixing}).

%%%%%%%%%%%%%%%  FIGURE %%%%%%%%%%%%%%%%%%%%%%%%%
%%%%%%%%%%%%%%%%%%%%%%%%%%%%%%%%%%%%%%%%%%%%%

If neutrinos are assumed to be massless, we can always redefine the
neutrino flavours, in such a way as to eliminate the
mixing in the lepton sector: $\overline{\mbox{\boldmath
$\nu$}}\hskip .7pt'_L\, \bml\hskip .7pt'_L =
\overline{\mbox{\boldmath $\nu$}}\hskip .7pt_L'\,
\bS^\dagger_\ell\,\bml^{}_L\equiv \overline{\mbox{\boldmath
$\nu$}}^{}_L\, \bml^{}_L$. Thus, we have lepton-flavour conservation
in the minimal SM without right-handed neutrinos.
If sterile $\nu^{}_R$ fields are included in the model, one
has an additional Yukawa term in Eq.~\eqn{eq:N_Yukawa}, giving rise
to a neutrino mass matrix \ $(\bM'_\nu)_{ij}\equiv c^{(\nu)}_{ij}\,
{v/\sqrt{2}}\,$. Thus, the model can accommodate non-zero neutrino
masses and lepton-flavour violation through a lepton mixing matrix
$\bV^{}_{\! L}$ analogous to the one present in the quark sector.
Note, however, that the total lepton number \ $\mathrm{L}\equiv \mathrm{L}_e + \mathrm{L}_\mu + \mathrm{L}_\tau$ \ is still conserved. We know experimentally that neutrino masses are tiny and, as shown in Table~\ref{table:LFV},
there are strong bounds on lepton-flavour violating decays. However, we do have a clear evidence of neutrino oscillation phenomena \cite{Tanabashi:2018oca}. Moreover, since right-handed neutrinos are singlets under
$SU(3)_C \otimes SU(2)_L \otimes U(1)_Y$, the SM gauge symmetry group
allows for a right-handed Majorana neutrino mass term, violating lepton number by two units. Non-zero neutrino masses clearly imply interesting new phenomena \cite{SM:11}.

%%%%%%%%%%%%%%%%%%%%%%%%%%%%%%%%%%%%%%%%%%%%%%%%%%%%%%%%%%%%%%%%%%%
\begin{table}[tb]\centering
\caption{Best published limits on lepton-flavour-violating transitions \cite{Tanabashi:2018oca}.}
%%%\cite{Decamp:1991uy,Abreu:1996mj,Adriani:1993sy,Akers:1995gz,Aaij:2013cby,Adam:2013mnn,Bellgardt:1987du,Bolton:1988af,Lees:2010ez,Miyazaki:2012mx,Bertl:2006up}.}
\label{table:LFV}
\renewcommand{\arraystretch}{1.1} % enlarge line spacing
\begin{tabular}{ll@{\hspace{1.2cm}}ll@{\hspace{1.1cm}}ll@{\hspace{1.2cm}}ll}
\hline \hline\\[-4mm]
\multicolumn{6}{l}{$\rm{Br}(\mu^-\to X^-)\cdot 10^{12}\qquad\quad (90\%\;\mathrm{CL})$}
\\[1mm] \hline \\[-4mm]
 $e^-\gamma$ & $0.42$ &
 $e^-2\gamma$ & $72$ &
 $e^-e^-e^+$ & $\phantom{1}1.0$ &&
\\[1mm] \hline\hline \\[-4mm]
\multicolumn{6}{l}{$\Gamma(\mu^- + N\to e^- + N)/\Gamma(\mu^- + N\to \mathrm{capture})\cdot 10^{12}\qquad\quad (90\%\;\mathrm{CL})$}
\\[1mm] \hline \\[-4mm]
 Au & $0.7$ & Ti & $4.3$ & Pb & 46 &  %%% S & $7$ & 
\\[1mm] \hline\hline \\[-4mm]
\multicolumn{6}{l}{$\rm{Br}(\tau^-\to X^-)\cdot 10^{8}\qquad\quad (90\%\;\mathrm{CL})$}
 \\ \\[-4mm] \hline \\[-4mm]
 $e^-\gamma$ & $3.3$ &
 $e^-e^+e^-$ & $\phantom{1}2.7$ &
 $e^-\mu^+\mu^-$ & $2.7$ &
 $e^-e^-\mu^+$ & $1.5$ 
 \\
 $\mu^-\gamma$ & $4.4$ &
 $\mu^-e^+e^-$ & $\phantom{1}1.8\quad\;$ &
 $\mu^-\mu^+\mu^-$ & $2.1$ &
 $\mu^-\mu^-e^+$ & $1.7$ 
 \\
 $e^-\pi^0$ & $8.0$ &
 $\mu^-\pi^0$ & $11$ &
 $e^-\phi$ & $3.1$  &
 $\mu^-\phi$ & $8.4$
 \\
 $e^-\eta$ & $9.2$ &
 $e^-\eta'$ & $16$ &
 $e^-\rho^0$ & $1.8$ &
 $e^-\omega$ & $4.8$ 
 \\
 $\mu^-\eta$ & $6.5$ &
 $\mu^-\eta'$ & $13$ &
 $\mu^-\rho^0$ & $1.2$ &
 $\mu^-\omega$ & $4.7$ 
 \\
 $e^-K_S$ & $2.6$ &
 $e^-K^{* 0}$ & $\phantom{1}3.2$ &
 $e^-\bar K^{* 0}$ & $3.4$ &
 $e^-K^+\pi^-$ & $3.1$ 
 \\
 $\mu^-K_S$ & $2.3$ &
 $\mu^-K^{*0}$ & $\phantom{1}5.9$ &
 $\mu^-\bar K^{*0}$ & $7.0$ &
 $\mu^-K^+\pi^-$ & $4.5$ 
 \\
 $e^-K_SK_S$ & $7.1$ &
 $e^-K^+K^-$ & $\phantom{1}3.4$ &
 $e^-\pi^+\pi^-$ & $2.3$ &
 $e^-\pi^+K^-$ & $3.7$
 \\
 $\mu^-K_SK_S$ & $8.0$ &
 $\mu^-K^+K^-$ & $\phantom{1}4.4$ &
 $\mu^-\pi^+\pi^-$ & $2.1$ &
 $\mu^-\pi^+K^-$ & $8.6$
\\
\multicolumn{3}{l}{$e^-f_0(980)\to e^-\pi^+\pi^-$} & $\phantom{1}3.2$ &
\multicolumn{3}{l}{$\!\!\!\mu^-f_0(980)\to \mu^-\pi^+\pi^-$} & $3.4$
\\[1mm] \hline\hline \\[-4mm]
\multicolumn{6}{l}{$\rm{Br}(Z\to X^0)\cdot 10^{6}\qquad\quad (95\%\;\mathrm{CL})$}
\\[1mm] \hline \\[-4mm]
 $e^\pm\mu^\mp$ & $0.75$ &
 $e^\pm\tau^\mp$ & $9.8$ &
 $\mu^\pm\tau^\mp$ & $12$ &
\\[1mm] \hline\hline \\[-4mm]
\multicolumn{6}{l}{$\rm{Br}(H\to X^0)\cdot 10^{3}\qquad\quad (95\%\;\mathrm{CL})$}
\\[1mm] \hline \\[-4mm]
 $e^\pm\mu^\mp$ & $0.061$ &
 $e^\pm\tau^\mp$ & $4.7$ &
 $\mu^\pm\tau^\mp$ & $2.5$ &
\\[1mm] \hline\hline
&&&\multicolumn{4}{l}{\hskip 1.2cm \vline \hskip .8cm} 
\\[-4mm] 
\multicolumn{3}{l}{$\rm{Br}(\pi^0\to X^0)\cdot 10^{9}\qquad\quad (90\%\;\mathrm{CL})$}
&\multicolumn{5}{l}{\hskip 1.2cm\vline\hskip .8cm $\rm{Br}(K^+\to X^+)\cdot 10^{11}\qquad\quad (90\%\;\mathrm{CL})$}
\\[1mm] \hline &&&\multicolumn{4}{l}{\hskip 1.2cm \vline \hskip .8cm}\\[-4mm]
$\mu^+ e^-$ & $0.38$ & $\mu^- e^+$ & 
\multicolumn{2}{l}{$3.4$ \hskip .61cm\vline\hskip .85cm
$\pi^+\mu^+ e^-$} & $1.3$ & $\pi^+\mu^- e^+$ & $52$
\\[1mm]\hline\hline \\[-4mm] 
\multicolumn{6}{l}{$\rm{Br}(K^0_L\to X^0)\cdot 10^{11}\qquad\quad (90\%\;\mathrm{CL})$}
\\[1mm] \hline \\[-4mm]
$e^\pm\mu^\mp$ & $0.47$ & $e^\pm e^\pm\mu^\mp\mu^\mp$ & $4.12$ &
$\pi^0\mu^\pm e^\mp$ & $7.6$ & $\pi^0\pi^0\mu^\pm e^\mp$ & $17$
\\[1mm] \hline\hline &&&\multicolumn{4}{l}{\hskip 1.2cm \vline \hskip .8cm} 
\\[-4mm]
\multicolumn{3}{l}{$\rm{Br}(B^0_{(s)}\to X^0)\cdot 10^{9}\qquad\quad (90\%\;\mathrm{CL})$}
&\multicolumn{5}{l}{\hskip 1.2cm\vline\hskip .8cm $\rm{Br}(B^+\to X^+)\cdot 10^{9}\qquad\quad (90\%\;\mathrm{CL})$}
\\[1mm] \hline &&&\multicolumn{4}{l}{\hskip 1.2cm \vline \hskip .8cm}
\\[-4mm] 
 $B^0\to e^\pm\mu^\mp$ & $1.0$ &
 $B^0_s\to e^\pm\mu^\mp$ & \multicolumn{2}{l}{$5.4$ 
\hskip .6cm\vline\hskip .85cm 
 $K^+ e^-\mu^+$} & $6.4$ &
 $K^+ e^+\mu^-$ & $7.0$ 
\\[1mm] \hline\hline \\[-4mm]
\end{tabular}\end{table}
%%%%%%%%%%%%%%%%%%%%%%%%%%%%%%%%%%%%%%%%%%%%%%%%%%%%%%%%%%%%%%%%%%%

The fermion masses and the quark mixing matrix $\bV$ are all
generated by the Yukawa couplings in Eq.~\eqn{eq:N_Yukawa}.
However, the complex coefficients $c_{ij}^{(f)}$ are not determined by the gauge symmetry; therefore, we have a large number of arbitrary parameters. A general $N_G\times N_G$
unitary matrix is characterized by $N_G^2$ real parameters: \ $N_G
(N_G-1)/2$ \ moduli and \ $N_G (N_G+1)/2$ \ phases. In the case of
$\,\bV$, many of these parameters are irrelevant because we can
always choose arbitrary quark phases. Under the phase redefinitions
\ $u_i\to \e^{i\phi_i}\, u_i$ \ and \ $d_j\to\e^{i\theta_j}\, d_j$,
the mixing matrix changes as \ $\bV_{\! ij}\to \bV_{\!
ij}\,\e^{i(\theta_j-\phi_i)}$; thus, $2 N_G-1$ phases are
unobservable. The number of physical free parameters in the
quark-mixing matrix then gets reduced to $(N_G-1)^2$: \
$N_G(N_G-1)/2$ moduli and $(N_G-1)(N_G-2)/2$ phases.

In the simpler case of two generations, $\bV$ is determined by a
single parameter. One then recovers the Cabibbo rotation matrix
\cite{cabibbo}
\bel{eq:cabibbo}
\bV\, = \,
\left(\bat \cos{\theta_C} &\sin{\theta_C} \\[2pt] -\sin{\theta_C}& \cos{\theta_C}\ea
\right)\, .
\ee
With $N_G=3$, the CKM matrix is described by three angles and one
phase. Different (but equivalent) representations can be found in
the literature. The Particle data Group \cite{Tanabashi:2018oca} advocates the use
of the following one as the `standard' CKM parametrization:
\beqn\label{eq:CKM_pdg}
\bV & = & 
\left[
\begin{array}{ccc} 1 & 0 & 0\\[2pt] 0 & c_{23} & s_{23}\\[2pt] 0 & -s_{23} & c_{23}\ea
\right]\cdot
\left[\begin{array}{ccc} c_{13} & 0 & s_{13}\, \e^{-i\delta_{13}}\\[2pt] 0 & 1 & 0\\[2pt] -s_{13}\,\e^{i\delta_{13}} & 0 & c_{13}\ea
\right]\cdot
\left[\begin{array}{ccc} c_{12} & s_{12} & 0\\[2pt] -s_{12} & c_{12} & 0\\[2pt] 0 & 0 & 1\ea
\right]
\no\\[5pt] & = &
\left[
\begin{array}{ccc}
c_{12}\, c_{13}  & s_{12}\, c_{13} & s_{13}\, \e^{-i\delta_{13}} \\[2pt]
-s_{12}\, c_{23}-c_{12}\, s_{23}\, s_{13}\, \e^{i\delta_{13}} &
c_{12}\, c_{23}- s_{12}\, s_{23}\, s_{13}\, \e^{i\delta_{13}} &
s_{23}\, c_{13}  \\[2pt]
s_{12}\, s_{23}-c_{12}\, c_{23}\, s_{13}\, \e^{i\delta_{13}} &
-c_{12}\, s_{23}- s_{12}\, c_{23}\, s_{13}\, \e^{i\delta_{13}} &
c_{23}\, c_{13}
\ea
\right] .
\eeqn
Here \ $c_{ij} \equiv \cos{\theta_{ij}}$ \ and \ $s_{ij} \equiv
\sin{\theta_{ij}}\,$, with $i$ and $j$ being generation labels
($i,j=1,2,3$). The real angles $\theta_{12}$, $\theta_{23}$ and
$\theta_{13}$ can all be made to lie in the first quadrant, by an
appropriate redefinition of quark field phases; then, \ $c_{ij}\geq
0\,$, $s_{ij}\geq 0$ \ and \ $0\leq \delta_{13}\leq 2\pi\,$.
Notice that $\delta_{13}$ is the only complex phase in the SM
Lagrangian. Therefore, it is the only possible source of $\cCP$-violation
phenomena. In fact, it was for this reason that the third generation
was assumed to exist \cite{KM:73},
before the discovery of the $b$ and the $\tau$.
With two generations, the SM could not explain the observed
$\cCP$ violation in the $K$ system.

\section{Lepton decays}

%%%%%%%%%%%%%%%  FIGURE %%%%%%%%%%%%%%%%%%%%%%%%%
\begin{figure}[tbh]\centering
\begin{minipage}[t]{.35\linewidth}\centering
\includegraphics[height=3cm]{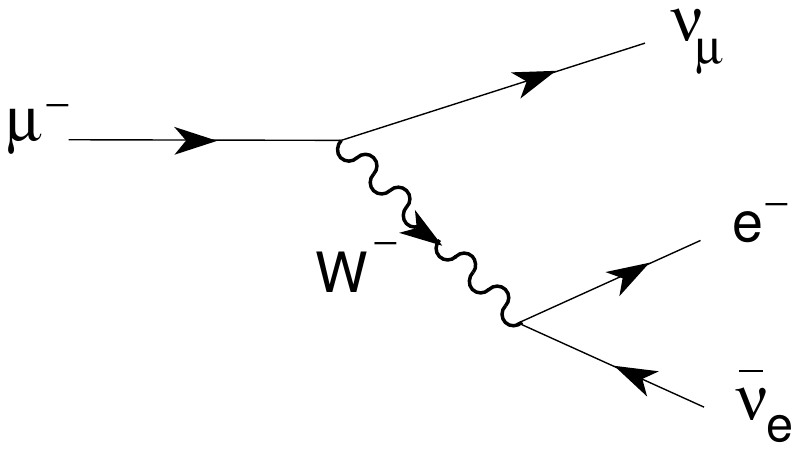} %%%ps}
\end{minipage}
\hskip 1.5cm
\begin{minipage}[t]{.45\linewidth}\centering
\includegraphics[height=3cm]{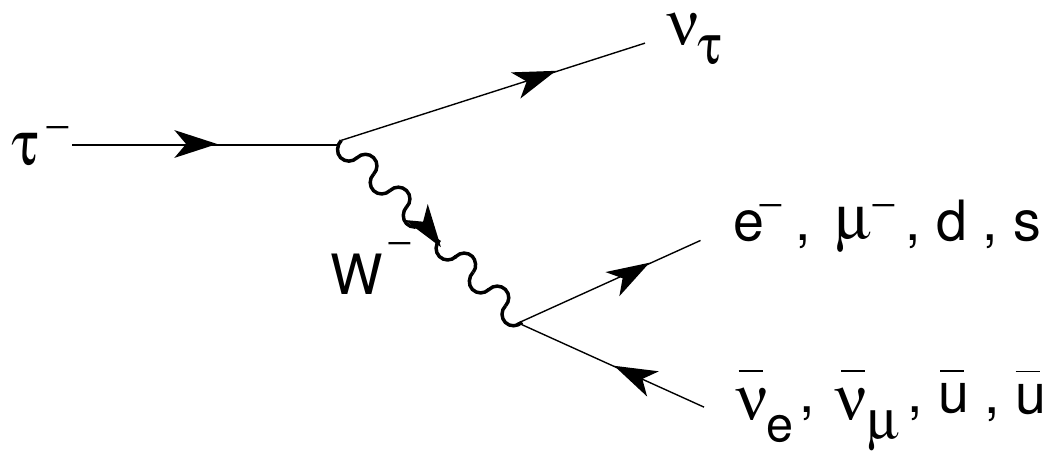} %%%ps}
\end{minipage}
\caption{Tree-level Feynman diagrams for \ $\mu^-\to e^-\bar\nu_e\,\nu_\mu$ \
and \ $\tau^-\to\nu_\tau X^-$ ($X^-=e^-\bar\nu_e,\, \mu^-\bar\nu_\mu,\,
d\bar u,\, s\bar u$).}
\label{fig:mu_decay}
\end{figure}
%%%%%%%%%%%%%%%%%%%%%%%%%%%%%%%%%%%%%%%%%%%%%

The simplest flavour-changing process is the leptonic
decay of the muon, which proceeds through the $W$-exchange
diagram shown in Fig.~\ref{fig:mu_decay}.
The momentum transfer carried by the intermediate $W$ is very small
compared to $M_W$. Therefore, the vector-boson propagator reduces
to a contact interaction,
\bel{eq:low_energy}
{-g_{\mu\nu} + q_\mu q_\nu/M_W^2 \over q^2-M_W^2}\quad\;
 \toLow\quad\; {g_{\mu\nu}\over M_W^2}\, .
\ee
The decay can then be described through an effective local
four-fermion Hamiltonian,
\bel{eq:mu_v_a}
\cH_{\mathrm{\scriptstyle eff}}\, = \,
{G_F \over\sqrt{2}}
\left[\bar e\gamma^\alpha (1-\gamma_5) \nu_e\right]\,
\left[ \bar\nu_\mu\gamma_\alpha (1-\gamma_5)\mu\right]\, ,
\ee
where
\bel{eq:G_F}
{G_F\over\sqrt{2}} \, =\, {g^2\over 8 M_W^2}\, =\,\frac{1}{2 v^2}
\ee
is called the Fermi coupling constant.
$G_F$ is fixed by the total decay width,
\bel{eq:mu_lifetime}
{1\over\tau_\mu}\, = \, \Gamma[\mu^-\to e^-\bar\nu_e\nu_\mu\, (\gamma)]
\, = \, {G_F^2 m_\mu^5\over 192 \pi^3}\,
\left( 1 + \delta_{\mathrm{\scriptstyle RC}}\right) \,
f\left(m_e^2/m_\mu^2\right) \, ,
\ee
where
$\, f(x) = 1-8x+8x^3-x^4-12x^2\ln{x}\, ,$
and \
$\delta_{\mathrm{\scriptstyle RC}} \approx {\alpha\over 2\pi} ({25\over 4}-\pi^2)$ \
%%
%\bel{eq:qed_corr}
%1+\delta_{\mathrm{\scriptstyle RC}} \, =\,
%\left[1+{\alpha\over 2\pi}\left({25\over 4}-\pi^2\right)\right]\,
%\left[ 1 +{3\over 5}{m_\mu^2\over M_W^2} - 2 {m_e^2\over M_W^2}\right]
%\, +\,\cdots
%% =  0.9958 \,
%\ee
%%
takes into account higher-order QED corrections, which are known
to $\cO(\alpha^2)$ \cite{MS:88,vRS:99,Pak:2008qt}. The tiny neutrino masses can be safely neglected. The measured lifetime \cite{Tishchenko:2012ie},
$\tau_\mu=(2.196\, 981\, 1\pm 0.000\, 002\, 2)\cdot 10^{-6}$ s,
implies the value
\bel{eq:gf}
G_F\, = \, (1.166\, 378\, 7\pm 0.000\, 000\, 6)\cdot 10^{-5} \:\mbox{\rm GeV}^{-2}
\,\approx\, {1\over (293 \:\mbox{\rm GeV})^2} \, .
\ee

The decays of the $\tau$ lepton proceed through the same $W$-exchange mechanism.
The only difference is that several final states are kinematically
allowed:
$\tau^-\to\nu_\tau e^-\bar\nu_e$,
$\tau^-\to\nu_\tau\mu^-\bar\nu_\mu$,
$\tau^-\to\nu_\tau d\bar u$ and $\tau^-\to\nu_\tau s\bar u$.
Owing to the universality of the $W$ couplings in $\cL_{\rms CC}$, all these
decay modes have equal amplitudes (if final fermion masses and
QCD interactions are neglected), except for an additional
$N_C |\bV_{\! ui}|^2$ factor ($i=d,s$) in the semileptonic
channels, where $N_C=3$ is the number of quark colours.
Making trivial kinematical changes in Eq.~\eqn{eq:mu_lifetime},
one easily gets the lowest-order prediction for the total
$\tau$ decay width:
\bel{eq:tau_decay_width}
{1\over\tau_\tau}\equiv\Gamma(\tau) \approx
\Gamma(\mu) \left({m_\tau\over m_\mu}\right)^5
\left\{ 2 + N_C
\left( |\bV_{\! ud}|^2 + |\bV_{\! us}|^2\right)\right\}
\approx {5\over\tau_\mu}\left({m_\tau\over m_\mu}\right)^5
  ,
\ee
where we have used the CKM unitarity relation
$|\bV_{\! ud}|^2 + |\bV_{\! us}|^2 = 1 - |\bV_{\! ub}|^2
\approx 1$
(we will see later that this is an excellent approximation).
From the measured muon lifetime, one has then
$\tau_\tau\approx 3.3\cdot 10^{-13}$~s, to be compared
with the experimental value
$\tau_\tau^{\mathrm{exp}} = (2.903\pm 0.005)\cdot 10^{-13}$~s \cite{Tanabashi:2018oca}.
The numerical difference is due to the effect of QCD corrections, which
enhance the hadronic $\tau$ decay width by about 20\%.
The size of these corrections has been accurately predicted
in terms of the strong coupling \cite{BNP:92}, allowing us
to extract from $\tau$ decays one of the most precise determinations of $\alpha_s$ \cite{Pich:2016bdg,Pich:2016yfh}.

In the SM all lepton doublets have identical couplings to the $W$ boson. Comparing
the measured decay widths of leptonic or semileptonic decays which only differ in the lepton flavour, one can test experimentally that
the W interaction is indeed the same, \ie that $g_e = g_\mu=g_\tau\equiv g$.
As shown in Table~\ref{tab:ccuniv}, the present data verify the universality of the leptonic charged-current couplings to the 0.2\% level.

%%%%%%%%%%%%%%%%%%%%  TABLES  %%%%%%%%%%%%%%%%%
\begin{table}[t]\centering
\caption{Experimental determinations of the ratios \ $g_\ell/g_{\ell'}$
\cite{Tanabashi:2018oca,Pich:2013lsa}.}
\renewcommand{\arraystretch}{1.2}
\begin{tabular}{c@{\hspace{0.9cm}}ccccc}
 \hline\hline &
%%% Present constraints on $|g_\mu/g_e|$
 $\Gamma_{\tau\to\mu}/\Gamma_{\tau\to e}$ &
 $\Gamma_{\pi\to\mu} /\Gamma_{\pi\to e}$ &
 $\Gamma_{K\to\mu} /\Gamma_{K\to e}$ &
 $\Gamma_{K\to\pi\mu} /\Gamma_{K\to\pi e}$ &
 $\Gamma_{W\to\mu} /\Gamma_{W\to e}$
 \\ \hline
 $|g_\mu/g_e|$
 & $1.0017\; (16)$ & $1.0010\; (9)$ & $0.9978\; (18)$ & $1.0010\; (25)$ & $0.993\; (7)$
 \\ \hline\hline &
%%% Present constraints on $|g_\tau/g_\mu|$
 $\Gamma_{\tau\to e}/\Gamma_{\mu\to e}$ &
 $\Gamma_{\tau\to\pi}/\Gamma_{\pi\to\mu}$ &
 $\Gamma_{\tau\to K}/\Gamma_{K\to\mu}$ &
 $\Gamma_{W\to\tau}/\Gamma_{W\to\mu}$
 \\ \hline
 %%%$\left|\frac{g_\tau}{g_\mu}\right|$
 $|g_\tau/g_\mu|$
 & $1.0011\; (14)$ & $0.9965\; (26)$ & $0.9860\; (73)$ & $1.034\; (13)$
 \\ \hline\hline &
%%% Present constraints on $|g_\tau/g_e|$
 $\Gamma_{\tau\to\mu}/\Gamma_{\mu\to e}$
 & $\Gamma_{W\to\tau}/\Gamma_{W\to e}$
 \\ \hline
 $|g_\tau/g_e|$
 & $1.0028\; (14)$ & $1.021\; (12)$
 \\ \hline\hline
\end{tabular}
\label{tab:ccuniv}
\end{table}
%%%%%%%%%%%%%%%%%%%%%%%%%%%%%%%%%%%%%%%%%%%%%%%%

\section{Quark mixing}

In order to measure the CKM matrix elements $\bV_{\! ij}$, one needs to study hadronic weak decays of the type
\ $H\to H'\, \ell^- \bar\nu_\ell$ \ or \ $H'\to H\, \ell^+ \nu_\ell$ that are
associated with the corresponding quark transitions $d_j\to u_i\,
\ell^-\bar\nu_\ell$ \  and \ $u_i\to d_j\, \ell^+\nu_\ell$\
(Fig.~\ref{fig:cDecay}). Since quarks are confined within hadrons,
the decay amplitude
\bel{eq:T_decay}
T[H\to H'\, \ell^- \bar\nu_\ell]\; = \; {G_F\over\sqrt{2}} \;\bV_{\! ij}\;\,
\langle H'|\, \bar u_i\, \gamma^\mu (1-\gamma_5)\, d_j\, | H\rangle \;\,
\left[\,\bar \ell \,\gamma_\mu (1-\gamma_5) \,\nu_\ell\,\right]
\ee
always involves an hadronic matrix element of the weak left current.
The evaluation of this matrix element is a non-perturbative QCD
problem, which introduces unavoidable theoretical uncertainties.

%%%%%%%%%%%%%%%  FIGURE %%%%%%%%%%%%%%%%%%%%%%%%%
\begin{figure}[tb]\centering
\includegraphics[width=12cm]{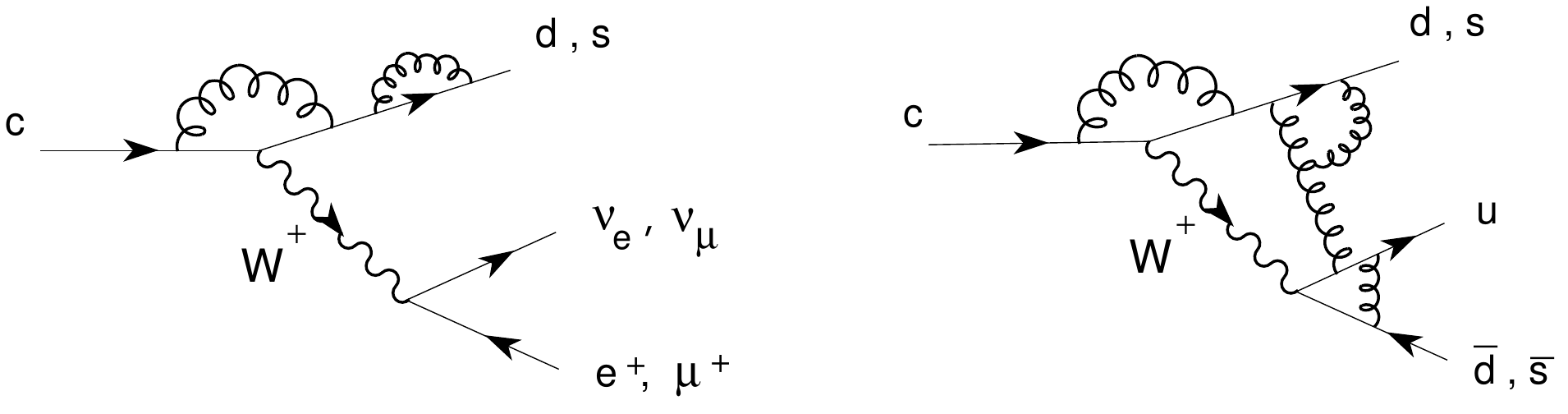} %%%ps}
\vskip -.3cm
\caption{
$\bV_{\! ij}$ are measured
in semileptonic decays (left), where
a single quark current is present.
Hadronic decays (right) involve two
different quark currents and are more affected
by QCD effects (gluons can couple everywhere).}
\label{fig:cDecay}
\end{figure}
%%%%%%%%%%%%%%%%%%%%%%%%%%%%%%%%%%%%%%%%%%%%%%%%%

One usually looks for a semileptonic transition where the matrix
element can be fixed at some specific kinematic point by a symmetry
principle. This has the virtue of reducing the theoretical
uncertainties to the level of symmetry-breaking corrections and
kinematic extrapolations. The standard example is a \ $0^-\to 0^-$
\ decay such as \ $K\to\pi \ell\nu_\ell\,$, $D\to K \ell\nu_\ell$ \ or \ $B\to D\ell\nu_\ell\,$, where, owing to parity (the vector and axial-vector currents have $J^{\mathcal{P}}=1^-$ and $1^+$, respectively), only the vector current contributes. The most general Lorentz decomposition of the hadronic matrix element contains two terms:
\bel{eq:vector_me}
\langle P'(k')| \,\bar u_i \,\gamma^\mu\, d_j\, | P(k)\rangle \; = \; C_{PP'}\,
\left\{\, (k+k')^\mu\, f_+(t)\, +\, (k-k')^\mu\, f_-(t)\,\right\}\, .
\ee
Here, $C_{PP'}$ is a Clebsh--Gordan factor relating $P\to P'$ transitions that only differ by the meson electromagnetic charges, and $t=(k-k')^2\equiv
q^2$ is the momentum transfer. The unknown strong dynamics is fully contained in the form factors $f_\pm(t)$. 

In the limit of equal quark masses,
$m_{u_i}=m_{d_j}$, the divergence of the vector current is zero.
Thus \ $q_\mu\left[\bar u_i \gamma^\mu d_j\right] = 0$, which
implies \ $f_-(t)=0$. Moreover, as shown in the appendix, $f_+(0)=1$ \ to all orders in the strong coupling because the associated flavour charge is a
conserved quantity.\footnote{
%%%%%%%% FOOTNOTE %%%%%%%%%%%%%%%%%%%%%%%
This is completely analogous to the electromagnetic charge conservation in QED.
The conservation of the electromagnetic current implies that the proton electromagnetic form factor does not get any QED or QCD correction at $q^2=0$ and, therefore,
$Q_p=2\, Q_u+Q_d=|Q_e|$. An explicit proof can be found in
Ref.~\cite{PI:96}.}
%%%%%%%% END OF FOOTNOTE %%%%%%%%%%%%%%%%%
Therefore, one only needs to estimate the corrections induced by the
quark mass differences.

Since $q_\mu\,\left[\bar \ell\gamma^\mu (1-\gamma_5)\nu_\ell\right]\sim m_\ell$, the contribution of $f_-(t)$ is kinematically suppressed in the electron and muon decay modes. The decay width can then be written as ($\ell=e,\mu$)
\bel{eq:decay_width}
\Gamma(P\to P' \ell \nu)
\; =\; {G_F^2 M_P^5\over 192\pi^3}\; |\bV_{\! ij}|^2\; C_{PP'}^2\;
|f_+(0)|^2\; \cI\; \left(1+\delta_{\rms RC}\right)\, ,
\ee
where $\delta_{\rms RC}$ is an electroweak radiative
correction factor and $\cI$ denotes a phase-space integral,
which in the limit $m_\ell=0$  takes the form
\bel{eq:ps_integral}
\cI\;\approx\;\int_0^{(M_P-M_{P'})^2} {dt\over M_P^8}\;
\lambda^{3/2}(t,M_P^2,M_{P'}^2)\;
\left| {f_+(t)\over f_+(0)}\right|^2\, .
\ee
The usual procedure to determine $|\bV_{\! ij}|$ involves three steps:
\begin{enumerate}
\item Measure the shape of the $t$ distribution. This fixes %the ratio
$|f_+(t)/f_+(0)|$ and therefore determines $\cI$.
\item Measure the total decay width $\Gamma$. Since $G_F$ is already known
from $\mu$ decay, one gets then an experimental value for the product $|f_+(0)\, \bV_{\! ij}|$, provided the radiative correction $\delta_{\rms RC}$ is known to the needed accuracy.
\item Get a theoretical prediction for $f_+(0)$.
\end{enumerate}
It is important to realize that theoretical input is always needed.
Thus, the accuracy of the $|\bV_{\! ij}|$ determination is limited
by our ability to calculate the relevant hadronic parameters and radiative corrections.

\subsection{Determination of $|\bV_{\! ud}|$ and $|\bV_{\! us}|$}

The conservation of the vector %%% and axial-vector
QCD currents in the massless quark limit allows for precise determinations of the light-quark mixings.
%%% $|\bV_{\! ud}|$ and $|\bV_{\! us}|$.
%
The most accurate measurement of $\bV_{\! ud}$ is done with
superallowed nuclear $\beta$ decays of the Fermi type ($0^+\to 0^+$),
where the nuclear matrix element
$\langle N'|\bar u\gamma^\mu d|N\rangle$
can be fixed by vector-current conservation.
The CKM factor is obtained through the relation \cite{Hardy:2014qxa}
%%%MS:06,CMS:04},
%
\bel{eq:ft_value}
|\bV_{\! ud}|^2\, =\,
\frac{\pi^3\ln{2}}{ft\, G_F^2 m_e^5\, (1+\delta_{\mathrm{\scriptstyle  RC}})}
\, = \,
\frac{(2984.430\pm 0.003)\,\mbox{\rm s}}{ft\,(1+\delta_{\mathrm{\scriptstyle  RC}})} \, = \,
\frac{(2984.430\pm 0.003)\,\mbox{\rm s}}{\mathcal{F}t\,(1+\Delta_R^V)}  
\, ,
\ee
where $ft$ denotes the product of a phase-space statistical decay-rate factor  and the measured half-life of the transition.
In order to determine $|\bV_{\! ud}|$, one needs to perform a careful
analysis of radiative corrections, including electroweak contributions, nuclear-structure corrections and isospin-violating nuclear effects.
The nucleus-dependent corrections, which are reabsorbed into an effective
nucleus-independent $\mathcal{F}t$-value, 
%%% are quite large ($\sim 3$--4\%)
%%% $\delta_{\mathrm{\scriptstyle RC}}\sim 3$--4\%, 
have a crucial role in bringing the results from different nuclei into good agreement. The weighted average of the fourteen most precise determinations yields $\mathcal{F}t= (3072.07\pm 0.63)$~s \cite{Hardy:2014qxa,Hardy:2018zsb}.
The remaining universal correction $\Delta_R^V$ is sizeable and its previously accepted value \cite{Marciano:2005ec} has been questioned by recent 
re-evaluations \cite{Seng:2018yzq,Seng:2020wjq}. Taking $\Delta_R^V = 0.02426 \pm 0.00032$ \cite{Czarnecki:2019mwq}, one gets 
\bel{eq:Vud}
|\bV_{\! ud}|\, =\, 0.97389\pm 0.00018 \, .
\ee
%
%The error is dominated by theoretical uncertainties stemming from nuclear Coulomb distortions and radiative corrections.

An independent determination of $|\bV_{\! ud}|$ can be obtained from neutron decay, $n\to p\, e^-\bar\nu_e$.
The axial current also contributes in this case; therefore, one needs
to use the experimental value of the axial-current matrix element at $q^2=0$,
$\langle p\, | \,\bar u\gamma^\mu\gamma_5 d\, |\, n\rangle  =
G_A \;\bar p \gamma^\mu n$, which can be extracted from the distribution of the neutron decay products. Using the current world averages,
$g_A\equiv G_A/G_V = -1.2732\pm 0.0023$ \ and \
$\tau_n = (879.4\pm 0.6)$~s  \cite{Tanabashi:2018oca}, and the estimated radiative corrections \cite{Czarnecki:2019mwq}, one gets
\bel{eq:n_decay}
|\bV_{\! ud}|\, =\,
\left\{ {(4906.4\pm 1.7)\,\mbox{\rm s}\over \tau_n\, (1+3 g_A^2)}\right\}^{1/2}
\, =\, 0.9755\pm 0.0015 \, ,
\ee
which is $1.1\,\sigma$ larger than \eqn{eq:Vud} but less precise. 
The uncertainty on the input value of $g_A$ has been inflated because the most recent and accurate measurements of $g_A$ disagree with the older experiments. Using instead the post-2002 average $g_A = -1.2762 \pm 0.0005$ \cite{Czarnecki:2019mwq}, results in $|\bV_{\! ud}| = 0.9736\pm 0.0005$; three times more precise and in better agreement with \eqn{eq:Vud}.

The pion $\beta$ decay $\pi^+\to\pi^0 e^+\nu_e$ offers a
cleaner way to measure
$|\bV_{\! ud}|$. It is a pure vector transition, with very small theoretical
uncertainties. At $q^2=0$, the hadronic matrix element
does not receive isospin-breaking contributions of first order in
$m_d-m_u$, \ie $f_+(0)= 1 + \cO[(m_d-m_u)^2]$ \cite{Ademollo:1964sr}.
The small available phase space makes it possible to theoretically
control the form factor with high accuracy over the entire kinematical
domain \cite{CKNP:03}; unfortunately, it also implies a very suppressed
branching fraction of $\cO(10^{-8})$.
From the currently measured value \cite{PiBeta:04},
%%% Br$(\pi^+\to\pi^0 e^+\nu_e)=(1.040\pm 0.006)\times 10^{-8}$, 
one gets $|\bV_{\! ud}| = 0.9749\pm 0.0026$ \cite{Tanabashi:2018oca}.
A tenfold improvement of the experimental accuracy would be needed to get
a determination competitive with \eqn{eq:Vud}.

%%% V_us

The standard determination of $|\bV_{\! us}|$ takes advantage of the
theoretically well-understood decay amplitudes in $K\to\pi\ell\nu_\ell$.
%%% $K_{\ell 3}$ 
The high accuracy achieved in high-statistics experiments \cite{Tanabashi:2018oca}, supplemented with theoretical calculations of electromagnetic and
isospin corrections \cite{CGN:08,KN:08}, allows us to extract the product
$|\bV_{\! us}\, f_+(0)| = 0.2165\pm 0.0004$ \cite{Antonelli:2010yf,Moulson:2014cra}, with $f_+(0)= 1 + \cO[(m_s-m_u)^2]$ the vector form factor of the
$K^0\to\pi^-\ell^+\nu_\ell$ decay  \cite{Ademollo:1964sr,BS:60}. The exact value of $f_+(0)$ has been thoroughly investigated since the first precise estimate by Leutwyler and Roos, $f_+(0) = 0.961\pm 0.008$ \cite{LR:84}. The most recent and precise lattice determinations exhibit a clear shift to higher values \cite{Bazavov:2013maa,Carrasco:2016kpy}, in agreement with the analytical chiral perturbation theory predictions at two loops \cite{BT:03,JOP:04,CEEKPP:05}. Taking the current lattice average (with $2+1+1$ active fermions), $f_+(0) = 0.9706\pm 0.0027$ \cite{Aoki:2019cca}, one obtains 
\bel{eq:Vus}
|\bV_{\! us}|\, =\, 0.2231\pm 0.0007 \, .
\ee

The ratio of radiative inclusive decay rates
$\Gamma[K\to\mu\nu (\gamma)]/\Gamma[\pi\to\mu\nu (\gamma)]$
provides also information on $\bV_{\! us}$ \cite{Antonelli:2010yf,MA:04}.
With a careful treatment ot electromagnetic and isospin-violating corrections, one extracts $|\bV_{\! us}/\bV_{\! ud}|\, |f_K/f_\pi| = 0.2760 \pm 0.0004$ \cite{Moulson:2014cra,CN:11,Cirigliano:2011ny}.
Taking for the ratio of meson decay constants the lattice average
$f_K/f_\pi = 1.1932 \pm 0.0019$ \cite{Aoki:2019cca}, one finally gets
\bel{eq:Vus_Vud}
\frac{|\bV_{\! us}|}{|\bV_{\! ud}|} \, =\, 0.2313 \pm 0.0005\, .
\ee
With the value of $|\bV_{\! ud}|$ in Eq.~\eqn{eq:Vud}, this implies
$|\bV_{\! us}| = 0.2253 \pm 0.0005$ that is $2.6\,\sigma$ larger than \eqn{eq:Vus}.

Hyperon decays are also sensitive to $\bV_{\! us}$ \cite{CSW:03}.
Unfortunately, in weak baryon decays the theoretical control on $SU(3)$-breaking corrections is not as good as for the meson case.
A conservative estimate of these effects leads to the result
$|\bV_{\! us}| = 0.226 \pm 0.005$ \cite{MP:05}.

The accuracy of all previous determinations is limited by theoretical uncertainties. The ratio of the inclusive $\Delta S = 0$
($\tau^-\to\nu_\tau \bar u d$) and $|\Delta S| = 1$ ($\tau^-\to\nu_\tau \bar u s$) tau decay widths provides a very clean observable to directly measure $|\bV_{\! us}|$ \cite{GJPPS:05,Pich:2013lsa}
because $SU(3)$-breaking corrections are suppressed by two powers of the
$\tau$ mass. The present $\tau$ decay data imply
$|\bV_{\! us}| = 0.2195 \pm 0.0019$ \cite{Amhis:2019ckw}, which is $1.8\,\sigma$ lower than \eqn{eq:Vus}, and  $3.0\,\sigma$ lower than the value extracted from Eqs.~\eqn{eq:Vus_Vud} and \eqn{eq:Vud}.

%the error being dominated by the experimental uncertainties.
%The central value has been shifted down by the BaBar and Belle measurements, which find branching ratios smaller than previous world averages for many $\tau$ decay modes \cite{Tanabashi:2018oca}. More precise data are needed to clarify this worrisome effect.
%From the measured $K^-\to\mu^-\bar\nu_\mu$ and $K\to\pi\ell\nu_\ell$ decay amplitudes, one actually predicts slightly higher $\tau^-\to\nu_\tau K^-$ and $\tau^-\to \nu_\tau (K\pi)^-$ branching ratios \cite{Pich:2013lsa,Antonelli:2013usa}. Since these channels are the largest contributions to the inclusive $|\Delta S| = 1$ tau decay width, their slight underestimate has a very significant effect on $|\bV_{\! us}|$. Using the kaon determinations for these $\tau$ branching ratios one gets instead $|\bV_{\! us}| = 0.2213 \pm 0.0023$.
%%If the strangeness-changing $\tau$ decay width is measured with a 1\% precision, the resulting $\bV_{\! us}$ uncertainty will get reduced to
%%around 0.6\%, \ie $\pm 0.0013$.

\subsection{Determination of $|\bV_{\! cb}|$ and $|\bV_{\! ub}|$}

In the limit of very heavy quark masses, QCD has additional flavour and spin
symmetries \cite{IW:89,GR:90,EH:90,GE:90} that can be used to make
precise determinations of $|\bV_{\! cb}|$, either from
exclusive semileptonic decays such as $B\to D \ell \bar\nu_\ell$ and $B\to D^* \ell\bar\nu_\ell$
\cite{NE:91,LU:90} or from the inclusive analysis of $b\to c\, \ell\,\bar\nu_\ell$ transitions. In the rest frame of a heavy-light meson $\bar Q q$, with $M_Q\gg (m_q, \Lambda_{\mathrm{QCD}})$, the heavy quark $Q$ is practically at rest and acts as a static source of gluons ($\lambda_Q\sim 1/M_Q\ll R_{\rms{had}}\sim 1/\Lambda_{\mathrm{QCD}}$). At $M_Q\to\infty$, the interaction becomes then independent of the heavy-quark mass and spin. Moreover, assuming that the charm quark is heavy enough, the $b\to c\ell\bar\nu_\ell$ transition within the meson does not modify the interaction with the light quark at zero recoil, \ie when the meson velocity remains unchanged ($v_D = v_B$).

Taking the limit $m_b > m_c\to\infty$, all form factors
characterizing the decays $B\to D \ell \bar\nu_\ell$ and $B\to D^* \ell\bar\nu_\ell$
reduce to a single function \cite{IW:89}, which depends on the product of the
four-velocities of the two mesons
$w\equiv v_B^{\phantom{*}}\cdot v_{D^{(*)}}
= (M_B^2 + M_{D^{(*)}}^2 - q^2) / (2 M_B M_{D^{(*)}})$.
Heavy quark symmetry determines the normalization of the rate at $w = 1$, the
maximum momentum transfer to the leptons, because the corresponding vector current
is conserved in the limit of equal $B$ and $D^{(*)}$ velocities.
The $B\to  D^*$ mode has the additional advantage that
corrections to the infinite-mass limit are of second order in $1/m_b - 1/m_c$
at zero recoil ($w = 1$) \cite{LU:90}.

The exclusive determination of $|\bV_{\! cb}|$ is obtained from an extrapolation of the measured spectrum to $w = 1$.
Using the CLN parametrization of the relevant form factors \cite{Caprini:1997mu}, which is based on heavy-quark symmetry and includes $1/M_Q$ corrections, the {\it Heavy Flavor Averaging group} (HFLAV) \cite{Amhis:2019ckw} %%%{Amhis:2019ckw} 
quotes the experimental value
$\eta_{\rms{EW}}\, {\cal F}(1)\, |\bV_{\! cb}| = (35.27\pm 0.38)\cdot 10^{-3}$ \ from $B\to D^* \ell\bar\nu_\ell$ data, while
the measured $B\to D \ell \bar\nu_\ell$ distribution results in
$\eta_{\rms{EW}}\, {\cal G}(1)\, |\bV_{\! cb}| = (42.00\pm 1.00)\cdot 10^{-3}$, where ${\cal F}(1)$ and ${\cal G}(1)$ are the corresponding form factors at $w=1$ and $\eta_{\rms{EW}}$ accounts for small electroweak corrections.
Lattice simulations are used to estimate the deviations from unity of the two form factors at zero recoil. Using 
$\eta_{\rms{EW}}\, {\cal F}(1) = 0.910\pm 0.013$ \cite{Aoki:2019cca} and
$\eta_{\rms{EW}}\, {\cal G}(1)= 1.061 \pm 0.010$ \cite{Lattice:2015rga},
one gets \cite{Amhis:2019ckw}
\bel{eq:Vcb_Excl}
|\bV_{\! cb}|\, =\, \left\{
\bat (38.76\pm 0.42_{\rms{exp}}\pm 0.55_{\rms{th}})\cdot 10^{-3} & \quad (B\to D^* \ell\bar\nu_\ell)\\[3pt]
(39.58\pm 0.94_{\rms{exp}}\pm 0.37_{\rms{th}})\cdot 10^{-3} & \quad (B\to D \ell\bar\nu_\ell) \ea\right.
\; =\; (39.02\pm 0.57)\cdot 10^{-3} \, .
\ee

It has been pointed out recently that the CLN parametrization is only valid within 2\% and this uncertainty has not been properly taken into account in the experimental extrapolations \cite{Bigi:2016mdz,Bigi:2017njr,Grinstein:2017nlq,Bernlochner:2017jka}. Using instead the more general BGL parametrization \cite{Boyd:1997kz}, combined with lattice and light-cone sum rules information, the analysis of the most recent $B\to D^* \ell\bar\nu_\ell$ Belle data \cite{Abdesselam:2017kjf,Waheed:2018djm} gives  
\cite{Gambino:2019sif} %%% \cite{Bigi:2017njr}
\bel{eq:Vcb_Excl2}
|\bV_{\! cb}|\, =\, (39.6\,{}^{+\: 1.1}_{-\: 1.0} )\cdot 10^{-3} \, ,
\ee
while a similar analysis of BaBar \cite{Aubert:2009ac} and Belle \cite{Glattauer:2015teq} $B\to D \ell\bar\nu_\ell$ data obtains \cite{Bigi:2016mdz}
\bel{eq:Vcb_Excl3}
|\bV_{\! cb}|\, =\, (40.49\pm 0.97)\cdot 10^{-3} \, .
\ee
These numbers are significantly higher than the corresponding HFLAV results in Eq.~\eqn{eq:Vcb_Excl} and indicate the presence of underestimated uncertainties. 

The inclusive determination of $|\bV_{\! cb}|$ uses the Operator Product Expansion \cite{BI:93,MW:94}
to express the total $b\to c\, \ell\,\bar\nu_\ell$ rate and moments of the
differential energy and invariant-mass spectra in a double expansion in powers
of $\alpha_s$ and $1/m_b$, which includes terms of $\cO(\alpha_s^2)$ and up to $\cO(1/m_b^5)$ \cite{Gremm:1996df,BBMU:03,GU:04,BLMT:04,BBU:05,BF:06,Pak:2008qt,Mannel:2010wj,Gambino:2011cq,Alberti:2012dn,Alberti:2013kxa,Mannel:2015jka}.  
The non-perturbative matrix elements of the corresponding local operators
are obtained from a global fit to experimental moments of inclusive
%%% $B\to X_c\, \ell\,\bar\nu_\ell$  
lepton energy and hadronic invariant mass distributions. 
%%%Some fits include in addition $B\to X_s\gamma$ (photon energy spectrum) . 
The most recent analyses find \cite{Alberti:2014yda,Gambino:2016jkc}
\bel{eq:Vcb_Incl}
|\bV_{\! cb}|\, =\, (42.00\pm 0.64)\cdot 10^{-3} \, .
\ee
This value, which we will adopt in the following, agrees within errors with the exclusive $B\to D$ determination in  \eqn{eq:Vcb_Excl3} and it is only
$1.9\,\sigma$ away from the $B\to D^*$ value in Eq.~\eqn{eq:Vcb_Excl2}. 

%At present there is a $2.1\,\sigma$ discrepancy between the exclusive and inclusive
%determinations. Following the PDG prescription \cite{Tanabashi:2018oca}, we average both values scaling the error by $\sqrt{\chi^2/\mathrm{dof}}=2.1$:
%%
%\bel{eq:Vcb}
%|\bV_{\! cb}|\, =\, (40.8\pm 1.1)\cdot 10^{-3} \, .
%\ee
%%

The presence of a light quark makes more difficult to control the theoretical uncertainties in the analogous determinations of $|\bV_{\! ub}|$. Exclusive $B\to\pi\ell\nu_\ell$ decays involve a non-perturbative form factor $f_+(t)$ which is estimated through light-cone sum rules
\cite{BZ:05,DKMMO:08,DMOW:11,Bharucha:2012wy} and lattice simulations \cite{Lattice:2015tia,Flynn:2015mha}.
The inclusive measurement requires the use of stringent experimental cuts to suppress the $b\to X_c\ell\nu_\ell$ background that has fifty times larger rates. This induces sizeable errors in the theoretical predictions \cite{Antonelli:2009ws,BLNP:05,AG:06,GGOU:07,ALFR:09,GNP:10,PA:09,LLM:10,GK:10},
which become sensitive to non-perturbative shape functions and depend
much more strongly on $m_b$. The HFLAV group quotes the values \cite{Amhis:2019ckw}
\bel{eq:Vub}
|\bV_{\! ub}|\, =\, \left\{
\bat (3.67\pm 0.09_{\rms{exp}}\pm 0.12_{\rms{th}})\cdot 10^{-3} & \quad (B\to \pi \ell\bar\nu_\ell)\\[3pt]
(4.32\pm 0.12_{\rms{exp}}\,{}^{+\: 0.12}_{-\: 0.13}{}_{\,\rms{th}})\cdot 10^{-3} & \quad (B\to X_u \ell\bar\nu_\ell) \ea\right.
\; =\;\; (3.95\pm 0.32)\cdot 10^{-3} \, .
\ee
Since the exclusive and inclusive determinations of $|\bV_{\! ub}|$ disagree, we have averaged both values scaling the error by $\sqrt{\chi^2/\mathrm{dof}}=2.8$.

LHCb has extracted $|\bV_{\! ub}|/|\bV_{\! cb}|$ from the measured ratio of high-$q^2$ events between the $\Lambda_b$ decay modes into \ $
p\mu\nu$ \ ($q^2>15\:\mathrm{GeV}^2$) \ and \ $\Lambda_{c\,}\mu\nu$ \ ($q^2>7\:\mathrm{GeV}^2$) \cite{Aaij:2015bfa,Amhis:2019ckw}:
\bel{eq:VubVcb}
\frac{|\bV_{\! ub}|}{|\bV_{\! cb}|}\, =\, 0.079\pm 0.004_{\rms{exp}} \pm 0.004_{\rms{FF}}\, ,
\ee
where the second error is due to the limited knowledge of the relevant form factors. This ratio is compatible with the values of $|\bV_{\! cb}|$ and $|\bV_{\! ub}|$ in Eqs.~\eqn{eq:Vcb_Incl} and \eqn{eq:Vub}, at the $1.6\,\sigma$ level.

$|\bV_{\! ub}|$ can be also extracted from the $B^-\to\tau^-\bar\nu_\tau$ decay width, taking the B-meson decay constant $f_B$ from lattice calculations \cite{Aoki:2019cca}. Unfortunately, the current tension between the BaBar \cite{Aubert:2007xj} and Belle \cite{Adachi:2012mm} measurements does not allow for a very precise determination. The particle data group quotes $|\bV_{\! ub}| = (4.01\pm 0.37)\cdot 10^{-3}$ \cite{Tanabashi:2018oca}, which agrees with either the exclusive or inclusive values in Eq.~\eqn{eq:Vub}.

\subsection{Determination of the charm and top CKM elements}

The analytic control of theoretical uncertainties is more difficult in semileptonic charm decays, because the symmetry arguments associated with the light and heavy quark limits get corrected by sizeable symmetry-breaking effects. The magnitude of $|\bV_{\! cd}|$ can be extracted from
$D\to\pi\ell\nu_\ell$ and $D\to\ell\nu_\ell$ decays, while $|\bV_{\! cs}|$ is obtained from $D\to K\ell\nu_\ell$ and $D_s\to \ell\nu_\ell$, using the lattice determinations of the relevant form factor normalizations and decay constants
\cite{Aoki:2019cca}. The HFLAV group quotes the averages \cite{Amhis:2019ckw}
\beqn\label{eq:Vcd_Vcs}
|\bV_{\! cd}|\, =\, 0.2204\pm 0.0040\, ,
\qquad\qquad\qquad
|\bV_{\! cs}|\, =\, 0.969\pm 0.010\, .
\eeqn

The difference of the ratio of double-muon to single-muon production by neutrino and antineutrino beams is proportional to the charm cross section off
valence $d$ quarks and, therefore, to $|\bV_{\! cd}|$ times the average semileptonic branching ratio of charm mesons. This allows for an independent determination of $|\bV_{\! cd}|$. Averaging data from several experiments, the PDG quotes \cite{Tanabashi:2018oca}
\bel{eq:Vcd_neutrino}
|\bV_{\! cd}|\, =\, 0.230\pm 0.011\, ,
\ee
which agrees with \eqn{eq:Vcd_Vcs} but has a larger uncertainty.
The analogous determination of $|\bV_{\! cs}|$
from $\nu s\to c X$ suffers from the uncertainty of the $s$-quark sea content.

The top quark has only been seen decaying into  bottom. From the ratio
of branching fractions
$\mathrm{Br}(t\to Wb)/\mathrm{Br}(t\to Wq)$, CMS has extracted \cite{Khachatryan:2014nda}
\bel{eq:Vtb_rat}
\frac{|\bV_{\! tb}|}{\sqrt{\sum_q |\bV_{\! tq}|^2}}\, >\, 0.975\quad \mathrm{(95\% CL)}
\, ,
\ee
where $q = b, s, d$. A more direct determination of $|\bV_{\! tb}|$
can be obtained from the single top-quark production cross section,
measured at the LHC and the Tevatron. The PDG quotes the world average \cite{Tanabashi:2018oca}
\bel{eq:Vtb}
|\bV_{\! tb}|\, =\, 1.019\pm 0.025 \, .
\ee

\subsection{Structure of the CKM matrix}

Using the previous determinations of CKM elements, we can check the unitarity of the quark mixing matrix. The most precise test involves the elements of
the first row:
\bel{eq:unitarity_test}
|\bV_{\! ud}|^2 + |\bV_{\! us}|^2 + |\bV_{\! ub}|^2 \, = \, 0.99825\pm 0.00047 \, ,
\ee
where we have taken as reference values the determinations in Eqs.~(\ref{eq:Vud}),
(\ref{eq:Vus}) and (\ref{eq:Vub}). Radiative corrections play a crucial role at the quoted level of uncertainty, while the $|\bV_{\! ub}|^2$ contribution is negligible. This relation exhibits a $3.7\,\sigma$ violation of unitarity, at the per-mill level, which calls for an independent re-evaluation of the very precise $|\bV_{\! ud}|$ value in Eq.~\eqn{eq:Vud} and improvements on the $|\bV_{\! us}|$ determination.

With the $|\bV_{\! cq}|^2$ values in Eqs.~\eqn{eq:Vcb_Incl} and \eqn{eq:Vcd_Vcs} we can also test the unitarity relation in the second row,
\bel{eq:unitarity_test2}
|\bV_{\! cd}|^2 + |\bV_{\! cs}|^2 + |\bV_{\! cb}|^2 \, = \, 0.989\pm 0.019 \, ,
\ee
and, adding the information on $|\bV_{\! tb}|$ in Eq.~\eqn{eq:Vtb}, the relation involving the third column,
\bel{eq:unitarity_test3}
|\bV_{\! ub}|^2 + |\bV_{\! cb}|^2 + |\bV_{\! tb}|^2 \, = \, 1.040\pm 0.051 \, .
\ee
The ratio of the total hadronic decay width of the $W$ to the leptonic one provides the sum \cite{LEPEWWG,LEPEWWG_SLD:06}
\bel{eq:unitarity_test4}
\sum_{j\,  =\,  d, s, b}\; \left( |\bV_{\! uj}|^2 + |\bV_{\! cj}|^2\right)
\; = \; 2.002\pm 0.027\, , \ee
which involves the first and second rows of the CKM matrix.
Although much less precise than Eq.~\eqn{eq:unitarity_test}, these three
results test unitarity at the 2\%, 5\% and 1.4\% level, respectively. 

From Eq.~\eqn{eq:unitarity_test4} one can also obtain an independent
estimate of $|\bV_{\! cs}|$, using the experimental knowledge
on the other CKM matrix elements, i.e., \ $|\bV_{\! ud}|^2 +
|\bV_{\! us}|^2 + |\bV_{\! ub}|^2 + |\bV_{\! cd}|^2 + |\bV_{\!
cb}|^2 = 1.0486\pm 0.0018\,$. This gives
\bel{eq:Vcs_LEP}
|\bV_{\! cs}| \, =\, 0.976 \pm 0.014\, ,
\ee
which agrees with the slightly more accurate direct determination in Eq.~(\ref{eq:Vcd_Vcs}).

The measured entries of the CKM matrix show a hierarchical pattern, with the diagonal elements being very close to one, the ones connecting the
first two generations having a size
\bel{eq:lambda} \lambda\approx |\bV_{\! us}| = 0.2231\pm 0.0007 \, ,
\ee
the mixing between the second and third families being of order
$\lambda^2$, and the mixing between the first and third quark generations
having a much smaller size of about $\lambda^3$.
It is then quite practical to use the
approximate parametrization \cite{WO:83}:

\bel{eq:wolfenstein}
\bV\; =\; \left[ \bath\displaystyle
1- {\lambda^2 \over 2} & \lambda & A\lambda^3 (\rho - i\eta)
\\[8pt]
-\lambda &\displaystyle 1 -{\lambda^2 \over 2} & A\lambda^2
\\[8pt]
A\lambda^3 (1-\rho -i\eta) & -A\lambda^ 2 &  1
\ea\right]\;\; +\;\; \cO\!\left(\lambda^4 \right) \, ,
\ee
where
\bel{eq:circle} A\approx {|\bV_{\! cb}|\over\lambda^2} = 0.844\pm
0.014 \, , \qquad\qquad \sqrt{\rho^2+\eta^2} \,\approx\,
\left|{\bV_{\! ub}\over \lambda \bV_{\! cb}}\right| \, =\, 0.422\pm
0.035 \, . \ee
Defining to all orders in $\lambda$ \cite{BLO:94}
$s_{12}\equiv\lambda$, $s_{23}\equiv A\lambda^2$ and $s_{13}\,
\e^{-i\delta_{13}}\equiv A\lambda^3 (\rho-i\eta)$,
Eq.~\eqn{eq:wolfenstein} just corresponds to a Taylor expansion of
Eq.~\eqn{eq:CKM_pdg} in powers of $\lambda$.

%%%%%%%%%%%%%%%%%%%%%%%%%%%% Mixing %%%%%%%%%%%%%%%%%%%%%%%%%%%%%%%%%%%%%%
\section{Meson-antimeson mixing}
\label{sec:BB_mixing}

Additional information on the CKM parameters can be obtained from
%%%flavour-changing neutral-current 
FCNC transitions, occurring at the one-loop level. An important example is provided by the mixing between the $B^0_d$ meson and its antiparticle.
This process occurs through  the box diagrams shown in Fig.~\ref{fig:Bmixing}, where two $W$ bosons are exchanged between a pair of quark lines.
The mixing amplitude is proportional to
\bel{eq:mixing}
\langle\bar B_d^0 | \cH_{\Delta B=2} |B_d^0\rangle\,\sim\,
\sum_{ij}\, \bV_{\! id}^{\phantom{*}}\bV_{\! ib}^*
\bV_{\! jd}^{\phantom{*}}\bV_{\! jb}^*\; S(r_i,r_j)
\,\sim\, \bV_{\! td}^2\; S(r_t,r_t) \, ,
\ee
where $S(r_i,r_j)$ is a loop function \cite{IL:81}
which depends on $r_i\equiv m_i^2/M_W^2$, with $m_i$ the masses of the up-type quarks running along the internal fermionic lines. Owing to the unitarity of the CKM matrix, the mixing vanishes for equal (up-type) quark masses (GIM mechanism \cite{GIM:70}); thus the flavour-changing transition is governed by the mass splittings between the $u$, $c$ and $t$ quarks.
Since the different CKM factors have all a similar size,
$\bV_{\! ud}^{\phantom{*}}\bV_{\! ub}^*\sim
\bV_{\! cd}^{\phantom{*}}\bV_{\! cb}^*\sim
\bV_{\! td}^{\phantom{*}}\bV_{\! tb}^*\sim A\lambda^3$,
the final amplitude is completely dominated by the top contribution.
This transition can then be used to perform
an indirect determination of $\bV_{\! td}$.

%%%%%%%%%%%%%%%  FIGURE %%%%%%%%%%%%%%%%%%%%%%%%%
\begin{figure}[tbh]\centering
\includegraphics[width=10cm]{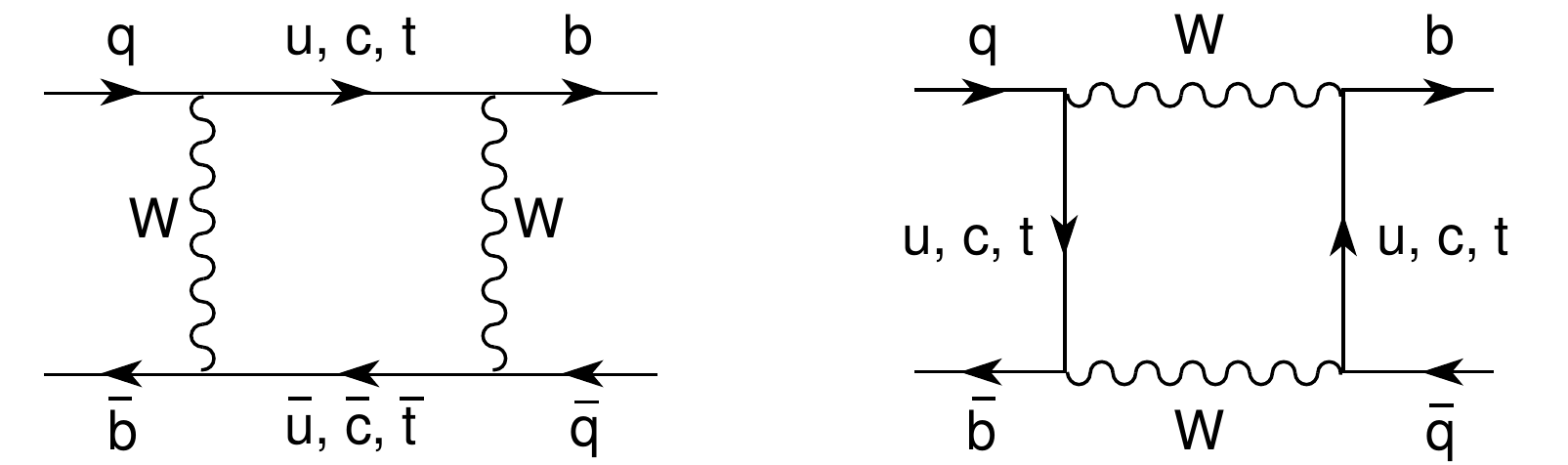}  %%%ps}
\caption{Box diagrams contributing to $B_q^0$--$\bar B_q^0$ mixing ($q=d,s$).}
\label{fig:Bmixing}
\end{figure}
%%%%%%%%%%%%%%%%%%%%%%%%%%%%%%%%%%%%%%%%%%%%%%%%%

Notice that this determination has a qualitatively different character
than the ones obtained before from tree-level weak decays.
Now, we are going to test the structure of the electroweak theory at the
quantum level.
This flavour-changing transition could then be sensitive to
contributions from new physics at higher energy scales.
Moreover, the mixing amplitude crucially depends on the
unitarity of the CKM matrix.
Without the GIM mechanism embodied in the CKM mixing structure, the
calculation of the analogous $K^0\to\bar K^0$ transition (replace
the $b$ by a strange quark $s$ in the box diagrams) would have failed
to explain the observed $K^0$--$\bar K^0$ mixing by several orders of
magnitude \cite{GL:74}.

\subsection{Mixing formalism}
\label{subsec:mixing_formalism}

Since weak interactions can transform a $P^0$ state ($P=K,\, D,\, B$)
into its antiparticle $\bar P^0$, these flavour eigenstates
are not mass eigenstates and do not follow an exponential decay law.
Let us consider an arbitrary mixture of the two flavour states,
\bel{eq:mixed_state}
|\psi(t)\rangle \, =\, a(t) \, |P^0\rangle + b(t) \, |\bar P^0\rangle
\,\equiv\, \left( \ba a(t) \\ b(t) \ea\right) \, ,
\ee
with the time evolution (in the meson rest frame)
\bel{eq:t_eq}
i\, {d\over dt}\, |\psi(t)\rangle \, =\, \cM\, |\psi(t)\rangle \, .
\ee
Assuming $\cCPT$ symmetry to hold, the $2\times 2$ mixing matrix can be written as
\be\label{eq:mass_matrix}
{\cal M} =
\left(   \begin{array}{cc} M & M_{12} \\ M_{12}^* & M \ea   \right)
- {i\over 2}
\left(   \begin{array}{cc} \Gamma & \Gamma_{12} \\
          \Gamma_{12}^* & \Gamma \ea   \right) \, .
\ee
The diagonal elements $M$ and $\Gamma$ are real parameters, which would correspond to the mass and width of the neutral mesons in the absence of mixing. The off-diagonal entries contain the $\Delta F=2$ transition amplitude ($F=S,C,B$):
\be\label{eq:M12_G12}
M_{12} -\frac{i}{2}\,\Gamma_{12} \; = \; \frac{1}{2 M}\left\{
\langle P^0| \cH_{\Delta F=2}(0) |\bar P^0\rangle
\; -\;\frac{1}{2}\,\int d^4x\; \langle P^0|\, T\left( \cH_{\Delta F=1}(x)  \,\cH_{\Delta F=1}(0)\right) |\bar P^0\rangle
%\; +\; \sum_f\; \frac{\langle P^0| \cH_{\Delta F=1} | f\rangle\,
%\langle f| \cH_{\Delta F=1} |\bar P^0\rangle}{M-E_f + i\,\varepsilon}
\right\}\, .
\ee
In addition to the short-distance $\Delta F=2$ Hamiltonian generated by the box diagrams, the mixing amplitude receives non-local contributions involving two $\Delta F=1$ transitions, which contain both {\it dispersive} and {\it absorptive} components, contributing to $M_{12}$ and $\Gamma_{12}$, respectively.
The {\it absorptive} contribution $\Gamma_{12}$ arises from on-shell intermediate states:
\be 
\Gamma_{12}\, =\, \frac{1}{2M}\; (2\pi)^4\,\sum_n\, \delta^{(4)}(p_{P^0}-p_n)\;
\langle P^0|\cH_{\Delta F=1}(0) |n\rangle\,\langle n|  \cH_{\Delta F=1}(0) |\bar P^0\rangle\, .
\ee
The sum extends over all possible intermediate states $|n\rangle$
into which the $|\bar P^0\rangle$ and $|P^0\rangle$ can both decay: $\bar P^0\to f\to P^0$.
%The sum extends over all possible intermediate states $|f\rangle$. Using the relation ($\mathscr{P}$ denotes principal part)
%%
%\be 
%\lim_{\varepsilon\to 0}\;\frac{1}{M-E_f + i\,\varepsilon}\, =\,
%\mathscr{P} 
%\left(\frac{1}{M-E_f}\right) - i\,\pi\,\delta(M-E_f)\, ,
%\ee
%%
%one can separate the {\it dispersive} and {\it absorptive} parts of \eqn{eq:M12_G12}. The {\it absorptive} contribution $\Gamma_{12}$ arises from on-shell intermediate states, \ie all states $|f\rangle$ into which the $|\bar P^0\rangle$ and $|P^0\rangle$ can both decay. 
In the SM, the $\Delta F=1$ Hamiltonian is generated through a single $W^\pm$ emission, as shown in Fig.~\ref{fig:cDecay} for charm decay. If $\cCP$ were an exact symmetry, $M_{12}$ and $\Gamma_{12}$ would also be real parameters.

The physical eigenstates of ${\cal M}$ are
\be\label{eq:eigenstates}
| P_\mp \rangle \, = \, {1\over\sqrt{|p|^2 + |q|^2}} \,
       \left[ p \, | P^0 \rangle \, \mp\, q \, | \bar P^0 \rangle \right] ,
\ee
with
\be\label{eq:q/p}
{q\over p} \, \equiv \, {1 - \bar\varepsilon \over
         1 + \bar\varepsilon} \, = \,
  \left( {M_{12}^* - {i\over 2}\Gamma_{12}^* \over
          M_{12} - {i\over 2}\Gamma_{12}} \right)^{1/2} .
\ee
The corresponding eigenvalues are
\bel{eq:MesonMixingEigenvalues}
M_{P_\mp} - \frac{i}{2}\,\Gamma_{P_\mp} \, =\,
\left( M \mp \frac{1}{2}\Delta M\right)
- \frac{i}{2}\left(\Gamma  \mp \frac{1}{2}\,\Delta\Gamma\right)\, ,
\ee
where\footnote{Be aware of the different sign conventions in the literature. Quite often, $\Delta M$ and $\Delta\Gamma$ are defined to be positive.}
\ $\Delta M \equiv M_{P_+}-M_{P_-}$ \ and \ $\Delta\Gamma\equiv\Gamma_{P_+}-\Gamma_{P_-}$ \ satisfy
\be
(\Delta M)^2 -\frac{1}{4}\, (\Delta \Gamma)^2\, =\, 
4\, |M_{12}|^2 - |\Gamma_{12}|^2
\ee
and
\be
\Delta M\; \Delta \Gamma\, =\, 
4\, \mathrm{Re}\, (M_{12} \Gamma_{12}^*)
\, =\,  4\, |M_{12}|\, |\Gamma_{12}|\,\cos{\phi}
\, ,
\ee
with $\phi\equiv\arg{(-M_{12}/\Gamma_{12})}$.

If $M_{12}$ and $\Gamma_{12}$ were real then $q/p = 1$ and the mass eigenstates
$| P_\mp \rangle $ would correspond to the
$\cCP$-even and $\cCP$-odd  states
(we use the phase convention\footnote{
%
%  FOOTNOTE
%
Since flavour is conserved by strong interactions, there is
some freedom in defining the phases of flavour eigenstates.
One could use
$\, |P^0_\zeta\rangle \equiv e^{-i\zeta} |P^0\rangle \, $ and
$|\bar P^0_\zeta\rangle \equiv e^{i\zeta} |\bar P^0\rangle$,
which satisfy
$\cC\cP\, |P^0_\zeta\rangle = - e^{-2i\zeta} |\bar P^0_\zeta\rangle$.
Both basis are trivially related:
$M_{12}^\zeta = e^{2i\zeta} M_{12}$,
$\Gamma_{12}^\zeta = e^{2i\zeta} \Gamma_{12}$ and
$(q/p)_\zeta = e^{-2i\zeta} (q/p)$.
Thus, $q/p\not=1$ does not necessarily imply $\cCP$ violation.
$\cCP$ is violated if $|q/p|\not=1$;
\ie $\mbox{\rm Re}(\bar\varepsilon)\not=0$ and
$\langle P_- |P_+\rangle \not= 0$.
Note that
$\langle P_- | P_+\rangle_\zeta =\langle P_- | P_+\rangle$.
Another phase-convention-independent quantity is
$(q/p) \, (\bar A_f/A_f)$,
where $A_f\equiv A(P^0\!\to\! f)$ and
$\bar A_f\equiv -A(\bar P^0\!\to\! f)$, for any final state $f$.}
%
%  END OF FOOTNOTE
%
\ $\cC\cP |P^0\rangle = - |\bar P^0\rangle$)
\bel{eq:CP_states}
|P_{1,2}\rangle\,\equiv\, \frac{1}{\sqrt{2}} \left( |P^0\rangle\mp
|\bar P^0\rangle\right)\, , \qquad \qquad\qquad
\cC\cP\, |P_{1,2}\rangle\, =\, \pm\, |P_{1,2}\rangle \, .
\ee
The two mass eigenstates are no longer orthogonal when $\cCP$ is violated:
\be \langle P_- | P_+\rangle\, =\, {|p|^2-|q|^2 \over |p|^2+|q|^2}\, = \,
{2\, \mathrm{Re}\, (\bar\varepsilon)\over (1 + |\bar\varepsilon|^2)} \, .
\ee

The time evolution of a state which was originally produced
as a $P^0$ or a  $\bar P^0$  is given by
\be\label{eq:evolution}
\left( \ba | P^0(t) \rangle  \\ | \bar P^0(t) \rangle \ea \right)
\; =\;
\left( \begin{array}{cc} g_1(t)  & {q\over p}\, g_2(t) \\[5pt]
     {p\over q}\, g_2(t) & g_1(t) \ea \right)\,
\left( \ba | P^0 \rangle  \\ | \bar P^0 \rangle \ea \right) \, ,
\ee
where 
\be\label{eq:g}
\left( \ba g_1(t) \\ g_2(t) \ea \right)\; =\;
\e^{-iMt}\, \e^{-\Gamma t/2}\,
\left( \ba \cos{[(x - i y) \Gamma t/2]} \\[5pt]
   -i \sin{[(x - i y) \Gamma t/2]} \ea \right) \, ,
\ee
with 
%%%\ $M = \frac{1}{2}\, (M_{P_+}+M_{P_-})$, \ $\Gamma =\frac{1}{2}\, (\Gamma_{P_+}+\Gamma_{P_-})$,
%
\bel{eq:x_y}
x\,\equiv\,\frac{\Delta M}{\Gamma}\, ,
\qquad\qquad\qquad
y\,\equiv\,\frac{\Delta \Gamma}{2\Gamma}\, .
\ee
%
%and\footnote{Be aware of the different sign conventions in the literature. Quite often, $\Delta M$ and $\Delta\Gamma$ are defined to be positive.}
%%
%\bel{eq:DM_DG}
%\Delta M \equiv M_{P_+}-M_{P_-} \, ,
%\qquad\qquad\qquad
%\Delta\Gamma\equiv\Gamma_{P_+}-\Gamma_{P_-}\, .
%\ee
%%

\subsection{Experimental measurements}
\label{subsec:exp_mixing}

The main difference between the $K^0$--$\bar K^0$ and
$B^0$--$\bar B^0$ systems stems from the different kinematics involved.
The light kaon mass only allows the hadronic decay modes $K^0\to 2\pi$ and
$K^0\to 3\pi$. Since $\cC\cP\, |\pi\pi\rangle = + |\pi\pi\rangle$, 
for both $\pi^0\pi^0$ and $\pi^+\pi^-$ final states, the
$\cCP$-even kaon state decays into $2\pi$ whereas the
$\cCP$-odd one decays into the phase-space-suppressed $3\pi$ mode.
Therefore, there is a large lifetime difference and we have
a short-lived
$|K_S\rangle \equiv |K_-\rangle \approx
|K_1\rangle + \bar\varepsilon_K |K_2\rangle $
and a long-lived
$|K_L\rangle \equiv |K_+\rangle \approx
|K_2\rangle + \bar\varepsilon_K |K_1\rangle $
kaon,
with $\Gamma_{K_L}\ll\Gamma_{K_S} \approx 2\, \Gamma_{K^0}$.
One finds experimentally that
$\Delta\Gamma_{K^0}\approx -\Gamma_{K_S}\approx -2\,\Delta M_{K^0}$ \cite{Tanabashi:2018oca}:
\bel{eq:Kmix}
\Delta M_{K^0} = (0.5293 \pm 0.0009)\cdot 10^{10}\:\mbox{\rm s}^{-1} \, ,
\qquad\qquad
\Delta \Gamma_{K^0} = -(1.1149 \pm 0.0005)\cdot 10^{10}\:\mbox{\rm s}^{-1} \, .
\ee
Thus, the two $K^0$--$\bar K^0$ oscillations parameters are sizeable and of similar magnitudes: $x_{K^0}\approx -y_{K^0}\approx 1$.

In the $B$ system, there are many open decay channels and a large part of them
are common to both mass eigenstates. Therefore, the $|B_\mp\rangle $ states
have a similar lifetime; \ie $|\Delta\Gamma_{B^0}|\ll\Gamma_{B^0}$.
Moreover, whereas the $B^0$--$\bar B^0$ transition
is dominated by the top box diagram, the decay amplitudes get
obviously their main contribution from the $b\to c$ process.
Thus, $|\Delta\Gamma_{B^0} / \Delta M_{B^0}| \sim m_b^2/ m_t^2 \ll 1$.
To experimentally measure the mixing transition requires the
identification of the $B$-meson flavour at both its production and decay time.
This can be done through flavour-specific decays such as
$B^0\to X \ell^+\nu_\ell$ and $\bar B^0\to X \ell^-\bar\nu_\ell$, where the lepton charge labels the initial $B$ meson.
In general, mixing is measured by studying pairs of $B$ mesons so that
one $B$ can be used to {\it tag} the initial flavour of the other meson.
For instance, in $e^+e^-$ machines one can look into the pair
production process
$e^+e^- \to B^0 \bar B^0 \to (X \ell\nu_\ell) \, (Y \ell \nu_\ell)$.
In the absence of mixing, the final leptons should have opposite charges;
the amount of like-sign leptons is then a clear signature of meson
mixing.

Evidence for a large $B_d^0$--$\bar B_d^0$ mixing was first reported in 1987 by ARGUS \cite{ARGUS:87}. This provided the first indication that the top quark was very heavy. Since then, many experiments have analysed the mixing probability. The present world-average values are \cite{Tanabashi:2018oca,Amhis:2019ckw}:
\bel{eq:BdMix}
\Delta M_{B^0_d} = (0.5065 \pm 0.0019)\cdot 10^{12}\:\mbox{\rm s}^{-1} \, ,
\qquad\qquad\qquad
x_{B^0_d} 
%%%\equiv {\Delta M_{B^0_d}\over\Gamma_{B^0_d}} 
= 0.769\pm 0.004
\, ,
\ee
while \ $y_{B^0_d} = 0.001\pm 0.005$ confirms the expected suppression of $\Delta\Gamma_{B^0_d}$.

The first direct evidence of $B_s^0$--$\bar B_s^0$ oscillations was obtained by CDF \cite{CDF:06}. The large measured mass difference reflects the CKM hierarchy
$|\bV_{\! ts}|^2 \gg |\bV_{\! td}|^2$, implying very fast oscillations
\cite{Tanabashi:2018oca,Amhis:2019ckw}:
$$\displaystyle
\Delta M_{B^0_s} = (17.757 \pm 0.021)\cdot 10^{12}\:\mbox{\rm s}^{-1} \, ,
\qquad\qquad\qquad
x_{B^0_s} 
%%%\equiv {\Delta M_{B^0_s}\over\Gamma_{B^0_s}} 
= 26.81\pm 0.08\, ,
$$
\bel{eq:BsMix}
\Delta \Gamma_{B^0_s} = -(0.090\pm 0.005)\cdot 10^{12}\:\mbox{\rm s}^{-1} \, ,
\qquad\qquad\qquad
y_{B^0_s} = - 0.068\pm 0.004\, .
\ee

Evidence of mixing has been also obtained in the $D^0$--$\bar D^0$ system. The present world averages \cite{Amhis:2019ckw},
\bel{eq:D0Mix}
x_{D^0} = -\left(0.39\, {}^{+\, 0.11}_{-\, 0.12}\right)\cdot 10^{-2}\, ,
\qquad\qquad\qquad
y_{D^0} = -\left( 0.65\, {}^{+\, 0.06}_{-\, 0.07}\right)\cdot 10^{-2}\, ,
\ee
confirm the SM expectation of a very slow oscillation, compared with the decay rate. Since the short-distance mixing amplitude originates in box diagrams with down-type quarks in the internal lines, it is very suppressed by the relevant combination of CKM factors and quark masses.

\subsection{Mixing constraints on the CKM matrix}
\label{subsec:mixing_constraints}

Long-distance contributions arising from intermediate hadronic states completely dominate the $D^0$--$\bar D^0$ mixing amplitude and are very sizeable for $\Delta M_{K^0}$,
%%%in the $K^0$--$\bar K^0$ case, 
making difficult to extract useful information on the CKM matrix. The situation is much better for $B^0$ mesons, owing to the dominance of the short-distance top contribution which is known to next-to-leading order (NLO) in the strong coupling
\cite{BJW:90,HN:94}. The main uncertainty stems from the
hadronic matrix element of the $\Delta B=2$ four-quark operator
\bel{eq:DB_op}
\langle\bar B^0_d\, |\, (\bar b\gamma^\mu(1-\gamma_5)d)\:
(\bar b\gamma_\mu(1-\gamma_5)d)\, |\, B^0_d\rangle \,\equiv\,
 {8\over 3} \, M_{B^0}^2\, \xi_{B}^2 \, ,
\ee
which is characterized through the non-perturbative parameter $\xi_B(\mu)\equiv f_{B} \sqrt{B_{B}(\mu)}$ \cite{PP:95}.
The current ($2+1$) lattice averages \cite{Aoki:2019cca} are
$\hat\xi_{B_d} = (225\pm 9)\:\mathrm{MeV}$,
$\hat\xi_{B_s}= (274\pm 8)\:\mathrm{MeV}$ and
$\hat\xi_{B_s}/\hat\xi_{B_d}= 1.206\pm 0.017$,
where
$\hat\xi_B\approx \alpha_s(\mu)^{-3/23} \xi_B(\mu)$
is the corresponding renormalization-group-invariant quantity.
Using these values, the measured mass differences in (\ref{eq:BdMix}) and (\ref{eq:BsMix}) imply
\bel{eq:V_td}
 |\bV_{\! tb}^* \bV_{\! td}|\, = \, 0.0080\pm 0.0003 \, ,
\qquad\;
 |\bV_{\! tb}^* \bV_{\! ts}|\, = \, 0.0388\pm 0.0012 \, ,
\qquad\;
 \frac{|\bV_{\! td}|}{|\bV_{\! ts}|}\, = \, 0.205\pm 0.003 \, .
\ee
The last number takes advantage of the smaller uncertainty in the ratio
$\hat\xi_{B_s}/\hat\xi_{B_d}$.
Since $|\bV_{\! tb}|\approx 1$, the mixing of $B^0_{d,s}$ mesons provides indirect determinations of
$|\bV_{\! td}|$ and $|\bV_{\! ts}|$. The resulting value of $|\bV_{\! ts}|$ is in agreement with Eq.~(\ref{eq:Vcb_Incl}), satisfying the unitarity constraint
$|\bV_{\! ts}|\approx |\bV_{\! cb}|$.
In terms of the $(\rho,\eta)$ parametrization of Eq.~\eqn{eq:wolfenstein},
one obtains
\bel{eq:circle_t}
\sqrt{(1-\rho)^2+\eta^2} \, = \, \left\{ \begin{array}{l}\displaystyle
\left|{\bV_{\! td}\over \lambda\bV_{\! cb}}\right|
\, = \, \phantom{0}0.86\pm 0.04
\\[12pt]\displaystyle
\left|{\bV_{\! td}\over \lambda\bV_{\! ts}}\right|
\, = \, 0.920\pm 0.013
\ea\right. \, .
\ee
%

%%%%%%%%%%%%%%%%%%%%%%%%%%%% CP %%%%%%%%%%%%%%%%%%%%%%%%%%%%%%%%%%%%%%

\section{$\cCP$ violation}
\label{subsec:CP-Violation}

While parity ($\cP$) and charge conjugation ($\cC$) are violated by the weak
interactions in a maximal way, the product of the two discrete
transformations is still a good symmetry of the gauge interactions (left-handed fermions
$\leftrightarrow$ right-handed antifermions). In fact, $\cCP$
appears to be a symmetry of nearly all observed phenomena. However,
a slight violation of the $\cCP$ symmetry at the level of $0.2\%$ is
observed in the neutral kaon system and more sizeable signals of
$\cCP$ violation have been established at the $B$ factories.
Moreover, the huge matter--antimatter asymmetry present in our
Universe is a clear manifestation of $\cCP$ violation and its
important role in the primordial baryogenesis.

The $\cCPT$ theorem guarantees that the product of the three
discrete transformations is an exact symmetry of any local and
Lorentz-invariant quantum field theory, preserving micro-causality.
A violation of $\cCP$ implies then a corresponding
violation of time reversal ($\cT$). Since $\cT$ is an antiunitary
transformation, this requires the presence of relative complex
phases between different interfering amplitudes.

The electroweak SM Lagrangian only contains a single complex phase
$\delta_{13}$ ($\eta$). This is the sole possible source of $\cCP$
violation and, therefore, the SM predictions for $\cCP$-violating
phenomena are quite constrained. The CKM mechanism requires several
necessary conditions in order to generate an observable
$\cCP$-violation effect. With only two fermion generations, the
quark mixing matrix cannot give rise to $\cCP$ violation;
therefore, for $\cCP$ violation to occur in a particular process,
all three generations are required to play an active role. In the
kaon system, for instance, $\cCP$ violation can only appear
at the one-loop level, where the top quark is present. In addition,
all CKM matrix elements must be non-zero and the quarks of a given
charge must be non-degenerate in mass. If any of these conditions
were not satisfied, the CKM phase could be rotated away by a
redefinition of the quark fields. $\cCP$-violation effects are then
necessarily proportional to the product of all CKM angles, and
should vanish in the limit where any two (equal-charge) quark masses
are taken to be equal. All these necessary conditions can be
summarized as a single requirement on the
original quark mass matrices $\bM'_u$ and $\bM'_d$ \cite{JA:85}:
\be
\cCP \:\mbox{\rm violation} \qquad \Longleftrightarrow \qquad
\mbox{\rm Im}\left\{\det\left[\bM_u^\prime \bM^{\prime\dagger}_u\, ,
  \,\bM^{\prime\phantom{\dagger}}_d \bM^{\prime\dagger}_d\right]\right\} \,\not=\, 0 \, .
\ee

Without performing any detailed calculation, one can make the
following general statements on the implications of the CKM mechanism
of $\cCP$ violation:

\bi
\item[--]
Owing to unitarity, for any choice of $i,j,k,l$ (between 1 and 3),
\beqn\label{eq:J_relation}
\mbox{\rm Im}\left[
\bV^{\phantom{*}}_{ij}\bV^*_{ik}\bV^{\phantom{*}}_{lk}\bV^*_{lj}\right]
\, =\, \cJ \,\sum_{m,n=1}^3 \epsilon_{ilm}\epsilon_{jkn}\, ,
\qquad\quad\\
\cJ \, =\, c_{12}\, c_{23}\, c_{13}^2\, s_{12}\, s_{23}\, s_{13}\, \sin{\delta_{13}}
\,\approx\, A^2\lambda^6\eta \, < \, 10^{-4}\, .
\eeqn
Any $\cCP$-violation observable involves the product $\cJ$
\cite{JA:85}. Thus, violations of the $\cCP$ symmetry are
necessarily small.
\item[--] In order to have sizeable $\cCP$-violating asymmetries
$\cA\equiv (\Gamma - \overline{\Gamma})/(\Gamma +
\overline{\Gamma})$, one should look for very suppressed decays,
where the decay widths already involve small CKM matrix elements.
\item[--] In the SM, $\cCP$ violation is a low-energy phenomenon,
in the sense that any effect should disappear when the quark mass
difference $m_c-m_u$ becomes negligible.
\item[--] $B$ decays are the optimal place for $\cCP$-violation signals to show up.
They involve small CKM matrix elements and are the lowest-mass
processes where the three quark generations play a direct
(tree-level) role. \ei

The SM mechanism of $\cCP$ violation is based on the unitarity of the
CKM matrix. Testing the constraints implied by unitarity
is then a way to test the source of $\cCP$ violation.
The unitarity tests in Eqs.~\eqn{eq:unitarity_test}, \eqn{eq:unitarity_test2}, \eqn{eq:unitarity_test3} and \eqn{eq:unitarity_test4} involve only the moduli of the CKM parameters, while $\cCP$ violation has to do with their phases.
More interesting are the off-diagonal unitarity conditions:
\beqn\label{eq:triangles}
\bV^\ast_{\! ud}\bV^{\phantom{*}}_{\! us} \, +\,
\bV^\ast_{\! cd}\bV^{\phantom{*}}_{\! cs} \, +\,
\bV^\ast_{\! td}\bV^{\phantom{*}}_{\! ts} & = & 0 \, ,
\nonumber\\[5pt]
\bV^\ast_{\! us}\bV^{\phantom{*}}_{\! ub} \, +\,
\bV^\ast_{\! cs}\bV^{\phantom{*}}_{\! cb} \, +\,
\bV^\ast_{\! ts}\bV^{\phantom{*}}_{\! tb} & = & 0 \, ,
\nonumber\\[5pt]
\bV^\ast_{\! ub}\bV^{\phantom{*}}_{\! ud} \, +\,
\bV^\ast_{\! cb}\bV^{\phantom{*}}_{\! cd} \, +\,
\bV^\ast_{\! tb}\bV^{\phantom{*}}_{\! td} & = & 0 \, .
\eeqn
These relations can be visualized by triangles in a complex
plane which, owing to Eq.~\eqn{eq:J_relation}, have the
same area $|\cJ|/2$.
In the absence of $\cCP$ violation, these triangles would degenerate
into segments along the real axis.

In the first two triangles, one side is much shorter than the other
two (the Cabibbo suppression factors of the three sides are
$\lambda$, $\lambda$ and $\lambda^5$ in the first triangle, and
$\lambda^4$, $\lambda^2$ and $\lambda^2$ in the second one). This is
why $\cCP$ effects are so small for $K$ mesons (first triangle), and
why certain  asymmetries in $B^0_s$ decays are predicted to be tiny
(second triangle).
The third triangle looks more interesting, since the three sides
have a similar size of about $\lambda^3$. They are small, which
means that the relevant $b$-decay branching ratios are small, but
once enough $B^0_d$ mesons have been produced, the $\cCP$-violation
asymmetries are sizeable. The present experimental constraints on
this triangle are shown in Fig.~\ref{fig:UTfit}, where it has been
scaled by dividing its sides by $\bV^\ast_{\!
cb}\bV^{\phantom{*}}_{\! cd}$. This aligns one side of the triangle
along the real axis and makes its length equal to 1; the coordinates
of the 3 vertices are then $(0,0)$, $(1,0)$ and
$(\bar\rho,\bar\eta)\approx (1-\lambda^2/2)\, (\rho,\eta)$.

%%%%%%%%%%%%%%%  FIGURE %%%%%%%%%%%%%%%%%%%%%%%%%
\begin{figure}[t]\centering
\includegraphics[width=11.cm,clip]{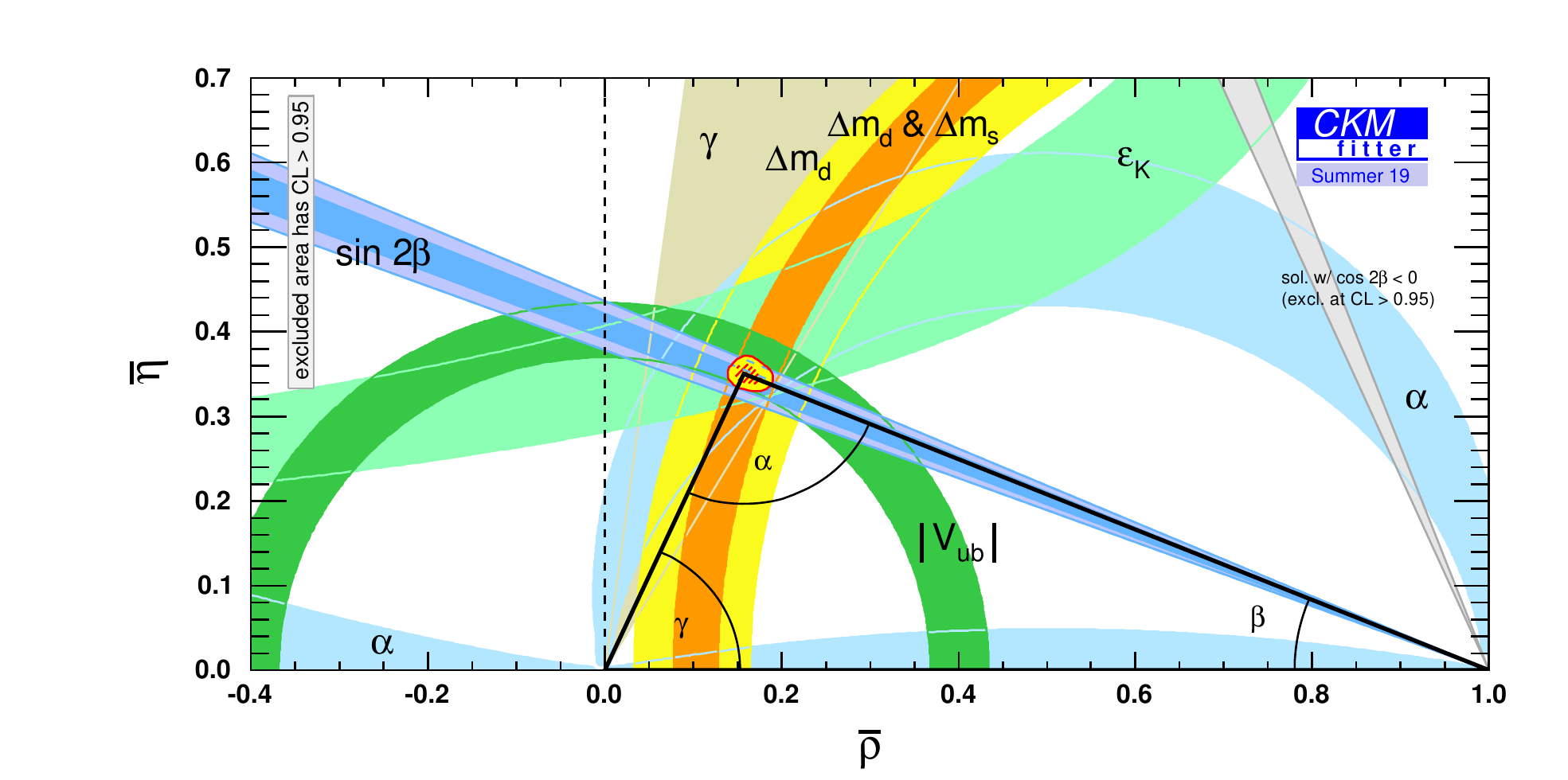}
\caption{Experimental constraints on the SM unitarity triangle
\cite{CKMfitter}.} \label{fig:UTfit}
\end{figure}
%%%%%%%%%%%%%%%%%%%%%%%%%%%%%%%%%%%%%%%%%%%%%%%%%

We have already determined the sides of the unitarity triangle
in Eqs.~\eqn{eq:circle} and (\ref{eq:circle_t}),
through two $\cCP$-conserving observables: $|\bV_{\! ub}/\bV_{\! cb}|$ and
$B^0_{d,s}$ mixing. This gives the circular rings shown in Fig.~\ref{fig:UTfit}, centered at the vertices $(0,0)$ and $(1,0)$. Their overlap at $\eta\not= 0$
establishes that $\cCP$ is violated (assuming unitarity).
More direct constraints on the parameter $\eta$ can be obtained from $\cCP$-violating observables, which provide sensitivity to the angles of the unitarity triangle
($\alpha + \beta + \gamma = \pi$):
\be\label{eq:angles}
\alpha\equiv\arg{\left[
   -{\bV^{\phantom{*}}_{\! td}\bV^*_{\! tb}\over
   \bV^{\phantom{*}}_{\! ud}\bV^*_{\! ub}} \right]} \, , \qquad\quad
\beta\equiv\arg{\left[
   -{\bV^{\phantom{*}}_{\! cd}\bV^*_{\! cb}\over
   \bV^{\phantom{*}}_{\! td}\bV^*_{\! tb}} \right]} \, , \qquad\quad
\gamma\equiv\arg{\left[
   -{\bV^{\phantom{*}}_{\! ud}\bV^*_{\! ub}\over
   \bV^{\phantom{*}}_{\! cd}\bV^*_{\! cb}} \right]}\, .
\ee

\subsection{Indirect and direct $\cCP$ violation in the kaon system}
\label{subsec:CP_kaon}

Any observable $\cCP$-violation effect is generated by the interference between
different amplitudes contributing to the same physical transition.
This interference can occur either through meson-antimeson mixing
or via final-state interactions, or by a combination of both effects.

The flavour-specific decays
$K^0\to\pi^- \ell^+\nu_\ell$ and $\bar K^0\to\pi^+ \ell^-\bar\nu_\ell$
provide a way to measure
the departure of the $K^0$--$\bar K^0$ mixing parameter
$|p/q|$ from unity.
In the SM, the decay amplitudes satisfy
$|A(\bar K^0\to\pi^+ \ell^-\bar\nu_\ell)| = |A(K^0\to\pi^- \ell^+\nu_\ell)|$;
therefore ($\ell=e,\mu$),
\bel{eq:deltaL}
%%%\delta_L 
A_L\,\equiv\,
{\Gamma(K_L\to\pi^- \ell^+\nu_\ell) - \Gamma(K_L\to\pi^+ \ell^-\bar\nu_\ell)\over
\Gamma(K_L\to\pi^- \ell^+\nu_\ell) + \Gamma(K_L\to\pi^+ \ell^-\bar\nu_\ell)}
\, =\, {|p|^2-|q|^2 \over |p|^2+|q|^2}
\,=\, {2\, \mathrm{Re}\, (\bar\varepsilon^{\phantom{'}}_K)\over
(1 + |\bar\varepsilon^{\phantom{'}}_K|^2)} \, .
\ee
The experimental measurement \cite{Tanabashi:2018oca},
$A_L = (3.32\pm 0.06)\cdot 10^{-3}$,
implies
\be\label{eq:Repsilon}
\mathrm{Re}\, (\bar\varepsilon^{\phantom{'}}_K)\, =\,
(1.66\pm 0.03)\cdot 10^{-3}\, ,
\ee
which establishes the presence of {\it indirect}\/ $\cCP$ violation
generated by the mixing amplitude.

If the flavour of the decaying meson $P$ is known, any observed difference between the decay rate $\Gamma(P\to f)$ and its $\cCP$ conjugate $\Gamma(\bar P\to \bar f)$ would indicate that $\cCP$ is directly violated in the decay amplitude. One could study, for instance,
$\cCP$ asymmetries in decays such as $K^\pm\to\pi^\pm\pi^0$
where the pion charges identify the kaon flavour; however,
no positive signals have been found in charged kaon decays.
Since at least two interfering contributions are needed,
let us write the decay amplitudes as
\be\label{eq:direct_b}
A[P \to f] \, = \, M_1 \, e^{i\phi_1}\, e^{i \delta_1}\,
   +\, M_2 \, e^{i\phi_2}\, e^{i \delta_2} \, ,
\qquad\;
A[\bar P \to \bar f] \, = \,
  M_1\, e^{-i\phi_1} e^{i \delta_1}\, +\, M_2\, e^{-i\phi_2}e^{i \delta_2} \, ,
\ee
where $\phi_i$ denote weak phases, $\delta_i$ strong final-state interaction phases and $M_i$ the moduli of the matrix elements. Notice that the weak phase gets reversed under $\cCP$, while the strong one remains of course invariant. The rate asymmetry is given by
\be\label{eq:direct_ratediff}
\mathcal{A}_{P \to f}^{\cCP}\,\equiv\,
{\Gamma[P \to f] - \Gamma[\bar P \to \bar f] \over
\Gamma[P \to f] + \Gamma[\bar P \to \bar f]} \; =\;
{-2 M_1 M_2 \,\sin{(\phi_1 - \phi_2)}\,
\sin{(\delta_1 - \delta_2)} \over
|M_1|^2 + |M_2|^2 + 2 M_1 M_2\, \cos{(\phi_1 - \phi_2)}\,
\cos{(\delta_1 - \delta_2)}} \, .
\ee
Thus, to generate a direct $\cCP$ asymmetry one needs:
1) at least two interfering amplitudes, which should be of comparable size
in order to get a sizeable asymmetry; 2) two different weak phases
[$\sin{(\phi_1 - \phi_2)}\not=0$], \ and \ 3) two different strong phases
[$\sin{(\delta_1 - \delta_2)}\not=0$].

Direct $\cCP$ violation has been searched for in decays of
neutral kaons, where $K^0$--$\bar K^0$ mixing is also involved. Thus,
both direct and indirect $\cCP$ violation need to be taken into account
simultaneously.
A $\cCP$-violation signal is provided by the ratios:
\bel{eq:etapm}
\eta_{+-} \,\equiv\, {A(K_L\to\pi^+\pi^-)\over A(K_S\to\pi^+\pi^-)}
\, =\, \varepsilon_K^{\phantom{'}} + \varepsilon_K'\, ,
\qquad\qquad
\eta_{00} \,\equiv\, {A(K_L\to\pi^0\pi^0)\over A(K_S\to\pi^0\pi^0)}
\, =\, \varepsilon_K^{\phantom{'}} - 2 \varepsilon_K'\, .
\ee
The dominant effect from $\cCP$ violation in $K^0$--$\bar{K}^0$ mixing
is contained in $\varepsilon_K^{\phantom{'}}$, while $\varepsilon_K'$ accounts for direct $\cCP$ violation in the decay amplitudes \cite{Cirigliano:2011ny}:
\bel{eq:eps_def}
\varepsilon_K^{\phantom{'}} = \bar\varepsilon_K^{\phantom{'}} + i \xi_0 \, ,
\qquad
\varepsilon_K' = {i\over\sqrt{2}} \,\omega\, (\xi_2 - \xi_0) \, ,
\qquad
\omega \equiv {\mathrm{Re}\, (A_2)\over\mathrm{Re}\, (A_0)}\,
   e^{i(\delta_2-\delta_0)} \, ,
\qquad
\xi_I \equiv  {\mathrm{Im}\, (A_I)\over\mathrm{Re}\, (A_I)} \, .
\ee
$A_I$ are the transition amplitudes into two pions with isospin $I=0,2$ (these are the only two values allowed by Bose symmetry for the final $2\pi$ state) and $\delta_I$ their corresponding strong phase shifts. Although $\varepsilon_K'$ is strongly suppressed by the small ratio
$|\omega|\approx 1/22$, a non-zero value has been established
through very accurate measurements, demonstrating the existence of direct
$\cCP$ violation in K decays \cite{NA48,KTeV,NA31,E731}:
\bel{eq:EpsExp}
\mathrm{Re} \left(\varepsilon_K'/\varepsilon_K^{\phantom{'}}\right) =
\frac{1}{3} \left( 1
  -\left|\frac{\eta_{_{00}}}{\eta_{_{+-}}}\right|\right) =\,
(16.6 \pm 2.3) \cdot  10^{-4} \, .
\ee
In the SM the necessary weak phases are generated through the gluonic and electroweak penguin diagrams shown in Fig.~\ref{fig:penguin}, involving virtual up-type quarks of the three generations in the loop. These short-distance contributions are known to NLO in the strong coupling
\cite{BJL:93,ciuc:93}. However, the theoretical prediction involves a delicate balance between the two isospin amplitudes and is sensitive to long-distance and isospin-violating effects. Using chiral perturbation theory techniques, one finds
$\mathrm{Re} \left(\varepsilon_K'/\varepsilon_K^{\phantom{'}}\right) =
(14 \pm 5) \cdot  10^{-4}$  \cite{PP:00,PPS:01,CPEN:03,Gisbert:2017vvj,Cirigliano:2019cpi}, in agreement with (\ref{eq:EpsExp}) but with a large uncertainty.\footnote{A very recent lattice calculation gives
$\mathrm{Re} \left(\varepsilon_K'/\varepsilon_K^{\phantom{'}}\right) =
(22 \pm 8) \cdot  10^{-4}$ \cite{Abbott:2020hxn}, with an even larger error. However, this result does not include yet important isospin-breaking corrections that are known to be negative \cite{Cirigliano:2019cpi}.}

%%%%%%%%%%%%%%%  FIGURE %%%%%%%%%%%%%%%%%%%%%%%%%
\begin{figure}[t]\centering
\includegraphics[width=4.6cm,clip]{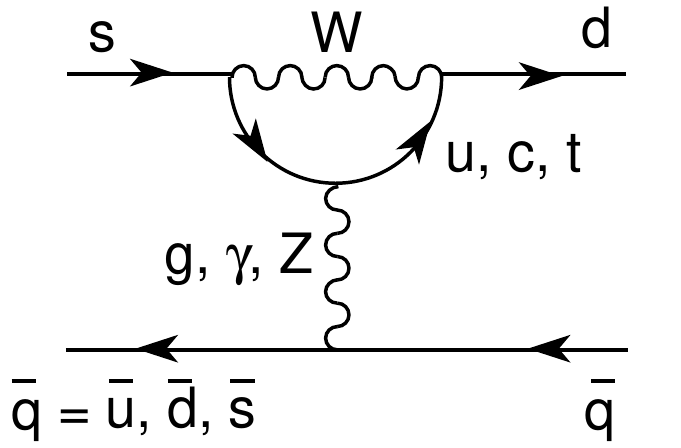}
\caption{$\Delta S=1$ penguin diagrams.}
\label{fig:penguin}
\end{figure}
%%%%%%%%%%%%%%%%%%%%%%%%%%%%%%%%%%%%%%%%%%%%%%%%%

Since $\mathrm{Re} \left(\varepsilon_K'/\varepsilon_K^{\phantom{'}}\right)\ll 1$, the ratios $\eta_{_{+-}}$ and $\eta_{_{00}}$
provide a measurement of
$\varepsilon_K^{\phantom{'}} = |\varepsilon_K^{\phantom{'}}|\,\e^{i\phi_\varepsilon}$ \cite{Tanabashi:2018oca}:
\bel{eq:eps_exp}
|\varepsilon_K^{\phantom{'}}| \, =\, \frac{1}{3} \left( 2 |\eta_{_{+-}}| + |\eta_{_{00}}|\right) \, =\,
(2.228\pm 0.011)\cdot 10^{-3} ,
\qquad\qquad
\phi_\varepsilon\, =\, (43.52\pm 0.05)^\circ\, ,
\ee
in perfect agreement with the semileptonic asymmetry $A_L$.
In the SM $\varepsilon_K^{\phantom{'}}$ receives short-distance contributions
from box diagrams involving virtual top and charm quarks, which are proportional
to
\bel{eq:epsilonK}
\varepsilon_K^{\phantom{'}}\,\propto\, \sum_{i,j=c,t}\,\eta_{ij}\;
\mathrm{Im}\!\left[\bV_{\! id}^{\phantom{*}}\bV_{\! is}^*
\bV_{\! jd}^{\phantom{*}}\bV_{\! js}^*\right]\, S(r_i,r_j)
\; \propto\; A^2\lambda^6 \bar\eta\,\left\{\eta_{tt}\, A^2 \lambda^4 (1-\bar\rho) + P_c\right\}
\, .
\ee
The first term shows the CKM dependence of the dominant top contribution,
$P_c$ accounts for the charm corrections \cite{Buchalla:1995vs} and
the short-distance QCD corrections $\eta_{ij}$ are known to NLO \cite{BJW:90,HN:94,BG:10}.
The measured value of $|\varepsilon_K^{\phantom{'}}|$
determines an hyperbolic constraint in the $(\bar\rho,\bar\eta)$ plane, shown in Fig.~\ref{fig:UTfit}, taking into account the theoretical uncertainty in
the hadronic matrix element of the $\Delta S=2$ operator \cite{Aoki:2019cca}.

%%%%%%%%%%%%%%%%%%%%%%%%%%%%%%%%%%%%%%%%%%%%%%%%%%%%%%%%%%%%%%%%%%%%%%%%%%%

\subsection{$\cCP$ asymmetries in B decays}

The semileptonic decays \
$B_q^0\to X^- \ell^+\nu_\ell$ \ and \ $\bar B^0_q\to X^+ \ell^-\bar\nu_\ell$ \ ($q=d,s$) provide the most direct way to measure the amount of $\cCP$ violation in the $B^0$--$\bar B^0$ mixing matrix, through
% an asymmetry analogous to Eq.~(\ref{eq:deltaL}):
%
\beqn\label{eq:aSL}
a_{\mathrm{sl}}^q & \equiv &
\frac{\Gamma(\bar B^0_q\to X^- \ell^+\nu_\ell) - \Gamma(B^0_q\to X^+ \ell^-\bar\nu_\ell)}{\Gamma(\bar B^0_q\to X^- \ell^+\nu_\ell) + \Gamma(B^0_q\to X^+ \ell^-\bar\nu_\ell)}
\; =\; {|p|^4-|q|^4 \over |p|^4+|q|^4}
\;\approx\, 4\; \mathrm{Re}\, (\bar\varepsilon_{B_q^0})
\no\\[5pt]
&\approx &\frac{|\Gamma_{12}|}{|M_{12}|}\,\sin{\phi_q}
\;\approx\;\frac{|\Delta\Gamma_{B^0_q}|}{|\Delta M_{B^0_q}|}\,\tan{\phi_q}
\, .
\eeqn
This asymmetry is expected to be tiny because
%%% $|\Delta\Gamma_{B^0}/\Delta M_{B^0}| \approx
$|\Gamma_{12}/M_{12}| \sim m_b^2/m_t^2 \ll 1$.
Moreover, there is an additional GIM suppression in the relative mixing phase
$\phi_q\equiv\arg{\left(-M_{12}/\Gamma_{12}\right)} \sim (m_c^2-m_u^2)/ m_b^2$,
implying a value of $|q/p|$  very close to 1.
Therefore, $a_{\mathrm{sl}}^q$ could be very sensitive to new sources of $\cCP$ violation beyond the SM, contributing to $\phi_q$.
The present measurements give \cite{Tanabashi:2018oca,Amhis:2019ckw}
\bel{eq:aSL_num}
\mathrm{Re}(\bar\varepsilon_{B_d^0})\, =\, (-0.5\pm 0.4)\cdot 10^{-3}\, ,
\qquad\qquad
\mathrm{Re}(\bar\varepsilon_{B_s^0})\, =\, (-0.15\pm 0.70)\cdot 10^{-3}\, .
\ee

The large $B^0$--$\bar B^0$ mixing provides a different way to generate the
required $\cCP$-violating interference.
There are quite a few nonleptonic final states which are reachable
both from a $B^0$ and a $\bar B^0$. For these flavour non-specific decays
the $B^0$ (or $\bar B^0$) can decay directly to the given final state $f$,
or do it after the meson has been changed to its antiparticle via the
mixing process; \ie there are two different amplitudes,
$A(B^0\to f)$ and $A(B^0\to\bar B^0\to f)$, corresponding to two possible
decay paths. $\cCP$-violating effects can then result from the interference
of these two contributions.

The time-dependent decay probabilities for the decay of a neutral
$B$ meson created at the time $t_0=0$ as a pure $B^0$
($\bar B^0$) into the final state $f$ \ ($\bar f\equiv \cCP\, f$) are:
\beqn
\label{eq:decay_b}
\Gamma[B^0(t)\to f] &\!\!\propto &\!\!
{1\over 2}\, e^{-\Gamma_{B^0} t}\, \left(|A_f|^2 + |\bar A_f|^2\right)\,
\left\{ 1 + C_f\, \cos{(\Delta M_{B^0} t)} - S_f\, \sin{(\Delta M_{B^0} t)} \right\}\, ,
\nonumber\\[7pt]
%\label{eq:decay_bbar}
\Gamma[\bar B^0(t)\to \bar f] &\!\!\propto &\!\!
{1\over 2}\, e^{-\Gamma_{B^0} t}\,\left( |\bar A_{\bar f}|^2 + |A_{\bar f}|^2\right)\,       \left\{ 1 - C_{\bar f}\, \cos{(\Delta M_{B^0} t)} + S_{\bar f}\, \sin{(\Delta M_{B^0} t)} \right\}\, ,
\quad\eeqn
where the tiny $\Delta\Gamma_{B^0}$ corrections have been neglected
and we have introduced the notation
$$
A_f \equiv A[B^0\to f] \, , \qquad\qquad
\bar A_f \equiv -A[\bar B^0\to f] \, , \qquad\qquad
\bar\rho_f\equiv \bar A_f / A_f \, ,
$$
\bel{eq:Bnotation}
A_{\bar f} \equiv A[B^0\to \bar f]\, , \qquad\qquad
\bar A_{\bar f} \equiv -A[\bar B^0\to \bar f] \, , \qquad\qquad
\rho_{\bar f}\equiv A_{\bar f} / \bar A_{\bar f}\, ,
\ee
$$
C_f \equiv \frac{1 - |\bar\rho_f|^2}{1 + |\bar\rho_f|^2} \, , \qquad\quad
S_f \equiv \frac{2 \,\mathrm{Im}\left( {q\over p} \,\bar\rho_f\right)}{1 + |\bar\rho_f|^2}
\, , \qquad\quad
C_{\bar f} \equiv - \frac{1 - |\rho_{\bar f}|^2}{1 + |\rho_{\bar f}|^2}\, , \qquad\quad
S_{\bar f} \equiv \frac{-2 \,\mathrm{Im}\left( {p\over q}\, \rho_{\bar f}\right)}{1 + |\rho_{\bar f}|^2}\, .
$$

$\cCP$ invariance demands the probabilities of $\cCP$-conjugate processes to be
identical. Thus, $\cCP$ conservation requires
$A_f = \bar A_{\bar f}$, $A_{\bar f} = \bar A_f$,
$\bar\rho_f = \rho_{\bar f}$ and
$\mathrm{Im}({q\over p}\, \bar\rho_f) = \mathrm{Im}({p\over q}\, \rho_{\bar f})$,
\ie $C_f = - C_{\bar f}$ and $S_f = - S_{\bar f}$.
Violation of any of the first three equalities would be a signal of
direct $\cCP$ violation. The fourth equality tests $\cCP$ violation generated by the interference of the direct decay $B^0\to f$ and the
mixing-induced decay $B^0\to\bar B^0\to f$.

For $B^0$ mesons
% $|\Gamma_{12}/M_{12}|<<1$, implying
%
\be
\left. {q\over p}\right|_{B^0_q} \;\approx\; \sqrt{{M_{12}^*\over M_{12}}} \;\approx\;
{\bV_{\!\! tb}^* \bV_{\!\! tq}^{\phantom{*}}
\over \bV_{\!\! tb}^{\phantom{*}} \bV_{\!\! tq}^*}
\;\equiv\; e^{-2 i \phi^M_q} \, ,
\ee
where $\phi^M_d =  \beta +\cO(\lambda^4)$ and $\phi^M_s = - \beta_s+\cO(\lambda^6)$.
The angle $\beta$ is defined in Eq.~(\ref{eq:angles}), while
$\beta_s\equiv \arg{\left[ -\left(\bV^{\phantom{*}}_{\! ts}\bV^*_{\! tb}\right)/
   \left(\bV^{\phantom{*}}_{\! cs}\bV^*_{\! cb}\right)\right]} = \lambda^2\eta +\cO(\lambda^4)$
is the equivalent angle in the $B^0_s$ unitarity triangle, which is predicted to be tiny.
Therefore, the mixing ratio $q/p$ is given by a known weak phase.

An obvious example of final states $f$ which can be reached both from the
$B^0$ and the $\bar B^0$ are $\cCP$ eigenstates; \ie states such that
$\bar f = \zeta_f f$ \ ($\zeta_f = \pm 1$).
In this case, $A_{\bar f} = \zeta_f A_f$,
$\bar A_{\bar f} = \zeta_f \bar A_f$,
$\rho_{\bar f}= 1/ \bar\rho_f$,
$C_{\bar f}= C_f$ and $S_{\bar f}= S_f$.
A non-zero value of $C_f$ or $S_f$ signals then $\cCP$ violation.
The ratios $\bar\rho_f$ and $\rho_{\bar f}$ depend in general on the
underlying strong dynamics. However,
for $\cCP$ self-conjugate final states, all dependence on the
strong interaction disappears if only one weak amplitude contributes to
the $B^0\to f$ and $\bar B^0\to f$ transitions \cite{CS:80,BS:81}.
In this case, we can write the decay amplitude as
$A_f = M e^{i \phi^D}\! e^{i \delta_s}$, with $M = M^*$ and $\phi^D$
and $\delta_s$  weak and strong phases.
The ratios $\bar\rho_f$ and $\rho_{\bar f}$ are then given  by
\be
\rho_{\bar f} \, =\, \bar\rho_f^*\, =\, \zeta_f\, e^{2i\phi^D} \, .
\ee
The modulus $M$ and the unwanted strong phase cancel out completely
from these two ratios; $\rho_{\bar f}$ and $\bar\rho_f$ simplify to
a single weak phase, associated with the underlying weak quark transition.
Since $|\rho_{\bar f}| = |\bar\rho_f| = 1$,
the time-dependent decay probabilities
become much simpler. In particular, $C_f=0$ and there is no
longer any dependence on \ $\cos{(\Delta M_{B^0} t)}$.
Moreover, the coefficients of the sinusoidal terms
are then fully known in terms of CKM mixing angles only:
$S_f = S_{\bar f} = -\zeta_f\sin{[2(\phi^M_q + \phi^D)]}
\equiv -\zeta_f\sin{(2\Phi)}$.
In this ideal case, the time-dependent $\cCP$-violating decay asymmetry
\bel{eq:CleanAsym}
\cA^{\cCP}_{\bar B^0\to \bar f}\,\equiv\,
{\Gamma[\bar B^0(t)\to\bar f] - \Gamma[B^0(t)\to f] \over
 \Gamma[\bar B^0(t)\to \bar f] + \Gamma[B^0(t)\to f]} \; = \; -
\zeta_f\sin{(2 \Phi)} \, \sin{(\Delta M_{B^0} t)}
\ee
provides a direct and clean measurement of  the CKM parameters
\cite{KLPS:88}.

When several decay amplitudes with different phases
contribute, $|\bar{\rho}_f|\not=1$ and the interference term will
depend both on CKM parameters and on the strong dynamics embodied
in $\bar{\rho}_f$.
The leading contributions to
$\bar b\to\bar q' q'\bar q$ are either the tree-level $W$ exchange
or penguin topologies generated by gluon ($\gamma$, $Z$) exchange.
Although of higher order in the strong (electroweak) coupling, penguin
amplitudes are logarithmically enhanced by the virtual $W$ loop and
are  potentially competitive. Table~\ref{tab:decays} contains the CKM
factors associated with the two topologies for
different $B$ decay modes into $\cCP$ eigenstates.
% Also shown is the relevant angle $\Phi$.

%%%%%%%%%%%%% Table B decays into CP eigenstates %%%%%%%%%%%%%%%%%%
\begin{table}[tbh]
\centering
\caption{CKM factors and relevant angle $\Phi$ for some $B$ decays into
$\cCP$ eigenstates.} \label{tab:decays}
\renewcommand{\arraystretch}{1.2}
\begin{tabular}{l@{\hspace{0.5cm}}l@{\hspace{0.5cm}}l@{\hspace{0.5cm}}l@{\hspace{0.5cm}}l}
\hline\hline
Decay & Tree-level CKM & Penguin CKM & Exclusive channels & $\Phi$ \\
\hline
$\bar b \to \bar c  c \bar s$ & $A \lambda^2$ & $-A \lambda^2$ &
$ B^0_d\to J/\psi\, K_S ,
J/\psi\, K_L$ & $\beta$ \\
&&& $ B^0_s\to D_s^+ D_s^-, J/\psi\,  \phi$ & $-\beta_s$ \\ \hline
$\bar b\to\bar s s \bar s$ &  & $-A \lambda^2$ &
$ B^0_d\to K_S\phi, K_L\phi$ &
$\beta$ \\
&&& $B^0_s\to\phi\phi$ & $-\beta_s$ \\ \hline
$\bar b\to\bar d d\bar s$ &  & $-A \lambda^2$ &
$ B^0_s\to K_S K_S, K_L K_L$ &
$-\beta_s$ \\ \hline
$\bar b\to\bar c c\bar d$ & $-A\lambda^3$ & $A\lambda^3 (1-\rho - i \eta)$ &
$ B^0_d\to D^+ D^- , J/\psi\, \pi^0$ & $\approx \beta$
\\ &&& $ B^0_s\to J/\psi\,  K_S, J/\psi\,  K_L$ & $\approx -\beta_s$ \\ \hline
$\bar b\to\bar u u\bar d$ & $A \lambda^3 (\rho + i \eta)$ &
$A\lambda^3 (1 - \rho - i \eta)$ &
$ B^0_d\to\pi^+\pi^- , \rho^0\pi^0 ,\omega\pi^0$
& $\approx \beta+\gamma$ \\
 & & & $ B^0_s\to\rho^0 K_{S,L} ,\omega K_{S,L} ,\pi^0 K_{S,L}$
  & $\not= \gamma -\beta_s$ \\
%%% & & & $\phantom{ B^0_s\to} \rho^0 K_L, \omega K_L, \pi^0 K_L$ &
\hline
$\bar b\to\bar s s\bar d$ &  & $A\lambda^3 (1 - \rho - i \eta)$ &
$ B^0_d\to K_S K_S, K_L K_L, \phi\pi^0$ & 0
\\ &&& $ B^0_s\to K_S\phi, K_L\phi$ & $-\beta-\beta_s$
\\ \hline\hline
\end{tabular}
\end{table}
%%%%%%%%%%%%%%%%%%%%% END TABLE %%%%%%%%%%%%%%%%%%%%%%%%%%%%%%%%%%%%%%%%%%

The gold-plated decay mode is $B^0_d\to J/\psi\,  K_S$. In addition of having a clean
experimental signature, the two topologies have the same (zero) weak phase. The
$\cCP$ asymmetry provides then a clean measurement of the mixing angle
$\beta$, without strong-interaction uncertainties.
Fig.~\ref{fig:Belle_asym} shows the Belle measurement \cite{Adachi:2012et} of
time-dependent $\bar b\to c\bar c \bar s$ asymmetries for $\cCP$-odd
($B^0_d\to J/\psi\,  K_S$, $B^0_d\to \psi' K_S$, $B^0_d\to \chi_{c1} K_S$)
and  $\cCP$-even ($B^0_d\to J/\psi\,  K_L$)
final states. A very nice oscillation is manifest, with opposite signs for the two
different choices of $\zeta_f=\pm 1$. Including the information obtained
from other $\bar b\to c\bar c \bar s$ decays, one gets the world average \cite{Amhis:2019ckw}:
\begin{equation}\label{eq:beta}
\sin{(2\beta)} = 0.699\pm 0.017\, .    
\end{equation}
Fitting an additional $\cos{(\Delta M_{B^0} t)}$ term in the measured asymmetries results in $C_f=-0.005\pm 0.015$ \cite{Amhis:2019ckw}, confirming the expected null result.
An independent measurement of $\sin{2\beta}$ can be obtained from
$\bar b\to s\bar s \bar s$ and $\bar b\to d\bar d \bar s$ decays, which only
receive penguin contributions and, therefore, could be more sensitive to new-physics corrections in the loop diagram. These modes give \
$\sin{(2\beta)} = 0.648\pm 0.038$ \cite{Amhis:2019ckw}, in good agreement with (\ref{eq:beta}).

%%%%%%%%%%%%%%%  FIGURE %%%%%%%%%%%%%%%%%%%%%%%%%
\begin{figure}[tb]\centering
\begin{minipage}[t]{.4\linewidth}\centering
\includegraphics[height=5.6cm,clip]{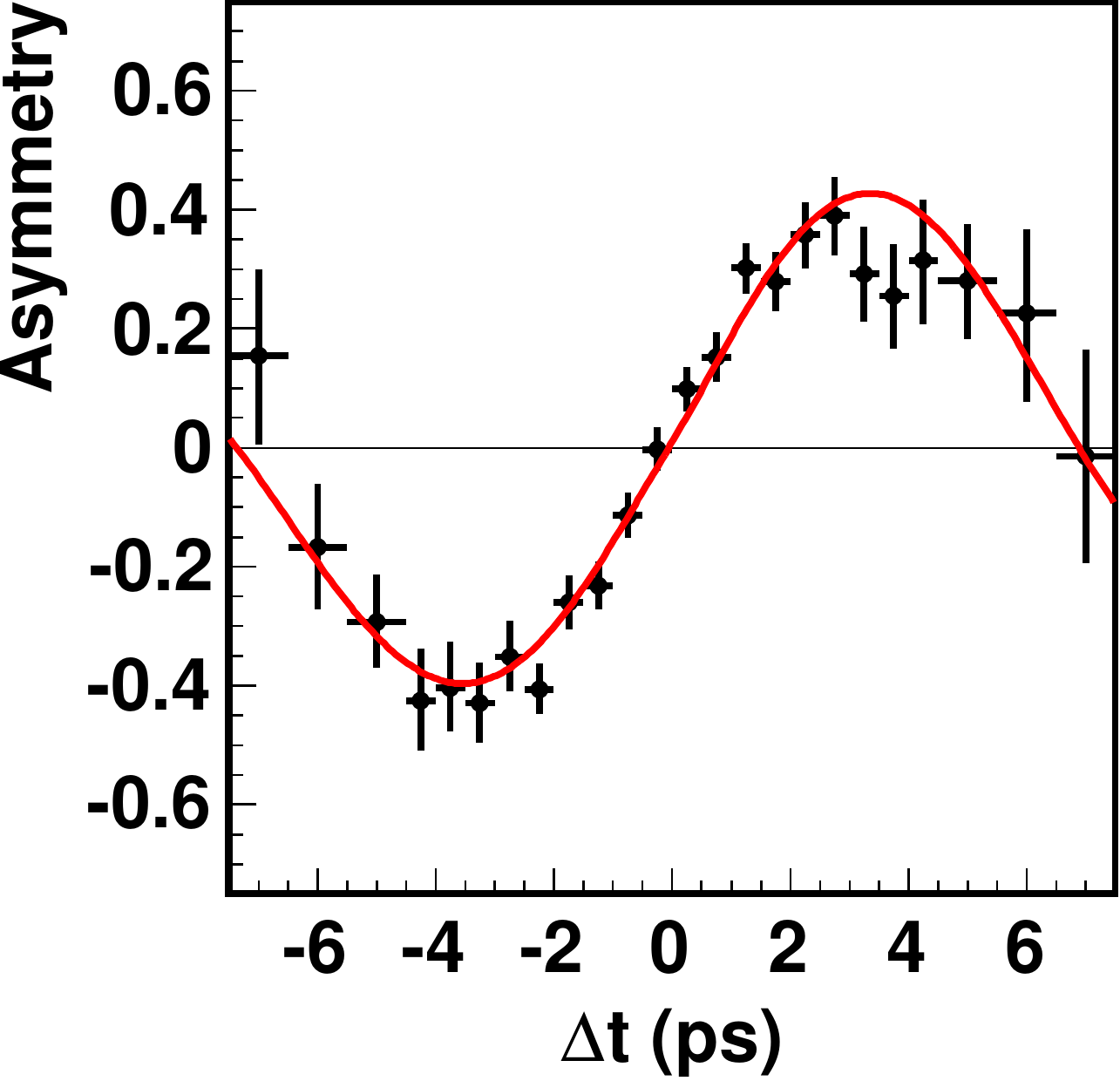}
\end{minipage}
\hskip 1.5cm
\begin{minipage}[t]{.4\linewidth}\centering
\includegraphics[height=5.6cm,clip]{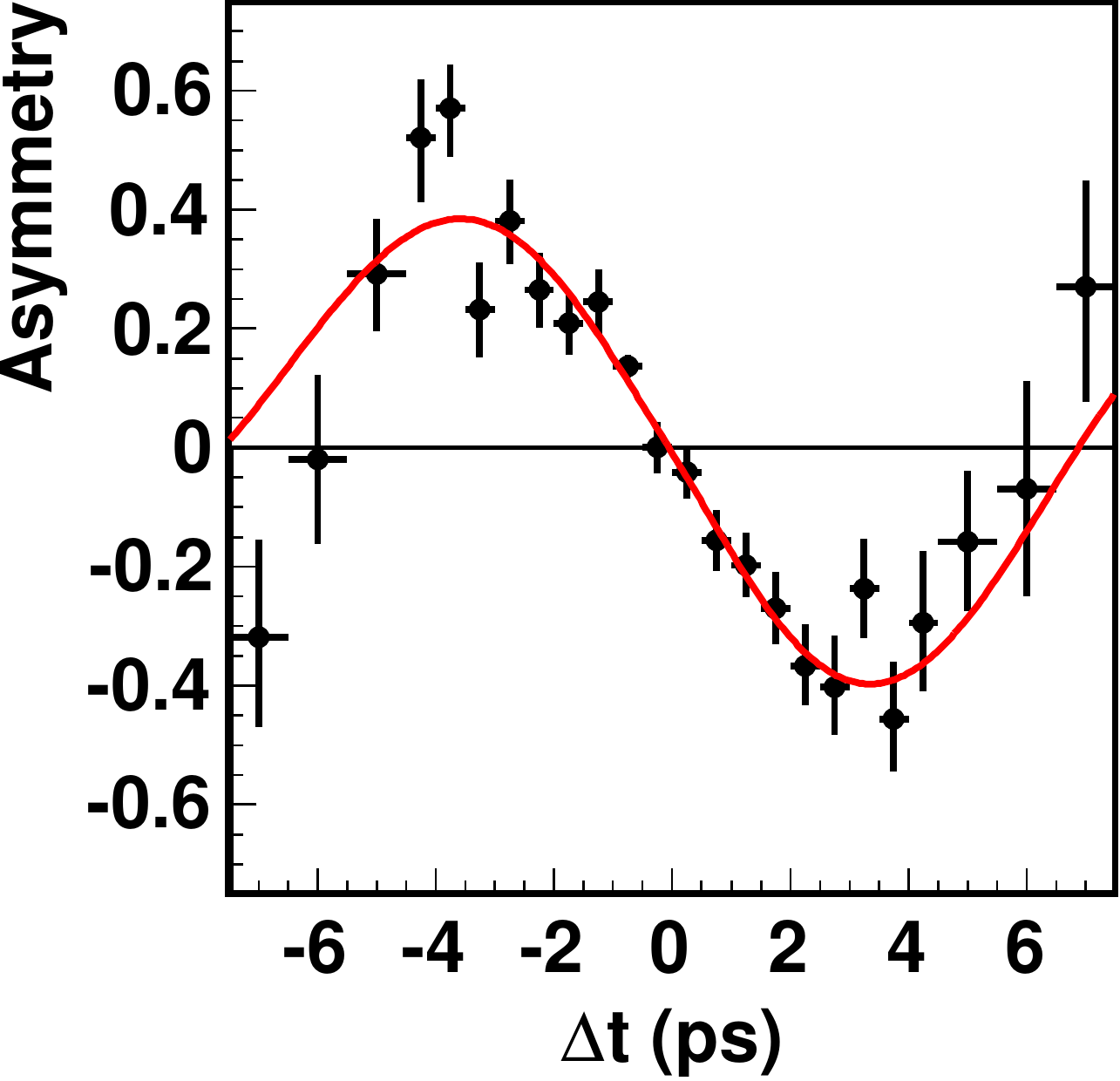}
\end{minipage}
\caption{Time-dependent asymmetries for $\cCP$-odd
($B^0_d\to J/\psi\, K_S$, $B^0_d\to \psi' K_S$, $B^0_d\to \chi_{c1} K_S$; $\zeta_f=-1$; left)
and  $\cCP$-even ($B^0_d\to J/\psi\, K_L$; $\zeta_f=+1$; right)
final states, measured by Belle \cite{Adachi:2012et}.}
\label{fig:Belle_asym}
\end{figure}
%%%%%%%%%%%%%%%%%%%%%%%%%%%%%%%%%%%%%%%%%%%%%

Eq. (\ref{eq:beta}) determines the angle $\beta$ up to a four-fold ambiguity: $\beta$, $\frac{\pi}{2}-\beta$, $\pi +\beta$ and $\frac{3\pi}{2}-\beta$.
The solution $\beta = (22.2\pm 0.7)^\circ$ is in remarkable agreement with the other phenomenological constraints on the unitarity triangle in Fig.~\ref{fig:UTfit}.
The ambiguity has been resolved through a time-dependent analysis of the Dalitz plot distribution in $B^0_d\to  D^{(*)} h^0$ decays ($h^0=\pi^0,\eta,\omega$), showing that $\cos{(2\beta)} = 0.91\pm 0.25$ and $\beta =(22.5\pm 4.6)^\circ$\cite{Adachi:2018itz}. This proves that $\cos{(2\beta)}$ is positive with a $3.7\,\sigma$ significance, while the multifold solution $\frac{\pi}{2}-\beta = (67.8\pm 0.7)^\circ$ is excluded at the $7.3\,\sigma$ level.

A determination of $\beta+\gamma =\pi-\alpha$ can be obtained from
$\bar b\to \bar u u\bar d$ decays, such as $B^0_d\to\pi\pi$ or $B^0_d\to\rho\rho$. However, the penguin contamination that carries a different weak phase can be sizeable. The time-dependent asymmetry in $B^0_d\to\pi^+\pi^-$ shows indeed a non-zero value for the $\cos{(\Delta M_{B^0} t)}$ term,
$C_f = -0.32\pm 0.04$ \cite{Amhis:2019ckw}, providing evidence of direct $\cCP$ violation and indicating the presence of an additional decay amplitude; therefore, $S_f = -0.63\pm 0.04 \not= \sin{2\alpha}$. One could still extract useful information on $\alpha$ (up to 16 mirror solutions),
using the isospin relations among the $B^0_d\to\pi^+\pi^-$, $B^0_d\to\pi^0\pi^0$ and $B^+\to\pi^+\pi^0$ amplitudes and their $\cCP$ conjugates \cite{GL:90}; however, only a loose constraint is obtained, given the limited experimental precision on $B^0_d\to\pi^0\pi^0$. Much stronger constraints are obtained from $B^0_d\to\rho^+\rho^-,\rho^0\rho^0$ because one can use the additional polarization information of two vectors in the final state to resolve the different contributions and, moreover, the small branching fraction
$\mathrm{Br}(B^0_d\to\rho^0\rho^0)=(9.6\pm 1.5)\cdot 10^{-7}$
\cite{Tanabashi:2018oca} implies a very small penguin contribution. Additional information can be obtained from $B^0_d,\bar B^0_d\to\rho^\pm\pi^\mp, a_1^\pm\pi^\mp$, although the final states are not $\cCP$ eigenstates. Combining all pieces of information results in \cite{Amhis:2019ckw}
\bel{eq:alpha}
\alpha \, =\, (84.9\,{}^{+\, 5.1}_{-\, 4.5})^\circ\, .
\ee

The angle $\gamma$ cannot be determined in $\bar b\to u\bar u\bar d$ decays such as $B_s^0\to\rho^0 K_S$ because the colour factors in the hadronic matrix element enhance the penguin amplitude with respect to the tree-level contribution. Instead, $\gamma$ can be measured through the tree-level decays $B\to D^{(*)} K^{(*)}$ ($\bar b\to \bar u c\bar s$) and $B\to \bar D^{(*)} K^{(*)}$ ($\bar b\to \bar c u\bar s$), using final states accessible in both $D^{(*)0}$ and $\bar D^{(*)0}$ decays and playing with the interference of both amplitudes \cite{GL:91,GW:91,AT:97}. The sensitivity can be optimized with Dalitz-plot analyses of the $D^0,\bar D^0$ decay products. The extensive experimental studies performed so far result in \cite{Amhis:2019ckw}
\bel{eq:gamma}
\gamma \, =\, (71.1\, {}^{+\, 4.6}_{-\, 5.3})^\circ\, .
\ee
Another ambiguous solution with $\gamma\leftrightarrow\gamma + \pi$ also exists.

%%%%%%%%%%%%%%%  FIGURE %%%%%%%%%%%%%%%%%%%%%%%%%
\begin{figure}[tb]\centering
\includegraphics[height=6.5cm,clip]{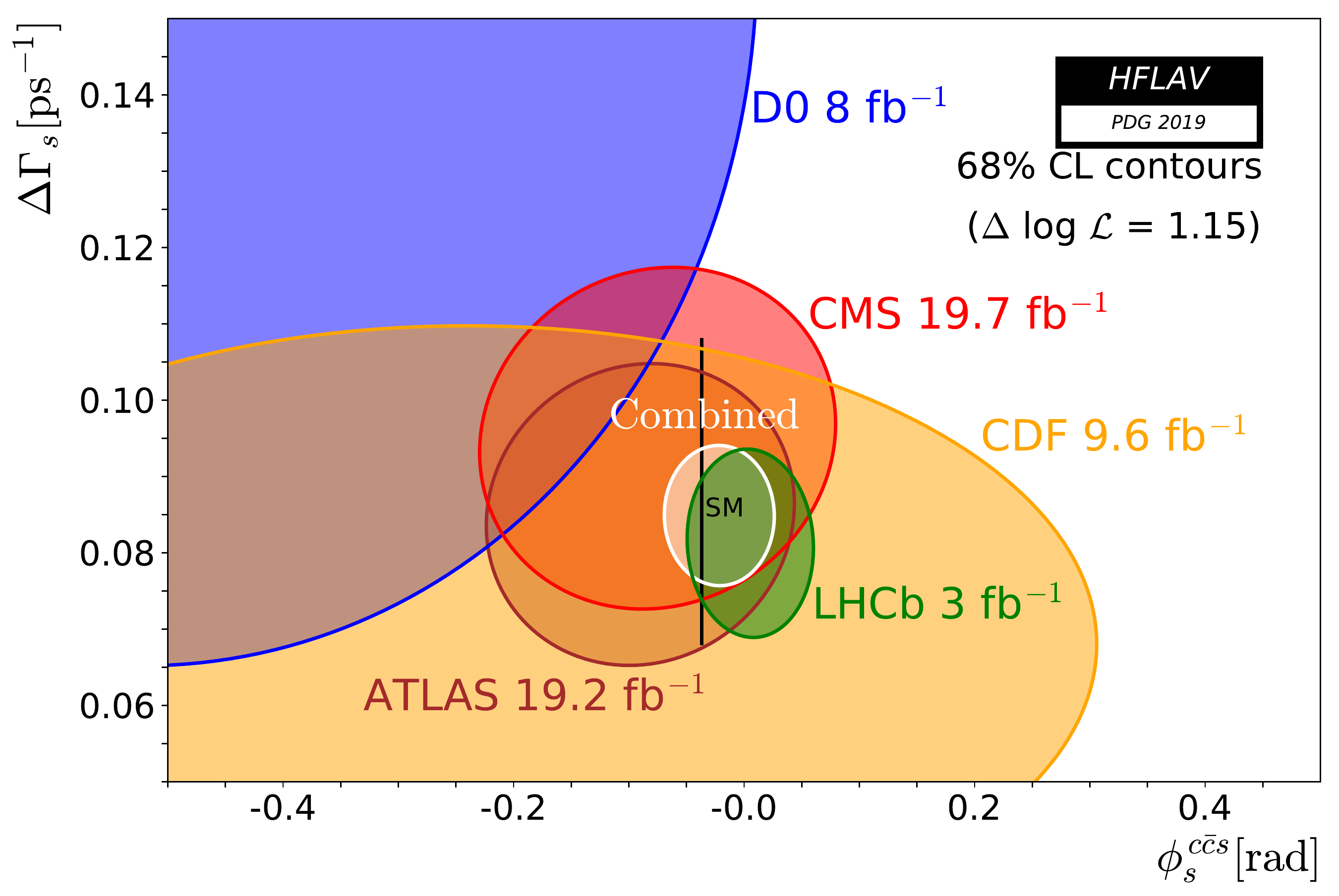}
\caption{68\% CL regions in $\Delta\Gamma_{B^0_s}$ and $\phi_s^{c\bar c s}$, extracted from $\,\bar b\to\bar c c\bar s$ $\,\cCP$ asymmetries of $B^0_s$ mesons \cite{Amhis:2019ckw}. The vertical black line shows the SM prediction \cite{Lenz:2006hd,Artuso:2015swg,Jubb:2016mvq}.}
\label{fig:Bs_asym}
\end{figure}
%%%%%%%%%%%%%%%%%%%%%%%%%%%%%%%%%%%%%%%%%%%%%

Mixing-induced $\cCP$ violation has been also searched for
in the decays of $B^0_s$ and $\bar B^0_s$ mesons into $J/\psi\, \phi, \psi(2\mathrm{S}) \phi, J/\psi\, K^+K^-, J/\psi\, \pi^+\pi^-$ and $D_s^+ D_s^-$. From the corresponding time-dependent $\cCP$ asymmetries,\footnote{The $\Delta\Gamma_{B^0_s}$ corrections to Eq.~\eqn{eq:decay_b} must be taken into account at the current level of precision.} one extracts correlated constraints on $\Delta\Gamma_s$ and the weak phase $\phi_s^{c\bar c s}\equiv 2 \Phi_s^{c\bar c s} \approx 2 \phi^M_s \approx -2\beta_s$ in Eq.~\eqn{eq:CleanAsym}, which are shown in Fig.~\ref{fig:Bs_asym}. 
They lead to \cite{Amhis:2019ckw}
\bel{eq:betas}
\beta_s\; = \; (0.60\pm 0.89)^\circ\, ,
\ee
in good agreement with the SM prediction
$\beta_s\approx \eta\lambda^2 \approx 1^\circ$.

\subsection{Global fit of the unitarity triangle}

The CKM parameters can be more accurately determined through a global fit to all available measurements, imposing the unitarity constraints and taking properly into account the theoretical uncertainties. The global fit shown in Fig.~\ref{fig:UTfit} uses frequentist statistics and gives \cite{CKMfitter}
\bel{eq:CKMfit}
\lambda\, =\, 0.22484\, {}^{+\, 0.00025}_{-\, 0.00006}\, ,
\qquad
A\, =\, 0.824\, {}^{+\, 0.006}_{-\, 0.015}\, ,
\qquad
\bar\rho\, =\, 0.157\, {}^{+\, 0.010}_{-\, 0.006}\, ,
\qquad
\bar\eta\, =\, 0.350\, {}^{+\, 0.008}_{-\, 0.007}\, .
\ee
This implies $\cJ = (3.06\, {}^{+\, 0.07}_{-\, 0.08})\cdot 10^{-5}$,
$\alpha = (91.7\, {}^{+\, 1.7}_{-\, 1.1})^\circ$,
$\beta = (22.6\, {}^{+\, 0.5}_{-\, 0.4})^\circ$ and
$\gamma = (65.8\, {}^{+\, 0.9}_{-\, 1.3})^\circ$.
Similar results are obtained by the UTfit group \cite{UTfit}, using instead a
Bayesian approach and a slightly different treatment of theoretical uncertainties.

\subsection{Direct $\cCP$ violation in $B$ decays}

The big data samples accumulated at the $B$ factories and the collider experiments have established the presence of direct $\cCP$ violation in
several decays of $B$ mesons. The most significative signals are \cite{Tanabashi:2018oca,Amhis:2019ckw}
$$
\cA^{\cCP}_{\bar B^0_d\to K^-\pi^+} = -0.083\pm 0.004\, ,
\qquad
\cA^{\cCP}_{\bar B^0_d\to \bar K^{*0}\eta} = 0.19\pm 0.05\, ,
\qquad
\cA^{\cCP}_{\bar B^0_d\to K^{*-}\pi^+} = -0.27\pm 0.04\, ,
$$
$$
C_{B^0_d\to\pi^+\pi^-} = -0.32\pm 0.04\, ,
\qquad\;
\gamma_{B\to D^{(*)}K^{(*)}}  = (71.1\, {}^{+\, 4.6}_{-\, 5.3})^\circ\, ,
\qquad\;
\cA^{\cCP}_{\bar B^0_s\to K^+\pi^-} = 0.221\pm 0.015\, ,
%%%\cA^{\cCP}_{B^-\to K^-D_{\cCP (+1)}} = 0.11\pm 0.04\, ,
$$
$$
\cA^{\cCP}_{B^-\to K^-\rho^0} = 0.37\pm 0.10\, ,
\qquad
\cA^{\cCP}_{B^-\to K^-\eta} = -0.37\pm 0.08\, ,
\qquad
\cA^{\cCP}_{B^-\to \pi^-\pi^+\pi^-} = 0.057\pm 0.013\, ,
%%%\cA^{\cCP}_{B^-\to K^-\pi^+\pi^-} = 0.027\pm 0.008\, ,
$$
$$
\cA^{\cCP}_{B^-\to K^-K^+K^-} = -0.033\pm 0.008\, ,
\qquad\qquad\qquad
\cA^{\cCP}_{B^-\to K^-K^+\pi^-} = -0.122\pm 0.021\, .
$$
\be
\cA^{\cCP}_{B^-\to K^-f_2(1270)} = -0.68\, {}^{+\, 0.19}_{-\,0.17}\, ,
\qquad\qquad\qquad
\cA^{\cCP}_{B^-\to \pi^- f_2(1270)} = 0.40\pm 0.06\, ,
%%%\cA^{\cCP}_{B^-\to \pi^-f_0(1370)} = 0.72\pm 0.22\, .
\ee
Another prominent observation of direct $\cCP$ violation has been done recently in charm decays \cite{Aaij:2019kcg}:
\be
\cA^{\mathrm{direct}\;\cCP}_{D^0\to K^-K^+}-\cA^{\mathrm{direct}\;\cCP}_{D^0\to \pi^-\pi^+}\, =\, (-15.7\pm 2.9)\cdot 10^{-4}\, ,
\ee
where the small contribution from $D^0$--$\bar D^0$ mixing has been subtracted, using the measured difference of reconstructed mean decay times of the two modes.
Unfortunately, owing to the unavoidable presence of strong
phases, a real theoretical understanding of the corresponding SM predictions is still lacking. Progress in this direction is needed to perform meaningful tests of the CKM mechanism of $\cCP$ violation and pin down any possible effects from new physics beyond the SM framework.

\section{Rare decays}

Complementary and very valuable information can be obtained from rare decays that in the SM are strongly suppressed by the GIM mechanism.
These processes are then sensitive to new-physics contributions with
a different flavour structure. Well-known examples are the $\bar K^0\to\mu^+\mu^-$ decay modes  \cite{Tanabashi:2018oca,Aaij:2020sbt},
\bel{eq:K2m_Br}
\mathrm{Br} (K_L\to\mu^+\mu^-)\, =\, (6.84\pm 0.11)\cdot 10^{-9}\, ,
\qquad\qquad
\mathrm{Br} (K_S\to\mu^+\mu^-)\, <\, 2.1\cdot 10^{-10}\quad (90\%\:\mathrm{CL})\, ,
\ee
which tightly constrain any hypothetical flavour-changing ($s\to d$) tree-level coupling of the $Z$ boson. In the SM, these decays receive short-distance contributions from the penguin and box diagrams displayed in Fig.~\ref{fig:K02m}. Owing to the unitarity of the CKM matrix, the resulting amplitude vanishes for equal up-type quark masses, which entails a heavy suppression:
\bel{eq:K2m_FlavStr}
\cM\,\propto\, \sum_{i=u,c,t} \bV_{\! is} \bV_{\! id}^*\; F(m_i^2/M_W^2)
\, =\, 
\bV_{\! cs} \bV_{\! cd}^*\; \tilde F(m_c^2/M_W^2) +
\bV_{\! ts} \bV_{\! td}^*\; \tilde F(m_t^2/M_W^2)\, ,
\ee
where $F(x)$ is a loop function and \ $\tilde F(x) \equiv F(x) - F(0)$. In addition to the loop factor $g^4/(16\pi^2)$, the charm contribution carries a mass suppression $\lambda m_c^2/M_W^2$, while the top term is proportional to $\lambda^5 A^2 (1-\rho+i\eta)\, m_t^2/M_W^2$. The large top mass compensates the strong Cabibbo suppression so that the top contribution is finally larger than the charm one. However, the total short-distance contribution to the $K_L$ decay, 
$\mathrm{Br} (K_L\to\mu^+\mu^-)_{\rms sd} = (0.79\pm 0.12) \cdot 10^{-9}$ \cite{Gorbahn:2006bm,Buchalla:1995vs}, is nearly one order of magnitude below the experimental measurement \eqn{eq:K2m_Br}, while $\mathrm{Br} (K_S\to\mu^+\mu^-)_{\rms sd} \approx 1.7\cdot 10^{-13}$ \cite{Cirigliano:2011ny}.

The decays $K_L\to\mu^+\mu^-$ and $K_S\to\mu^+\mu^-$ are actually dominated by long-distance contributions, through \ $K_L\to \pi^0,\eta,\eta'\to \gamma\gamma\to\mu^+\mu^-$ \ and \ $K_S\to \pi^+\pi^-\to \gamma\gamma\to\mu^+\mu^-$. The absorptive component from two on-shell intermediate photons nearly saturates the measured $\mathrm{Br} (K_L\to\mu^+\mu^-)$ \cite{GomezDumm:1998gw}, while the $K_S$ decay rate is predicted to be  $\mathrm{Br} (K_S\to\mu^+\mu^-) = 5.1\cdot 10^{-12}$ \cite{Cirigliano:2011ny,Ecker:1991ru}. These decays can be rigorously analised with chiral perturbation theory techniques \cite{Cirigliano:2011ny}, but the strong suppression of their short-distance contributions does not make possible to extract useful information on the CKM parameters. Nevertheless, it is possible to predict the longitudinal polarization $P_L$ of either muon in the $K_L$ decay, a $\cCP$-violating observable which in the SM is dominated by indirect 
$\cCP$ violation from $K^0$--$\bar K^0$ mixing: $|P_L| = (2.6\pm 0.4)\cdot 10^{-3}$ \cite{Ecker:1991ru}.

%%%%%%%%%%%%%%%  FIGURE %%%%%%%%%%%%%%%%%%%%%%%%%
\begin{figure}[t]\centering
\includegraphics[width=10cm,clip]{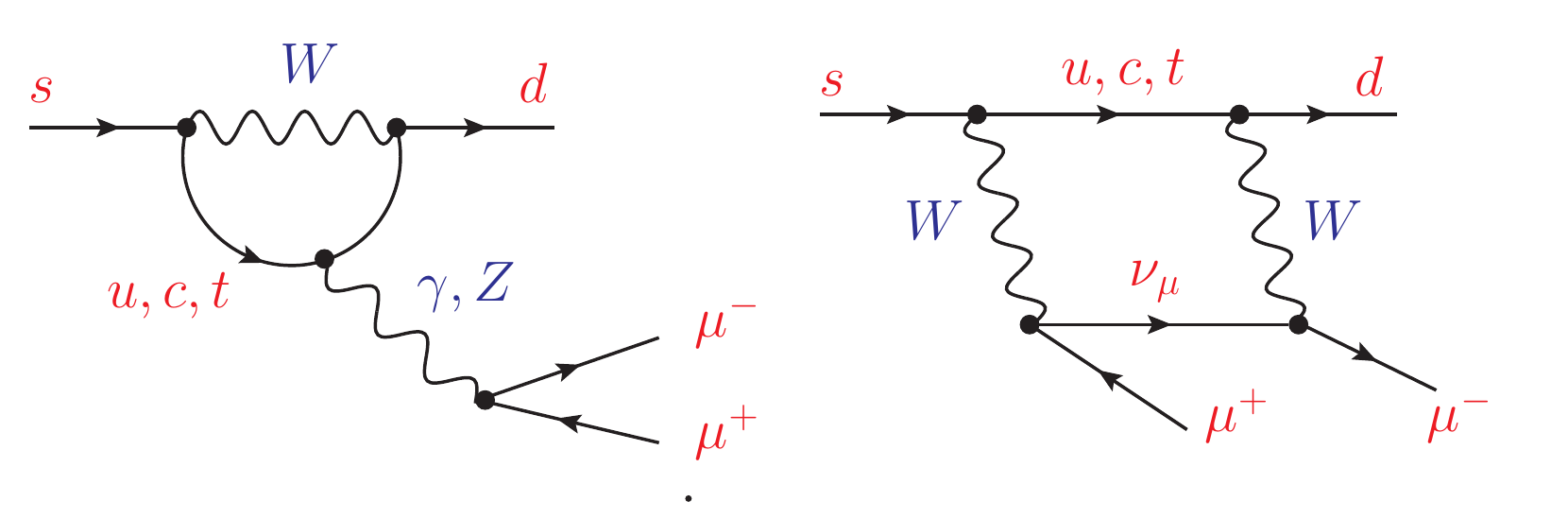}
\caption{Short-distance penguin (left) and box (right) SM contributions to $\bar K^0\to\mu^+\mu^-$.}
\label{fig:K02m}
\end{figure}
%%%%%%%%%%%%%%%%%%%%%%%%%%%%%%%%%%%%%%%%%%%%%

Other interesting kaon decay modes such as $K^0\to \gamma\gamma$, $K\to \pi\gamma\gamma$ and $K\to \pi\ell^+\ell^-$ are also governed by long-distance physics \cite{Cirigliano:2011ny,Ecker:1987hd}. Of particular interest is the decay $K_L\to \pi^0 e^+ e^-$, since it is sensitive to new sources of $\cCP$ violation. At lowest order in $\alpha$ the decay proceeds through $K_2^0\to\pi^0\gamma^*$ that violates $\cCP$, while the $\cCP$-conserving contribution through $K_L^0\to\pi^0\gamma^*\gamma^*$ is suppressed by an additional power of $\alpha$ and it is found to be below $10^{-12}$ \cite{Cirigliano:2011ny}.
The $K_L\to \pi^0 e^+ e^-$ transition is then dominated by the
$\cO(\alpha)$ $\cCP$-violating contributions \cite{Ecker:1987hd}, both from $K^0$--$\bar K^0$ mixing and direct $\cCP$ violation. The estimated rate $\mathrm{Br}(K_L\to \pi^0 e^+ e^-) =(3.1\pm 0.9)\cdot 10^{-11}$ \cite{Cirigliano:2011ny,Buras:1994qa,Buchalla:2003sj} is only a factor ten smaller than the present (90\% CL) upper bound of $2.8\cdot 10^{-10}$ \cite{AlaviHarati:2003mr} and should be reachable in the near future.

The decays $K^\pm\to\pi^\pm\nu\bar\nu$ and
$K_L\to\pi^0\nu\bar\nu$ provide a more direct access to CKM information because long-distance effects play a negligible role. The decay amplitudes are dominated by $Z$-penguin and $W$-box loop diagrams of the type shown in Fig.~\ref{fig:K02m} (replacing the final muons by neutrinos), and are proportional to the hadronic $K\pi$ matrix element of the $\Delta S=1$ vector current, which (assuming isospin symmetry) is extracted from $K\to\pi\ell\nu_\ell$ decays. The small long-distance and isospin-violating corrections can be estimated with chiral perturbation theory.
The neutral decay is $\cCP$ violating and proceeds almost entirely through direct $\cCP$ violation (via interference with mixing). Taking the CKM inputs from other observables, the predicted SM rates are \cite{Buras:2005gr,Brod:2010hi,Gorbahn:KAON2019}:
\be
\mathrm{Br} (K^+  \to \pi^+ \nu \bar{\nu})_{\mathrm{th}}
 = (8.5 \pm 0.6) \cdot 10^{-11}\, ,
\qquad\qquad
\mathrm{Br} (K_L \to \pi^0 \nu \bar{\nu})_{\mathrm{th}}
 =  (2.9 \pm 0.3) \cdot 10^{-11}\, .
\ee
The uncertainties are largely parametrical, due to CKM input,
the charm and top masses and $\alpha_s(M_Z)$.
On the experimental side, the current upper bounds on the charged \cite{Ruggiero:KAON2019} and neutral \cite{Ahn:2018mvc} modes are
%charged kaon mode was already observed
%\cite{E949:08}, while only an upper bound on the neutral mode
%has been achieved \cite{E391a:08}:
%
\be
\mathrm{Br} (K^+  \to \pi^+ \nu \bar{\nu}) < 1.85\cdot 10^{-10}\, ,
%%% =  (1.73^{+1.15}_{-1.05}) \cdot 10^{-10}\, ,
\qquad\qquad
\mathrm{Br} (K_L \to \pi^0 \nu \bar{\nu})
<   3.0  \cdot 10^{-9}  \quad  (90 \% \, {\rm CL}).
\ee
The ongoing CERN NA62 experiment aims to reach $\cO(100)$ $K^+  \to \pi^+ \nu \bar{\nu}$ events (assuming SM rates), while increased sensitivities on the $K_L  \to \pi^0 \nu \bar{\nu}$ mode are expected to be achieved by the KOTO experiment at J-PARC. These experiments will start to seriously probe the new-physics potential of these decays.

%is currently collecting $K^+ \to \pi^+ \nu \bar{\nu}$ data with the goal of measuring the branching ratio with an accuracy of 10\% \cite{Anelli:2005ju}. The neutral mode is being searched for by the KOTO experiment at J-PARC \cite{Beckford:2017gsf}, which aims to reach the SM predicted branching ratio.

The inclusive decay $\bar B\to X_s \gamma$ provides another powerful test of the SM flavour structure at the quantum loop level. It proceeds through a $b\to s \gamma$ penguin diagram with an on-shell photon. The present experimental world average, 
$\mathrm{Br}(\bar B\to X_s \gamma)_{E_\gamma\ge 1.6~\mathrm{GeV}} =
(3.32\pm 0.15)\cdot 10^{-4}$ \cite{Amhis:2019ckw},
agrees very well with the SM theoretical prediction,
$\mathrm{Br}(\bar B\to X_s \gamma)_{E_\gamma\ge 1.6~\mathrm{GeV}}^{\mathrm{th}} = (3.40\pm 0.17)\cdot 10^{-4}$ \cite{Misiak:2015xwa,Czakon:2015exa,Misiak:2020vlo}, obtained  at the next-to-next-to-leading order.

%The strongest bound on FCNC transitions in charm decays is 
%$\mathrm{Br}(D^0\to\mu^+\mu^-) < 6.2 \cdot 10^{-9}$ (90\% CL) \cite{Aaij:2013cza}.

The LHC experiments have recently reached the SM sensitivity for the $B_d^0$ and $B_s^0$ decays into $\mu^+\mu^-$ pairs \cite{CMS:2014xfa,Aaij:2017vad,Aaboud:2018mst}. The current world averages \cite{Tanabashi:2018oca},\footnote{$\overline{\mathrm{Br}}$ is the time-integrated branching ratio, which for $B_s^0$ is slightly affected by the sizeable value of $\Delta\Gamma_s$ \cite{DeBruyn:2012wk}.}
\bel{eq:B2m}
\overline{\mathrm{Br}} (B^0_d\to\mu^+\mu^-)\, =\, (1.4\, {}^{+\, 1.6}_{-\, 1.4})\cdot 10^{-10}\, ,
\qquad\qquad
\overline{\mathrm{Br}} (B_s^0\to\mu^+\mu^-)\, =\, (3.0\pm 0.4)\cdot 10^{-9}\, ,
\ee
agree with the SM predictions
$\overline{\mathrm{Br}} (B_d^0\to\mu^+\mu^-) = (1.06\pm 0.09)\cdot 10^{-10}$
and $\overline{\mathrm{Br}} (B_s^0\to\mu^+\mu^-) = (3.65\pm 0.23)\cdot 10^{-9}$ \cite{Bobeth:2013uxa}.
Other interesting FCNC decays with $B$ mesons are $\bar B\to K^{(*)} l^+l^-$
%%% $\bar B^0\to l^+l^-$ 
and $\bar B\to K^{(*)}\nu\bar\nu$ \cite{Antonelli:2009ws}.

\section{Flavour constraints on new physics}

The CKM matrix provides a very successful description of flavour phenomena, as it is clearly exhibited in the unitarity test of Fig.~\ref{fig:UTfit}, showing how very different observables converge into a single choice of CKM parameters. This is a quite robust and impressive result. One can perform separate tests with different subsets of measurements, according to their $\cCP$-conserving or $\cCP$-violating nature, or splitting them into tree-level and loop-induced processes. In all cases, one finally gets a closed triangle and similar values for the fitted CKM entries \cite{CKMfitter,UTfit}. However, the SM mechanism of flavour mixing and $\cCP$ violation is conceptually quite unsatisfactory because it does not provide any dynamical understanding of the numerical values of fermion masses, and mixings. We completely ignore the reasons why the fermion spectrum contains such a hierarchy of different masses, spanning many orders of magnitude, and which fundamental dynamics is behind the existence of three flavour generations and their observed mixing structure.

The phenomenological success of the SM puts severe constraints on possible scenarios of new physics. The absence of any clear signals of new phenomena in the LHC searches is pushing the hypothetical new-physics scale at higher energies, above the TeV. The low-energy implications of new dynamics beyond the SM can then be analysed in terms of an effective Lagrangian, containing only the known SM fields:
\bel{eq:EffLagNP}
\cL_{\rms eff}\, =\, L_{\rms SM}\, +\,\sum_{d>4} \sum_k\,\frac{c_k^{(d)}}{\Lambda_{\rms NP}^{d-4}}\; O_k^{(d)}\, .
\ee
The effective Lagrangian is organised as an expansion in terms of dimension-$d$ operators $O_k^{(d)}$, invariant under the SM gauge group, suppressed by the corresponding powers of the new-physics scale $\Lambda_{\rms NP}$. The dimensionless couplings $c_k^{(d)}$ encode information on the underlying dynamics. The lowest-order term in this dimensional expansion is the SM Lagrangian that contains all allowed operators of dimension 4. 

At low energies, the terms with lower dimensions dominate the physical transition amplitudes. There is only one operator with $d=5$ (up to Hermitian conjugation and flavour assignments), which violates lepton number by two units and is then related with the possible existence of neutrino Majorana masses \cite{Weinberg:1979sa}. Taking $m_\nu\gap 0.05$~eV, one estimates a very large lepton-number-violating scale $\Lambda_{\rms NP}/c^{(5)}\lap 10^{15}$~GeV \cite{Pich:2013lsa}. Assuming lepton-number conservation, the first signals of new phenomena should be associated with $d=6$ operators.

One can easily analyse the possible impact of $\Delta F=2$ ($F=S,C,B$) four-quark operators ($d=6$), such as the SM left-left operator in Eq.~\eqn{eq:DB_op} that induces a $\Delta B=2$ transition. Since the SM box diagram provides an excellent description of the data, hypothetical new-physics contributions can only be tolerated within the current uncertainties, which puts stringent upper bounds on the corresponding Wilson coefficients $\tilde c_k\equiv c_k^{(6)}/\Lambda_{\rms NP}^2$.
For instance, $\Delta M_{B^0_d}$ and $S_{J/\psi K_S}$ imply that the real ($\cCP$-conserving) and imaginary ($\cCP$-violating) parts of $\tilde c_k$ must be below $10 ^{-6}\;\mathrm{TeV}^{-2}$ for a $(\bar b_L\gamma^\mu d_L)^2$ operator, and nearly one order of magnitude smaller ($10^{-7}$) for $(\bar b_R d_L)(\bar b_L d_R)$ \cite{Isidori:2010kg}. Stronger bounds are obtained in the kaon system from $\Delta M_{K^0}$ and $\varepsilon_K$. For the $(\bar s_L\gamma^\mu d_L)^2$ operator the real (imaginary) coefficient must be below $10^{-6}$ ($3\cdot 10^{-9}$), while the corresponding bounds for $(\bar s_R d_L)(\bar s_L d_R)$ are $7\cdot 10^{-9}$ ($3\cdot 10^{-11}$), in the same TeV${}^{-2}$ units \cite{Isidori:2010kg}. Taking the coefficients $c_k^{(6)}\sim\cO(1)$, this would imply $\Lambda_{\rms NP} > 3\cdot 10^{3}$~TeV  for $B^0_d$ 
and $\Lambda_{\rms NP} > 3\cdot 10^5$~TeV for $K^0$. Therefore, two relevant messages emerge from the data:
\begin{enumerate}
\item A generic flavour structure with coefficients $c_k^{(6)}\sim\cO(1)$ is completely ruled out at the TeV scale.
\item New flavour-changing physics at $\Lambda_{\rms NP}\sim 1$~TeV could only be possible if the corresponding Wilson coefficients $c_k^{(6)}$ inherit the strong SM suppressions generated by the GIM mechanism.
\end{enumerate}

The last requirement can be satisfied by assuming that the up and down Yukawa matrices are the only sources of quark-flavour symmetry breaking (minimal flavour violation) \cite{Hall:1990ac,Chivukula:1987py,DAmbrosio:2002vsn}. In the absence of Yukawa interactions, the SM Lagrangian has a $\cG\equiv U(3)_{L_L}\otimes U(3)_{Q_L} \otimes U(3)_{\ell_R} \otimes U(3)_{u_R} \otimes U(3)_{d_R}$ global flavour symmetry, because one can rotate arbitrarily in the 3-generation space each one of the five SM fermion components in Eq.~\eqn{eq:structure}. The Yukawa matrices are the only explicit breakings of this large symmetry group. Assuming that the new physics does not introduce any additional breakings of the flavour symmetry $\cG$ (beyond insertions of Yukawa  matrices), one can easily comply with the flavour bounds.
Otherwise,  flavour data  provide  very  strong  constraints  on  models  with
additional  sources  of  flavour  symmetry  breaking  and probe  physics  at  energy  scales  not  directly  accessible at accelerators.

The subtle SM cancellations suppressing FCNC transitions could be easily destroyed in the presence of new physics contributions. To better appreciate the non-generic  nature of the flavour structure, let us analyse the simplest extension of the SM scalar sector with a second Higgs doublet, which increases the number of quark Yukawas:
\bel{eq:2HDM_Yukawa}
\cL_Y\, =\, -\sum_{a=1}^2\left\{ \bar Q_L'\, \cY^{(a)\raisebox{3pt}{$\scriptscriptstyle\prime$}}_d \phi_a\, d'_R
+ \bar Q_L'\, \cY^{(a)\raisebox{3pt}{$\scriptscriptstyle\prime$}}_u \phi^c_a\, u'_R
+ \bar L'_L\, \cY^{(a)\raisebox{3pt}{$\scriptscriptstyle\prime$}}_\ell \phi_a\, \ell'_R
\right\} \, +\, \mathrm{h.c.}\, ,
\ee
where $\phi_a^T = (\phi_a^{(+)},\phi_a^{(0)})$ are the two scalar doublets,
$\phi_a^c$ their $\cC$-conjugate fields, and $Q'_L$ and $L'_L$ the left-handed quark and lepton doublets, respectively. All fermion fields are written as three-dimensional flavour vectors and $\cY^{(a)\raisebox{3pt}{$\scriptscriptstyle\prime$}}_f$ are $3\times 3$ complex matrices in flavour space. With an appropriate scalar potential, the neutral components of the scalar doublets acquire vacuum expectation values
$\langle 0|\phi_a^{(0)}|0\rangle = v_a\,\e^{i\theta_a}$. It is convenient to make a $U(2)$ transformation in the space of scalar fields, $(\phi_1,\phi_2)\to (\Phi_1,\Phi_2)$, so that only the first doublet has a non-zero vacuum expectation value $v = (v_1^2+v_2^2)^{1/2}$. $\Phi_1$ plays then the same role as the SM Higgs doublet, while $\Phi_2$ does not participate in the electroweak symmetry breaking. In this scalar basis the Yukawa interactions become more transparent:
\bel{eq:Yukawa_Higgs_basis}
\sum_{a=1}^2 \cY^{(a)\raisebox{3pt}{$\scriptscriptstyle\prime$}}_{d,\ell}\, \phi_a\, =\,\sum_{a=1}^2 Y^{(a)\raisebox{3pt}{$\scriptscriptstyle\prime$}}_{d,\ell}\, \Phi_a\, ,
\qquad\qquad\qquad
\sum_{a=1}^2 \cY^{(a)\raisebox{3pt}{$\scriptscriptstyle\prime$}}_{u}\, \phi^c_a\, =\,\sum_{a=1}^2 Y^{(a)\raisebox{3pt}{$\scriptscriptstyle\prime$}}_{u}\, \Phi^c_a\, .
\ee
The fermion masses originate from the $\Phi_1$ couplings, because $\Phi_1$ is the only field acquiring a vacuum expectation value:
\bel{eq:MassMatrix}
M_f'\, =\, Y^{(1)\raisebox{3pt}{$\scriptscriptstyle\prime$}}_{f}\;\frac{v}{\sqrt{2}}\, .
\ee
The diagonalization of these fermion mass matrices proceeds in exactly the same way as in the SM, and defines the fermion mass eigenstates $d_i$, $u_i$, $\ell_i$, with diagonal mass matrices $M_f$, as described in Section~\ref{sec:flavour}. However, in general, one cannot diagonalize simultaneously all Yukawa matrices, \ie in the fermion mass-eigenstate basis the matrices $Y_f^{(2)}$ remain non-diagonal, giving rise to dangerous flavour-changing transitions mediated by neutral scalars. The appearance of FCNC interactions represents a major phenomenological shortcoming, given the very strong experimental bounds on this type of phenomena. 

To avoid this disaster, one needs to implement ad-hoc dynamical restrictions to guarantee the suppression of FCNC couplings at the required level. Unless the Yukawa couplings are very small or the scalar bosons very heavy, a specific flavour structure is required by the data. The unwanted non-diagonal neutral couplings can be eliminated requiring the alignment in flavour space of the Yukawa matrices \cite{Pich:2009sp}:
\bel{eq:alignment}
Y^{(2)}_{d,\ell}\, =\, \varsigma_{d,\ell}\, Y^{(1)}_{d,\ell}
\, =\, \frac{\sqrt{2}}{v}\,\varsigma_{d,\ell}\,  M_{d,\ell}\, ,
\qquad\qquad\qquad
Y^{(2)}_{u}\, =\, \varsigma_{u}^*\, Y^{(1)}_{u}
\, =\, \frac{\sqrt{2}}{v}\,\varsigma_{u}^*\,  M_{u}\, ,
\ee
with $\varsigma_f$ arbitrary complex proportionality parameters.\footnote{Actually, since one only needs that $Y^{(1)\raisebox{3pt}{$\scriptscriptstyle\prime$}}_f$ and $Y^{(2)\raisebox{3pt}{$\scriptscriptstyle\prime$}}_f$ can be simultaneously diagonalized, in full generality the factors $\varsigma_f$ could be 3-dimensional diagonal matrices in the fermion mass-eigenstate basis (generalized alignment) \cite{Penuelas:2017ikk}. The fashionable models of types I, II, X and Y, usually considered in the literature, are particular cases of the flavour-aligned Lagrangian with all $\varsigma_f$ parameters real and fixed in terms of $\tan{\beta} = v_2/v_1$ \cite{Pich:2009sp}.}

Flavour alignment constitutes a very simple implementation of minimal flavour violation. It results in a very specific dynamical structure, with all fermion-scalar interactions being proportional to the corresponding fermion masses. The Yukawas are fully characterized by the three complex alignment parameters $\varsigma_f$, which introduce new sources of $\cCP$ violation. The aligned two-Higgs doublet model Lagrangian satisfies the flavour constraints \cite{Jung:2010ik,Jung:2012vu,Jung:2013hka,Li:2014fea,Ilisie:2015tra,Cherchiglia:2016eui,Chang:2015rva,Hu:2016gpe,Cho:2017jym,Han:2015yys,Enomoto:2015wbn}, and leads to a rich collider phenomenology with five physical scalar bosons \cite{Celis:2013rcs,Abbas:2015cua,Ilisie:2014hea,Altmannshofer:2012ar,Bai:2012ex,Duarte:2013zfa,Ayala:2016djv,Wang:2013sha}: $h$, $H$, $A$ and $H^\pm$.

\section{Flavour anomalies}

The experimental data exhibit a few deviations from the SM predictions \cite{Pich:2019pzg}. For instance, Table~\ref{tab:ccuniv} shows a  $2.6\,\sigma$ 
%%% ($2.4\,\sigma$) 
violation of lepton universality in $|g_\tau/g_\mu|$ 
%%% ($|g_\tau/g_e|$) 
at the 1\% level, from $W\to\ell\nu$ decays, that is difficult to reconcile with the precise 0.15\% limits extracted from virtual $W^\pm$ transitions, shown also in the same table. In fact, it has been demonstrated that it is not possible to accommodate this deviation from universality with an effective Lagrangian and, therefore, such a signal could only be explained by the introduction of new light degrees of freedom that so far remain undetected \cite{Filipuzzi:2012mg}.
Thus, the most plausible explanation is a small problem (statistical fluctuation or underestimated systematics) in the LEP-2 measurements that will remain unresolved until more precise high-statistics $W\to\ell\nu$ data samples become available.

Some years ago BaBar reported a non-zero $\cCP$ asymmetry in $\tau^\pm\to\pi^\pm K_S\nu$ decays at the level of $3.6\cdot 10^{-3}$ \cite{BABAR:2011aa},  the same size than the SM expectation from $K^0$--$\bar K^0$ mixing \cite{Bigi:2005ts,Grossman:2011zk} but with the opposite sign, which represents a $2.8\,\sigma$ anomaly. So far, Belle has only reached a null result with a smaller $10^{-2}$ sensitivity and, therefore, has not been able to either confirm or refute the asymmetry. Nevertheless, on very general grounds, it has been shown that the BaBar signal is incompatible with other sets of data ($K^0$ and $D^0$ mixing, neutron electric dipole moment) \cite{Cirigliano:2017tqn}. 

Another small flavour anomaly was triggered by the unexpected large value of $\mathrm{Br}(B^-\to\tau^-\bar\nu)$, found in 2006 by Belle \cite{Ikado:2006un} and later confirmed by BaBar \cite{Aubert:2007xj}, which implied values of $|\bV_{\! ub}|$ higher than the ones measured in semileptonic decays or extracted from global CKM fits.
While the BaBar results remain unchanged, the reanalysis of the full Belle data sample resulted in a sizeable $\sim 40\%$ reduction of the measured central value \cite{Adachi:2012mm}, eliminating the discrepancy with the SM but leaving a pending disagreement with the BaBar results.

In the last few years a series of anomalies have emerged in $b\to c\tau\nu$ and $b\to s\mu^+\mu^-$ transitions. Given the difficulty of the experimental analyses, the results should be taken with some caution and further studies with larger data sets are still necessary. Nevertheless, these anomalies exhibit a quite consistent pattern that makes them intriguing.

\subsection{\boldmath $b\to c\tau\nu$ decays}

In 2012 the BaBar collaboration \cite{Lees:2012xj} observed an excess in $B\to D^{(*)}\tau\nu_\tau$ decays with respect to the SM predictions \cite{Fajfer:2012vx}, indicating a violation of lepton-flavour universality at the 30\% level. The measured observables are the ratios
\bel{eq:RD}
R(D)\,\equiv\,\frac{\mathrm{Br}(B\to D\tau\nu)}{\mathrm{Br}(B\to D\ell\nu)}\, ,
\qquad\qquad\qquad
R(D^*)\,\equiv\, \frac{\mathrm{Br}(B\to D^*\tau\nu)}{\mathrm{Br}(B\to D^*\ell\nu)}\, ,
\ee
with $\ell= e,\mu$, where many sources of experimental and theoretical errors cancel. The effect has been later confirmed by LHCb \cite{Aaij:2015yra} ($D^*$ mode only) and Belle \cite{Huschle:2015rga} (Fig.~\ref{fig:RD-RDs}). Although the results of the last two experiments are slightly closer to the SM expected values, 
$R(D) = 0.302\pm 0.004$ 
%%%\cite{Aoki:2019cca,Lattice:2015rga,Na:2015kha} 
and $R(D^*) = 0.258\, {}^{+\, 0.006}_{-\, 0.005}$ \cite{Bigi:2016mdz,Bernlochner:2017jka,Bigi:2017jbd,Jaiswal:2017rve,Jung:2018lfu,Murgui:2019czp,Mandal:2020htr},
%%% \cite{Fajfer:2012vx}, 
the resulting world averages \cite{Amhis:2019ckw}
\bel{eq:RD2}
R(D)
%\,\equiv\, \frac{\mathrm{Br}(B\to D\tau\nu)}{\mathrm{Br}(B\to D\ell\nu)}
\, =\, 0.340\pm 0.027\pm 0.013\, ,
\qquad\qquad
R(D^*)
%\,\equiv\, \frac{\mathrm{Br}(B\to D^*\tau\nu)}{\mathrm{Br}(B\to D^*\ell\nu)}
\, =\, 0.295\pm 0.011\pm 0.008\, ,
\ee
deviate at the $3.2\,\sigma$ level (considering their correlation of $-0.38$)
from the SM predictions, which is a very large effect for a tree-level SM transition.

%%%%%%%%%%%%%%%  FIGURE %%%%%%%%%%%%%%%%%%%%%%%%%
\begin{figure}[tbh]\centering
\includegraphics[width=10cm,clip]{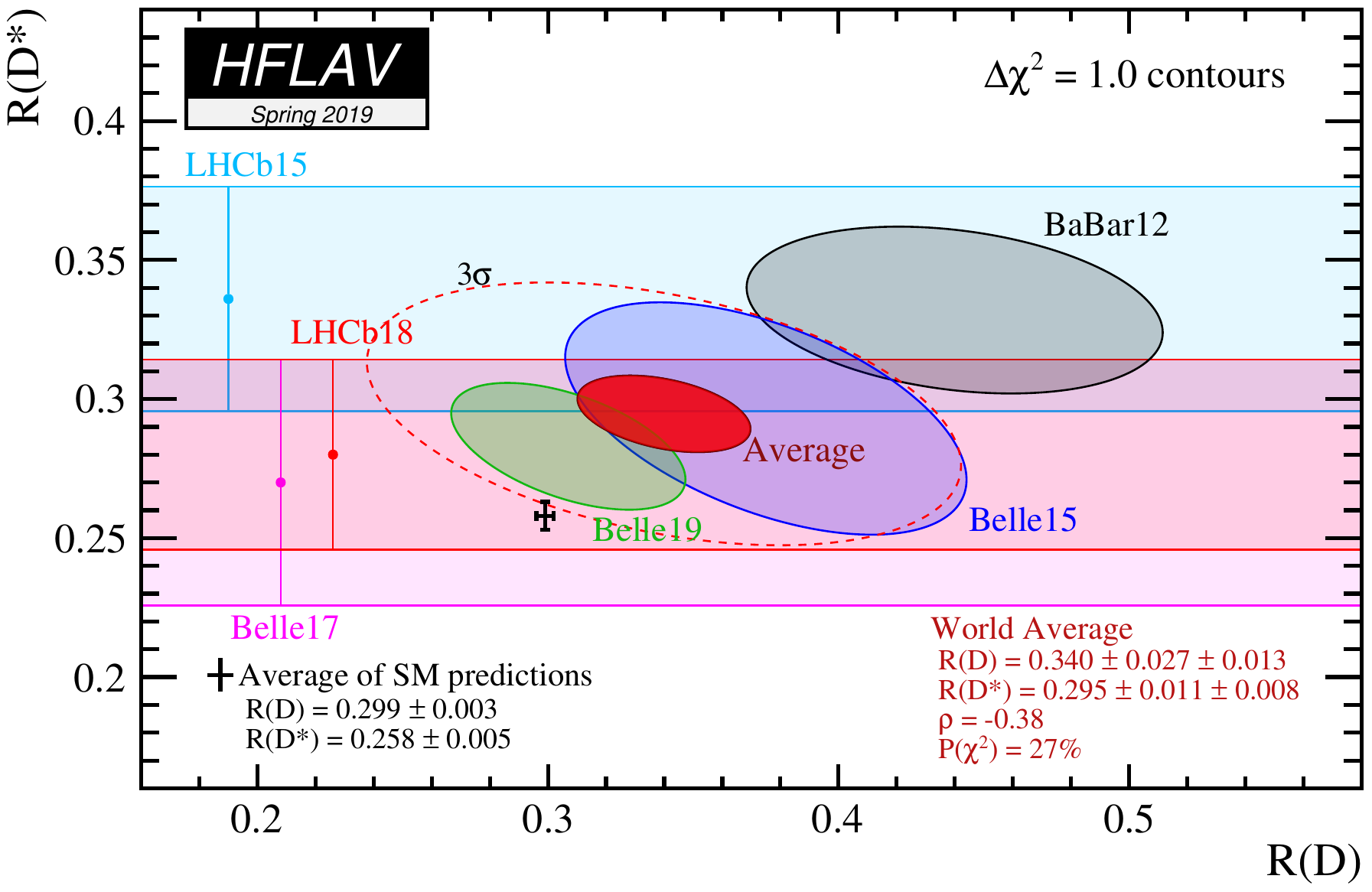}
\caption{Measurements of $R(D)$ and $R(D^*)$ and their average compared with the SM predictions. Filled contours correspond to $\Delta\chi^2=1$ (68\% CL for the bands and 39\% CL for the ellipses), while the dashed ellipse displays the $3\,\sigma$ region \cite{Amhis:2019ckw}.}
\label{fig:RD-RDs}
\end{figure}
%%%%%%%%%%%%%%%%%%%%%%%%%%%%%%%%%%%%%%%%%%%%%

%A worrisome aspect to keep in mind is that the latest $R(D^*)$  measurements of both LHCb and Belle, with the $\tau$ reconstructed with hadronic instead of leptonic decay modes, exhibit a shift to smaller values, approaching the SM prediction. On the theoretical side, the recent re-evaluation of form-factor extrapolations related to the exclusive $|\bV_{\! cb}|$ determination favours a slightly larger prediction $R(D^*) = 0.260\pm 0.008$ \cite{Bigi:2017njr}, with an enlarged uncertainty, while a smaller error is assigned to the other ratio, 
%$R(D) = 0.299\pm 0.003$ \cite{Bigi:2016mdz}.

Different new-physics explanations of the anomaly have been put forward: new charged vector or scalar bosons, leptoquarks, right-handed neutrinos, etc.\footnote{A long, but not exhaustive, list of relevant references is given in Refs.~\cite{Murgui:2019czp,Mandal:2020htr}.}
The normalized $q^2$ distributions measured by BaBar \cite{Lees:2012xj} and Belle \cite{Huschle:2015rga} do not favour large deviations from the SM \cite{Celis:2016azn}.  One must also take into account that the needed enhancement of the $b\to c\tau\nu$ transition is constrained by the cross-channel $b\bar c\to \tau\nu$.   A conservative (more stringent) upper bound $\mathrm{Br}(B_c\to\tau\nu)<30\%$ (10\%) can be extracted from the $B_c$ lifetime \cite{Celis:2016azn,Alonso:2016oyd} (LEP data \cite{Akeroyd:2017mhr}).
A global fit to the data, using a generic low-energy effective Hamiltonian with four-fermion effective operators \cite{Murgui:2019czp,Mandal:2020htr}, finds several viable possibilities. However, while the fitted results clearly indicate that new-physics contributions are needed (much lower $\chi^2$ than in the SM), they don't show any strong preference for a particular Wilson coefficient. 
The simplest solution would be 
%
%\cite{Celis:2012dk}.}
%A natural candidate for an additional tree-level contribution to these decays would be a charged scalar boson, coupling to fermions proportionally to their masses, as predicted in flavour-aligned two-Higgs-doublet models \cite{Pich:2009sp,Jung:2010ik}. This could easily increase the value of $R(D)$, since the scalar form factor contribution to the $\tau$ mode is already sizeable in the SM. However, to accommodate a large increase of $R(D^*)$, which is sensitive to several vector and axial-vector form factors, requires too big scalar contributions that are in tension with the measured $q^2$ distribution. The usual two-Higgs doublet models (types I, II, X and Y) cannot be made compatible with the data, but a (generalized) flavour-aligned scenario with the simultaneous presence of left- and right-handed couplings to the quarks improves considerably the global fit \cite{Celis:2012dk}. Taking into account the known constraints from the $B_c$ decay width and the inclusive $b\to X_c\tau\nu$ branching ratio, an explanation of the anomaly in terms of an intermediate charged scalar would require values of $R(D^*)$ $1\,\sigma$ smaller than the present experimental central value. The scalar contribution could be disentangled through measurements of the differential distributions and the polarization of the final $\tau$ and/or $D^*$ \cite{Celis:2012dk}.
%
%A different possibility is the existence of 
some new-physics contribution that only manifests in the Wilson coefficient of the SM operator $(\bar c_L\gamma^\mu b_L)(\bar\tau_L\gamma_\mu\nu_L)$. This would imply a universal enhancement of all $b\to c\tau\nu$ transitions,
%%
%\be 
%\frac{R(D)}{R(D)_{\rms SM}} \, =\, \frac{R(D^*)}{R(D^*)_{\rms SM}} \, =\, \frac{R(J/\psi)}{R(J/\psi)_{\rms SM}} \, ,
%\ee
%%
%%%
%%\be 
%%R(D)/R(D)_{\rms SM} \, =\, R(D^*)/R(D^*)_{\rms SM} \, =\, R(J/\psi)/R(J/\psi)_{\rms SM} \, ,
%%\ee
%%
in agreement with 
%%%the experimental ratios $R(D)$ and $R(D^*)$, and 
the recent LHCb observation of the $B_c\to J/\psi\,\tau\nu$ decay \cite{Aaij:2017tyk},
\be 
R(J/\psi)\,\equiv\,\frac{\mathrm{Br}(B_c\to J/\psi\,\tau\nu)}{\mathrm{Br}(B_c\to J/\psi\,\ell\nu)}\, =\, 0.71\pm 0.17\pm 0.18\, ,
\ee
which is $2\,\sigma$ above the SM expected value \ $R(J/\psi)\sim 0.25$--0.28 \cite{Anisimov:1998uk,Kiselev:2002vz,Ivanov:2006ni,Hernandez:2006gt}. Writing the four-fermion left-left operator in terms of $SU(2)_L\otimes U(1)_Y$ invariant operators at the electroweak scale, and imposing that the experimental bounds on $\mathrm{Br}(b\to s\nu\bar\nu)$ are satisfied, this possibility would imply rather large rates in $b\to s\tau^+\tau^-$
transitions \cite{Alonso:2015sja,Crivellin:2017zlb,Capdevila:2017iqn}, but still safely below the current upper limits \cite{Bobeth:2011st}.

\subsection{\boldmath $b\to s\ell^+\ell^-$ decays}

The rates of several $b\to s\mu^+\mu^-$ transitions have been found at LHCb to be consistently lower than their SM predictions: 
$B^+ \to K^+ \mu^+ \mu^-$ \cite{Aaij:2014pli,Aaij:2012vr},
$B^+ \to K^{*+} \mu^+ \mu^-$ \cite{Aaij:2014pli},
$B^0_d \to K^0 \mu^+ \mu^-$ \cite{Aaij:2014pli},
$B^0_d \to  K^{*0}\mu^{+}\mu^{-}$ \cite{Aaij:2016flj}, $B_s^0\to\phi\mu^+\mu^-$ \cite{Aaij:2015esa} and
$\Lambda^{0}_{b} \rightarrow \Lambda \mu^+\mu^-$ \cite{Aaij:2015xza}.
The angular and invariant-mass distributions of the final products in $B \to  K^{*}\mu^{+}\mu^{-}$ have been also studied by 
ATLAS \cite{Aaboud:2018krd}, BaBar \cite{Lees:2015ymt}, 
Belle \cite{Wehle:2016yoi}, CDF \cite{Aaltonen:2011ja},
CMS \cite{Sirunyan:2017dhj} and LHCb \cite{Aaij:2016flj}. The rich variety of angular dependences in the four-body $K\pi\mu^+\mu^-$ final state allows one to disentangle different sources of dynamical contributions. Particular attention has been put in the so-called {\it optimised observables} $P'_i(q^2)$, where $q^2$ is the dilepton invariant-mass squared, which are specific combinations of angular observables that are free from form-factor uncertainties in the heavy quark-mass limit \cite{DescotesGenon:2012zf}. A sizeable discrepancy with the SM prediction \cite{Descotes-Genon:2014uoa,Altmannshofer:2014rta,Straub:2015ica}, shown in Fig.~\ref{fig:P5}, has been identified in two adjacent bins of the $P'_5$ distribution, 
just below the $J/\Psi$ peak.
%%% at large $K^*$ recoil \cite{Aaij:2016flj,Aaboud:2018krd,Wehle:2016yoi,Sirunyan:2017dhj}. 
Belle has also included $K^*e^+e^-$ final states in the analysis, but the results for this electron mode are compatible with the SM expectations \cite{Wehle:2016yoi}. 
%The muon $P'_5$ anomaly seems to be also present in the most recent ATLAS  \cite{Aaboud:2018krd} and CMS \cite{Sirunyan:2017dhj} studies, although with lower statistical significance. 
%The hadronic uncertainties of the SM predictions are not easy to quantify, but have been scrutinised by several groups \cite{Capdevila:2017ert,Jager:2014rwa,Ciuchini:2015qxb,Chobanova:2017ghn,Bobeth:2017vxj,Blake:2017fyh}.

%%%%%%%%%%%%%%%  FIGURE %%%%%%%%%%%%%%%%%%%%%%%%%
\begin{figure}[t]\centering
\includegraphics[height=7.cm]{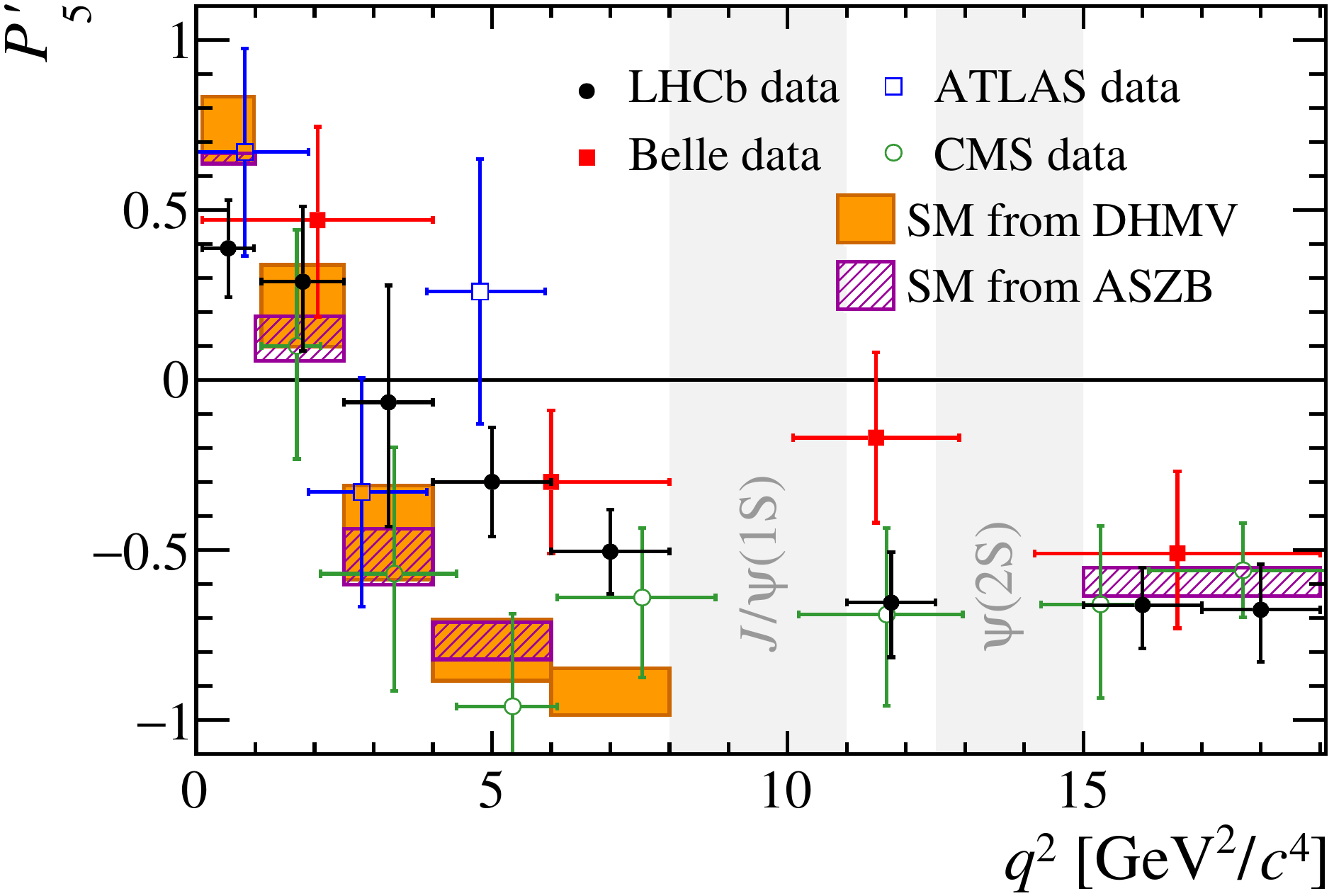}
\caption{Comparison between the SM predictions for $P'_5$ \cite{Descotes-Genon:2014uoa,Altmannshofer:2014rta,Straub:2015ica} and the experimental measurements \cite{Dettori:2018igw}.}
%\begin{minipage}[t]{.45\linewidth}\centering
%\includegraphics[height=5cm]{Figures/p_p4_dhmv.pdf} 
%\end{minipage}
%\hskip 1.5cm
%\begin{minipage}[t]{.45\linewidth}\centering
%\includegraphics[height=5cm]{Figures/p_p5_dhmv.pdf}
%\end{minipage}
%\caption{Comparison between the SM predictions for $P'_4$ and $P'_5$ \cite{Descotes-Genon:2014uoa} and their experimental values
%\cite{Abdesselam:2016llu}.}
\label{fig:P5}
\end{figure}
%%%%%%%%%%%%%%%%%%%%%%%%%%%%%%%%%%%%%%%%%%%%%

The SM predictions for the previous observables suffer from hadronic uncertainties that are not easy to quantify. However, LHCb has also reported sizeable violations of lepton flavour universality, at the $2.1$-$2.6\,\sigma$ level, through the ratios \cite{Aaij:2017vbb}
\bel{eq:RKs}
R_{K^{*0}}\,\equiv\, \frac{\mathrm{Br}(B^0_d \to K^{*0} \mu^+ \mu^-)}{B^0_d \to K^{*0} e^+ e^-)}\, =\, \left\{ \begin{array}{ccc}
0.66 \,{}^{+\, 0.11}_{-\,0.07} \pm 0.03\, , &\qquad & q^2\in\, [0.045\, ,\, 1.1]\;\mathrm{GeV}^2\, ,
\\[5pt]
0.69 \,{}^{+\, 0.11}_{-\,0.07} \pm 0.05\, , && q^2\in\, [1.1\, ,\, 6.0]\;\mathrm{GeV}^2\, ,
\ea\right.
\ee
and \cite{Aaij:2019wad}
\bel{eq:RK}
R_K\,\equiv\, \left.\frac{\mathrm{Br}(B^+ \to K^+ \mu^+ \mu^-)}{\mathrm{Br}(B^+ \to K^+ e^+ e^-)}\right|_{q^2\in\, [1.1\, ,\, 6.0]\;\mathrm{GeV}^2}
\, =\, 0.846\,{}^{+\, 0.060}_{-\,0.054} \,{}^{+\, 0.016}_{-\,0.014}\, .
\ee
These observables constitute a much cleaner probe of new physics because most theoretical uncertainties cancel out \cite{Bobeth:2011nj,Hiller:2003js,Bobeth:2007dw,Bouchard:2013mia}. In the SM, the only difference between the muon and electron channels is the lepton mass. The SM theoretical predictions, $R_K = 1.00\pm 0.01_{\rms QED}$, $R_{K^{*0}}[0.045,1.1] = 0.906\pm 0.028_{\rms th}$  and $R_{K^{*0}}[1.1,6] =1.00\pm 0.01_{\rms QED}$ \cite{Bordone:2016gaq}, deviate from these experimental measurements by $2.4\,\sigma$, $2.1\,\sigma$ and $2.6\,\sigma$, respectively. Owing to their larger uncertainties, the recent Belle measurements of $R_{K^*}$ \cite{Abdesselam:2019wac} and $R_K$ \cite{Abdesselam:2019lab} are compatible with the SM as well as with LHCb.

Global fits to the $b\to s\ell^+\ell^-$ data with an effective low-energy Lagrangian
\bel{eq:Leff}
\cL_{\rms eff}\, =\, \frac{G_F}{\sqrt{2}}\, \bV_{\!tb}\bV_{\!ts}^*\,\frac{\alpha}{\pi}\;\sum_{i,\ell}\, C_{i,\ell}\, O_i^\ell
\ee
show a clear preference for new-physics contributions to the operators
$O_9^\ell = (\bar s_L\gamma_\mu b_L)(\bar \ell\gamma^\mu \ell)$ and $O_{10}^\ell = (\bar s_L\gamma_\mu b_L)(\bar \ell\gamma^\mu\gamma_5 \ell)$, with $\ell=\mu$ 
\cite{Aebischer:2019mlg,Alguero:2019ptt,Ciuchini:2019usw,Datta:2019zca,Kowalska:2019ley,Arbey:2019duh,Alok:2019ufo}.
Although the different analyses tend to favour slightly different solutions, two main common scenarios stand out: either $\delta C_{9,\mu}^{\mathrm{NP}}\approx -0.98$ or
$\delta C_{9,\mu}^{\mathrm{NP}} = - \delta C_{10,\mu}^{\mathrm{NP}}\approx -0.46$. Both constitute large shifts ($-24\%$ and $-11\%$, respectively) from the SM values: $C_{9,\mu}^{\mathrm{SM}}(\mu_b) \approx 4.1$ and $C_{10,\mu}^{\mathrm{SM}}(\mu_b) \approx -4.3$, at $\mu_b = 4.8$~GeV. The first possibility is slightly preferred by the global analysis of all data, while the left-handed new-physics solution accommodates better the lepton-flavour-universality-violating observables \cite{Alguero:2019ptt}.

The left-handed scenario is theoretically appealing because it can be easily generated through $SU(2)_L\otimes U(1)_Y$-invariant effective operators at the electroweak scale that, moreover, could also provide an explanation to the $b\to c\tau\nu$ anomaly. This possibility emerges naturally from the so-called $U_1$ vector leptoquark model \cite{Angelescu:2018tyl}, and can be tested experimentally, since it implies a $b\to s\tau^+\tau^-$ rate three orders of magnitude larger than the SM expectation \cite{Capdevila:2017iqn}. For a recent review of theoretical models with a quite complete list of references, see Ref.~\cite{Bifani:2018zmi}.

%%
%%%%%%%%%%%%%%%%  FIGURE %%%%%%%%%%%%%%%%%%%%%%%%%
%\begin{figure}[hbt]\centering
%\begin{minipage}[t]{.45\linewidth}\centering
%\includegraphics[height=7cm]{Figures/C9C10-Total-Plot.pdf} 
%\end{minipage}
%\hskip 1.cm
%\begin{minipage}[t]{.45\linewidth}\centering
%\includegraphics[height=7cm]{Figures/C9eeC9mumu-Total-Plot.pdf}
%\end{minipage}
%\caption{Allowed 1, 2 and 3 $\sigma$ regions in the $(C_{9,\mu}^{\rms NP},C_{10,\mu}^{\rms NP})$
%and $(C_{9,\mu}^{\rms NP},C_{9,e}^{\rms NP})$ planes, from a global fit to $b\to s\ell^+\ell^-$ data \cite{Capdevila:2017bsm}. The contours from specific experiments are show at $3 \,\sigma$.}
%\label{fig:O9-O10_Fit}
%\end{figure}
%%%%%%%%%%%%%%%%%%%%%%%%%%%%%%%%%%%%%%%%%%%%%%
%%
%of the order of $\sim -25\%$ of the SM contribution to $C_{9,\mu}$ \cite{Capdevila:2017bsm}. The statistical significance for non-zero values of $C_{10,\mu}^{\rms NP}$, $C_{9,e}^{\rms NP}$ and $C_{10,e}^{\rms NP}$ is low, although a reasonable fit to the data could be obtained with 
%$C_{9,\mu}^{\rms NP}-C_{9,e}^{\rms NP}-C_{10,\mu}^{\rms NP}+C_{10,e}^{\rms NP}\sim -1.4$ \cite{Altmannshofer:2017yso}.
%Many attempts to explain these results within specific scenarios of new physics have been put forward (new gauge bosons, leptoquarks, new fermions and/or scalars, etc.). A representative list of references can be found in Refs.~\cite{Fajfer:2018bfj,Marzocca:2018wcf,Camargo-Molina:2018cwu}.

\section{Discussion}

The flavour structure of the SM is one of the main pending questions
in our understanding of weak interactions. Although we do not know the reasons of the observed family replication, we have learnt experimentally that the number of SM generations is just three (and no more). Therefore, we must study as precisely as possible the few existing flavours, to get some hints on the
dynamics responsible for their observed structure.

In the SM all flavour dynamics originate in the fermion mass matrices,
which determine the measurable masses and mixings. Thus, flavour is related through the Yukawa interactions with the scalar sector, the part of the SM Lagrangian that is more open to theoretical speculations. At present, we totally ignore the underlying dynamics responsible for the vastly different scales exhibited by the fermion spectrum and the particular values of the measured mixings. The SM Yukawa matrices are just a bunch of arbitrary parameters to be fitted to data, which is conceptually unsatisfactory.

The SM incorporates a mechanism to generate $\cCP$ violation, through the single phase naturally occurring in the CKM matrix. This mechanism, deeply rooted into the unitarity structure of $\bV$, implies very specific requirements for $\cCP$ violation to show up.
The CKM matrix has been thoroughly investigated in dedicated experiments
and a large number of $\cCP$-violating processes have been studied in detail.
The flavour data seem to fit into the SM framework, confirming that the fermion mass matrices are the dominant source of flavour-mixing phenomena. However, a fundamental explanation of the flavour dynamics is still lacking.

At present, a few flavour anomalies have been identified in $b\to c\tau\nu$ and $b\to s\mu^+\mu^-$ transitions. Whether they truly represent the first signals of new phenomena, or just result from statistical fluctuations and/or underestimated systematics remains to be understood. New experimental input from LHC and Belle-II should soon clarify the situation. Very valuable information on the flavour dynamics is also expected from BESS-III and from several kaon (NA62, KOTO) and muon (MEG-II, Mu2e, Mu3e, COMET) 
%%% , PRISM/PRIME 
experiments, complementing the high-energy searches for new phenomena at LHC.
 Unexpected surprises may well be discovered, probably giving hints of new physics at higher scales and offering clues to the problems of fermion mass generation, quark mixing and family replication.

%***********************************************
%
%The dynamics of flavour is a broad and fascinating subject, which is
%closely related to the so far untested scalar sector of the SM.
%
%
%Thus, flavour phenomena imposes severe restrictions on possible extensions of the SM.

\section*{Acknowledgements}

I want to thank the organizers for the charming atmosphere of this
school and all the students for their many interesting questions and
comments. This work has been supported in part by the Spanish Government and ERDF funds from the EU Commission [Grants FPA2017-84445-P and FPA2014-53631-C2-1-P], by Generalitat Valenciana [Grant Prometeo/2017/053] and by the Spanish
Centro de Excelencia Severo Ochoa Programme [Grant SEV-2014-0398]. 

\appendix

\section{Conservation of the vector current}

In the limit where all quark masses are equal, the QCD Lagrangian remains invariant under global $SU(N_f)$ transformations of the quark fields in flavour space, where $N_f$ is the number of (equal-mass) quark flavours. This guarantees the conservation of the corresponding Noether currents \
$V^\mu_{ij} = \bar u_i\gamma^\mu d_j$. In fact, using the QCD equations of motion, one immediately finds that
\bel{eq:VCC}
\partial_\mu V^\mu_{ij}\,\equiv\,\partial_\mu\left(\bar u_i\gamma^\mu d_j\right) \, =\,i\, \bigl( m_{u_i} - m_{d_j} \bigr)\, \bar u_i d_j\, ,
\ee
which vanishes when $m_{u_i} = m_{d_j}$.
In momentum space, this reads \ $q_\mu V^\mu_{ij} = \cO(m_{u_i} - m_{d_j})$, with $q_\mu$ the corresponding momentum transfer.

Let us consider a $0^-(k)\to 0^-(k')$ weak transition mediated by the vector current 
$V^\mu_{ij}$. The relevant hadronic matrix element is given in Eq.~\eqn{eq:vector_me} and contains two form factors $f_\pm(q^2)$.
%%%, with $q^2=(k'-k)^2$. 
The conservation of the vector current implies that $f_-(q^2)$ is identically zero when $m_{u_i} = m_{d_j}$. Therefore,
\bel{eq:CVC-matrix}
\langle P'_i(k')| V^\mu_{ij}(x) | P_j(k)\rangle\, =\, \e^{iq\cdot x}\; C_{PP'}\; (k+k')^\mu\; f_+(q^2)\, .
\ee
We have made use of translation invariance to write \ $V^\mu_{ij}(x) = 
\e^{iP\cdot x}\, V^\mu_{ij}(0)\, \e^{-iP\cdot x}$, with $P^\mu$ the four-momentum operator.
This determines the dependence of the matrix element on the space-time coordinate, with $q^\mu=(k'-k)^\mu$.

The conserved Noether charges
\bel{eq:NoetherCharges}
\cN_{ij}\, =\, \int d^3x\; V^0_{ij}(x)\, =\, \int d^3x\;  u_i^\dagger(x)\, d_j(x)\, ,
\ee
annihilate one quark $d_j$ and create instead one $u_i$ (or replace $\bar u_i$ by $\bar d_j$), transforming the meson $P$ into $P'$ (up to a trivial Clebsh-Gordan factor $C_{PP'}$ that, for light-quarks, takes the value $1/\sqrt{2}$ when $P'$ is a $\pi^0$ and is 1 otherwise). Thus,
\bel{eq:CVC1} 
\langle P'(k') | \cN_{ij} | P(k)\rangle\, =\, C_{PP'}\; \langle P'(k') | P'(k)\rangle\, =\, C_{PP'}\; (2\pi)^3\; 2 k^0\;\delta^{(3)}(\vec k'- \vec k\, )\, .
\ee
On the other side, inserting in this matrix element the explicit expression for $\cN_{ij}$ in \eqn{eq:NoetherCharges} and making use of \eqn{eq:CVC-matrix},
\bel{eq:CVC2}
\langle P'(k') | \cN_{ij} | P(k)\rangle\, =\,
\int d^3x\; \langle P'(k') |  V^0_{ij}(x) | P(k)\rangle\, =\,
C_{PP'}\; (2\pi)^3\; 2 k^0\;\delta^{(3)}(\vec q\, )\; f_+(0)\, .
\ee
Comparing Eqs.~\eqn{eq:CVC1} and \eqn{eq:CVC2}, one finally obtains the result
\be 
f_+(0)\, =\, 1\, .
\ee
Therefore, the flavour symmetry $SU(N_f)$ determines the normalization of the form factor at $q^2=0$, when $m_{u_i} = m_{d_j}$. It is possible to prove that the deviations from 1 are of second order in the symmetry-breaking quark mass difference, \ie
$f_+(0) = 1 + \cO[(m_{u_i} - m_{d_j})^2]$ \cite{Ademollo:1964sr}.


\begin{thebibliography}{99}

\bibitem{GL:61} S.L. Glashow, {\it Nucl. Phys.} {\bf 22} (1961) 579.

\bibitem{WE:67} S. Weinberg, {\it Phys. Rev. Lett.} {\bf 19} (1967) 1264.

\bibitem{SA:69} A. Salam, in {\em Elementary Particle Theory},
ed. N. Svartholm (Almquist and Wiksells, Stockholm, 1969), p. 367.

\bibitem{SM:11}
%Updated version of
%the lectures given at the 2004 (San Feliu de Guixols, Spain) and 2006 (Aronsborg, Sweden)
%European Schools of High-Energy Physics:
A. Pich, {\it The Standard Model of Electroweak Interactions},
Reports CERN-2006-003 and CERN-2007-005, ed. R. Fleischer, p. 1; arXiv:hep-ph/0502010, arXiv:0705.4264 [hep-ph].

%\cite{Pich:2011nh}
\bibitem{Pich:2011nh}
  A.~Pich, {\it Flavour Physics and CP Violation},
%%%  doi:10.5170/CERN-2013-003.119
Proceedings 6th CERN -- Latin-American School of High-Energy Physics (CLASHEP 2011, Natal, Brazil, March 33 - April 5, 2011),
Report CERN-2013-003, p.119, ed. C. Grojean, M. Mulders and M. Spiropulu,
  arXiv:1112.4094 [hep-ph].


\bibitem{cabibbo}  N. Cabibbo, {\it Phys. Rev. Lett.} {\bf 10} (1963) 531.

\bibitem{KM:73} M. Kobayashi and T. Maskawa, {\it Prog. Theor. Phys.} {\bf 42}
  (1973) 652.

\bibitem{GIM:70} S.L. Glashow, J. Iliopoulos and L. Maiani,
{\it Phys. Rev.} {\bf D2} (1970) 1285.


%\cite{Tanabashi:2018oca}
\bibitem{Tanabashi:2018oca}
M.~Tanabashi \textit{et al.} [Particle Data Group],
%``Review of Particle Physics,''
Phys.\ Rev.\ D \textbf{98} (2018) no.3, 030001;
%%% doi:10.1103/PhysRevD.98.030001
%4305 citations counted in INSPIRE as of 11 Apr 2020
and 2019 update at http://pdg.lbl.gov.


%%\cite{Patrignani:2016xqp}
%\bibitem{Patrignani:2016xqp}
%  C.~Patrignani {\it et al.} [Particle Data Group],
%  %``Review of Particle Physics,''
%  {\it Chin. Phys.} {\bf C40} (2016)  100001; http://pdg.lbl.gov.
  
\bibitem{MS:88}
W.J. Marciano and A. Sirlin, {\it Phys. Rev. Lett.} {\bf 61} (1988) 1815.

\bibitem{vRS:99}
T. van Ritbergen and R.G. Stuart, {\it Phys. Rev. Lett.} {\bf 82} (1999)
488; {\it Nucl. Phys.} {\bf B564} (2000) 343.

%\cite{Pak:2008qt}  
\bibitem{Pak:2008qt}
  A.~Pak and A.~Czarnecki,
  %``Mass effects in muon and semileptonic b ---> c decays,''
  {\it Phys. Rev. Lett.} {\bf 100} (2008) 241807.
 % doi:10.1103/PhysRevLett.100.241807
 % [arXiv:0803.0960 [hep-ph]].
  %%CITATION = doi:10.1103/PhysRevLett.100.241807;%%
  %77 citations counted in INSPIRE as of 30 Apr 2018


%\cite{Tishchenko:2012ie}
\bibitem{Tishchenko:2012ie} MuLan Collaboration,
%%% V.~Tishchenko \textit{et al.} [MuLan],
%``Detailed Report of the MuLan Measurement of the Positive Muon Lifetime and Determination of the Fermi Constant,''
{\it Phys. Rev.} {\bf D87} (2013) %%%no.5, 
052003.
%doi:10.1103/PhysRevD.87.052003
%[arXiv:1211.0960 [hep-ex]].
%65 citations counted in INSPIRE as of 12 Apr 2020

\bibitem{BNP:92} E. Braaten, S. Narison and A. Pich, {\it Nucl. Phys.} {\bf B373} (1992) 581.

%\cite{Pich:2016bdg}
\bibitem{Pich:2016bdg}
  A.~Pich and A.~Rodr\'{\i}guez-S\'anchez,
  %``Determination of the QCD coupling from ALEPH $\tau$ decay data,''
 {\it Phys. Rev.} {\bf D94} (2016) 034027.
 % doi:10.1103/PhysRevD.94.034027
 % [arXiv:1605.06830 [hep-ph]].
  %%CITATION = doi:10.1103/PhysRevD.94.034027;%%
  %27 citations counted in INSPIRE as of 28 Apr 2018

%\cite{Pich:2016yfh}
\bibitem{Pich:2016yfh}
  A.~Pich,
  %``Precision physics with QCD,''
  EPJ Web Conf.\  {\bf 137} (2017) 01016.
%  doi:10.1051/epjconf/201713701016
 % [arXiv:1612.05010 [hep-ph]].
  %%CITATION = doi:10.1051/epjconf/201713701016;%%
  %2 citations counted in INSPIRE as of 28 Apr 2018

%\cite{Pich:2013lsa}
\bibitem{Pich:2013lsa}
  A.~Pich,
  %``Precision Tau Physics,''
  {\it Prog. Part. Nucl. Phys.}  {\bf 75} (2014) 41.
%  doi:10.1016/j.ppnp.2013.11.002
%  [arXiv:1310.7922 [hep-ph]].
  %%CITATION = doi:10.1016/j.ppnp.2013.11.002;%%
  %106 citations counted in INSPIRE as of 28 Apr 2018


\bibitem{PI:96} A. Pich, {\it Flavourdynamics}, arXiv:hep-ph/9601202.


%\cite{Hardy:2014qxa}
\bibitem{Hardy:2014qxa}
  J.~C.~Hardy and I.~S.~Towner,
  %``Superallowed $0^+\to 0^+$ nuclear β decays: 2014 critical survey, with precise results for $V_{ud}$ and CKM unitarity,''
  {\it Phys. Rev.} {\bf C91} (2015)  025501.
 % doi:10.1103/PhysRevC.91.025501
 % [arXiv:1411.5987 [nucl-ex]].
  %%CITATION = doi:10.1103/PhysRevC.91.025501;%%
  %114 citations counted in INSPIRE as of 28 Apr 2018

%\cite{Hardy:2018zsb}
\bibitem{Hardy:2018zsb}
J.~Hardy and I.~Towner,
%``Nuclear Beta Decays and CKM Unitarity,''
arXiv:1807.01146 [nucl-ex].
%13 citations counted in INSPIRE as of 12 Apr 2020

%\cite{Marciano:2005ec}
\bibitem{Marciano:2005ec}
W.~J.~Marciano and A.~Sirlin,
%``Improved calculation of electroweak radiative corrections and the value of V(ud),''
{\it Phys. Rev. Lett.} {\bf 96} (2006) 032002.
%doi:10.1103/PhysRevLett.96.032002
%[arXiv:hep-ph/0510099 [hep-ph]].
%229 citations counted in INSPIRE as of 12 Apr 2020

%\bibitem{MS:06} W.J. Marciano and A. Sirlin, {\it Phys. Rev. Lett.}
%   {\bf 96} (2006) 032002;   %%% [hep-ph/0510099]
%   {\bf 71} (1993) 3629.
%
%\bibitem{CMS:04} A. Czarnecki, W.J. Marciano and A. Sirlin, {\it Phys. Rev.}
%   {\bf D70} (2004) 093006.  %%% [hep-ph/0406324].

%\cite{Seng:2018yzq}
\bibitem{Seng:2018yzq}
C.~Seng, M.~Gorchtein, H.~H.~Patel and M.~J.~Ramsey-Musolf,
%``Reduced Hadronic Uncertainty in the Determination of $V_{ud}$,''
{\it Phys. Rev. Lett.} {\bf 121} (2018) 241804.
%doi:10.1103/PhysRevLett.121.241804
%[arXiv:1807.10197 [hep-ph]].
%48 citations counted in INSPIRE as of 12 Apr 2020

%\cite{Seng:2020wjq}
\bibitem{Seng:2020wjq}
C.~Seng, X.~Feng, M.~Gorchtein and L.~Jin,
%``Joint lattice QCD - dispersion theory analysis confirms the top-row CKM unitarity deficit,''
arXiv:2003.11264 [hep-ph].
%1 citations counted in INSPIRE as of 15 Apr 2020
 
%\cite{Czarnecki:2019mwq}
\bibitem{Czarnecki:2019mwq}
A.~Czarnecki, W.~J.~Marciano and A.~Sirlin,
%``Radiative Corrections to Neutron and Nuclear Beta Decays Revisited,''
{\it Phys. Rev.} {\bf D100} (2019) 073008.
%doi:10.1103/PhysRevD.100.073008
%[arXiv:1907.06737 [hep-ph]].
%19 citations counted in INSPIRE as of 12 Apr 2020  
  
  
\bibitem{Ademollo:1964sr} M. Ademollo and R. Gatto, {\it Phys. Rev. Lett.} {\bf 13} (1964) 264.

\bibitem{CKNP:03}	
% The Pionic beta decay in chiral perturbation theory.
V. Cirigliano, M. Knecht, H. Neufeld and H. Pichl,
{\it Eur. Phys. J.} {\bf C27} (2003) 255.
% e-Print: arXiv:hep-ph/0209226 [hep-ph]

\bibitem{PiBeta:04}
D. Pocanic \etal, {\it Phys. Rev. Lett.} {\bf 93} (2004) 181803.

\bibitem{CGN:08} V. Cirigliano, M. Giannotti and H. Neufeld, {\it JHEP} {\bf 0811} (2008) 006.  %%% arXiv:0807.4507 [hep-ph].

\bibitem{KN:08} A. Kastner and H. Neufeld, {\it Eur. Phys. J.} {\bf C57} (2008) 541.
%%%   arXiv:0805.2222 [hep-ph].

%\cite{Antonelli:2010yf}
\bibitem{Antonelli:2010yf}
  M.~Antonelli {\it et al.} [FlaviaNet Working Group on Kaon Decays],
  %``An Evaluation of $|V_{us}|$ and precise tests of the Standard Model from world data on leptonic and semileptonic kaon decays,''
  {\it Eur. Phys. J.} {\bf C69} (2010) 399.
 % doi:10.1140/epjc/s10052-010-1406-3
 % [arXiv:1005.2323 [hep-ph]].
  %%CITATION = doi:10.1140/epjc/s10052-010-1406-3;%%
  %235 citations counted in INSPIRE as of 28 Apr 2018
  
%\cite{Moulson:2014cra}
\bibitem{Moulson:2014cra}
  M.~Moulson,
  %``Experimental determination of $V_{us}$ from kaon decays,''
  arXiv:1411.5252 [hep-ex].
  %%CITATION = ARXIV:1411.5252;%%
  %31 citations counted in INSPIRE as of 28 Apr 2018

\bibitem{BS:60} R.E. Behrends and A. Sirlin, {\it Phys. Rev. Lett.} {\bf 4} (1960) 186.

\bibitem{LR:84} H. Leutwyler and M. Roos, {\it Z. Phys.} {\bf C25} (1984) 91.

%\cite{Bazavov:2013maa}
\bibitem{Bazavov:2013maa}
  A.~Bazavov {\it et al.},
  %``Determination of $|V_{us}|$ from a lattice-QCD calculation of the $K\to\pi\ell\nu$ semileptonic form factor with physical quark masses,''
  {\it Phys. Rev. Lett.}  {\bf 112} (2014) 112001.
 % doi:10.1103/PhysRevLett.112.112001
 % [arXiv:1312.1228 [hep-ph]].
  %%CITATION = doi:10.1103/PhysRevLett.112.112001;%%
  %36 citations counted in INSPIRE as of 29 Apr 2018

%\cite{Carrasco:2016kpy}
\bibitem{Carrasco:2016kpy}
  N.~Carrasco {\it et al.},
  %%% , P.~Lami, V.~Lubicz, L.~Riggio, S.~Simula and C.~Tarantino,
  %``$K \to \pi$ semileptonic form factors with $N_f=2+1+1$ twisted mass fermions,''
  {\it Phys. Rev.} {\bf D93} (2016) 114512.
 % doi:10.1103/PhysRevD.93.114512
 % [arXiv:1602.04113 [hep-lat]].
  %%CITATION = doi:10.1103/PhysRevD.93.114512;%%
  %18 citations counted in INSPIRE as of 29 Apr 2018
  

\bibitem{BT:03} J. Bijnens and P. Talavera, {\it Nucl. Phys.} {\bf B669} (2003) 341.

\bibitem{JOP:04} M. Jamin, J.A. Oller and A. Pich, {\it JHEP} {\bf 0402} (2004) 047.
% Order p6 chiral couplings from the scalar K-pi form-factor.
% e-Print: hep-ph/0401080

\bibitem{CEEKPP:05} V. Cirigliano \etal, {\it JHEP} {\bf 0504} (2005) 006.
% arXiv:hep-ph/0503108 [hep-ph]

%%\cite{Aoki:2016frl}
%\bibitem{Aoki:2016frl}
%  S.~Aoki {\it et al.} [FLAG Working Group],
%  %``Review of lattice results concerning low-energy particle physics,''
%  {\it Eur. Phys. J.} {\bf C77} (2017) 112.
% % doi:10.1140/epjc/s10052-016-4509-7
% % [arXiv:1607.00299 [hep-lat]].
%  %%CITATION = doi:10.1140/epjc/s10052-016-4509-7;%%
%  %279 citations counted in INSPIRE as of 28 Apr 2018

%\cite{Aoki:2019cca}
\bibitem{Aoki:2019cca}
S.~Aoki \textit{et al.} [Flavour Lattice Averaging Group],
%``FLAG Review 2019,''
{\it Eur. Phys. J.}  \textbf{C80} (2020) 113.
%doi:10.1140/epjc/s10052-019-7354-7
%[arXiv:1902.08191 [hep-lat]].
%202 citations counted in INSPIRE as of 15 Apr 2020


  

\bibitem{MA:04} W.J. Marciano, {\it Phys. Rev. Lett.} {\bf 93} (2004) 231803.
% [hep-ph/0402299].

\bibitem{CN:11} V. Cirigliano and H. Neufeld, {\it Phys. Lett.} {\bf B700} (2011) 7.
% arXiv:1102.0563 [hep-ph]

%\cite{Cirigliano:2011ny}
\bibitem{Cirigliano:2011ny}
  V.~Cirigliano, G.~Ecker, H.~Neufeld, A.~Pich and J.~Portol\'es,
  %``Kaon Decays in the Standard Model,''
  {\it Rev. Mod. Phys.} {\bf 84} (2012) 399.
 % doi:10.1103/RevModPhys.84.399
 % [arXiv:1107.6001 [hep-ph]].
  %%CITATION = doi:10.1103/RevModPhys.84.399;%%
  %155 citations counted in INSPIRE as of 29 Apr 2018

\bibitem{CSW:03} N. Cabibbo, E.C. Swallow and R. Winston, {\it Annu. Rev. Nucl. Part. Sci.} {\bf 53} (2003) 39; {\it Phys. Rev. Lett.} {\bf 92} (2004) 251803.

\bibitem{MP:05} V. Mateu and A. Pich, {\it JHEP} {\bf 0510} (2005) 041.
% R. Flores-Mendieta, Phys. Rev. D70 (2004) 114036.

\bibitem{GJPPS:05} E. G\'amiz, M. Jamin, A. Pich, J. Prades and F. Schwab,
{\it Phys. Rev. Lett.} {\bf 94} (2005) 011803; JHEP {\bf 0301} (2003) 060.

%%\cite{Amhis:2016xyh}
%\bibitem{Amhis:2016xyh}
%  Y.~Amhis {\it et al.} [HFLAV Collaboration],
%  %``Averages of $b$-hadron, $c$-hadron, and $\tau$-lepton properties as of summer 2016,''
% {\it  Eur. Phys. J.} {\bf C77} (2017) 895;
% % doi:10.1140/epjc/s10052-017-5058-4
% % [arXiv:1612.07233 [hep-ex]].
%  %%CITATION = doi:10.1140/epjc/s10052-017-5058-4;%%
%  %297 citations counted in INSPIRE as of 29 Apr 2018
%https://hflav.web.cern.ch/.

%\cite{Amhis:2019ckw}
\bibitem{Amhis:2019ckw}
Y.~S.~Amhis \textit{et al.} [HFLAV Collaboration],
%``Averages of $b$-hadron, $c$-hadron, and $\tau$-lepton properties as of 2018,''
arXiv:1909.12524 [hep-ex].
%63 citations counted in INSPIRE as of 15 Apr 2020
https://hflav.web.cern.ch/.

%%\cite{Antonelli:2013usa}
%\bibitem{Antonelli:2013usa}
%  M.~Antonelli, V.~Cirigliano, A.~Lusiani and E.~Passemar,
%  %``Predicting the $\tau$ strange branching ratios and implications for $V_{us}$,''
%  {\it JHEP} {\bf 1310} (2013) 070.
% % doi:10.1007/JHEP10(2013)070
% % [arXiv:1304.8134 [hep-ph]].
%  %%CITATION = doi:10.1007/JHEP10(2013)070;%%
%  %26 citations counted in INSPIRE as of 29 Apr 2018
  

\bibitem{IW:89}
 N. Isgur and M. Wise, {\it Phys. Lett.} {\bf B232} (1989) 113; {\bf B237} (1990)
 527.

\bibitem{GR:90} B. Grinstein, {\it Nucl. Phys.} {\bf B339} (1990) 253.

\bibitem{EH:90} E. Eichten and B. Hill, {\it Phys. Lett.} {\bf B234} (1990) 511.

\bibitem{GE:90} H. Georgi, {\it Phys. Lett.} {\bf B240} (1990) 447.

\bibitem{NE:91} M. Neubert, {\it Phys. Lett.} {\bf B264} (1991) 455.

\bibitem{LU:90} M. Luke, {\it Phys. Lett.} {\bf B252} (1990) 447.

%\cite{Caprini:1997mu}
\bibitem{Caprini:1997mu}
  I.~Caprini, L.~Lellouch and M.~Neubert,
  %``Dispersive bounds on the shape of anti-B ---> D(*) lepton anti-neutrino form-factors,''
  {\it Nucl. Phys.} {\bf B530} (1998) 153.
 % doi:10.1016/S0550-3213(98)00350-2
 % [hep-ph/9712417].
  %%CITATION = doi:10.1016/S0550-3213(98)00350-2;%%
  %382 citations counted in INSPIRE as of 29 Apr 2018

%%\cite{Bailey:2014tva}
%\bibitem{Bailey:2014tva}
%  J.~A.~Bailey {\it et al.} [Fermilab Lattice and MILC Collaborations],
%  %``Update of $|V_{cb}|$ from the $\bar{B}\to D^*\ell\bar{\nu}$ form factor at zero recoil with three-flavor lattice QCD,''
%  {\it Phys. Rev.} {\bf D89} (2014) 114504.
% % doi:10.1103/PhysRevD.89.114504
% % [arXiv:1403.0635 [hep-lat]].
%  %%CITATION = doi:10.1103/PhysRevD.89.114504;%%
%  %102 citations counted in INSPIRE as of 29 Apr 2018

%\cite{Lattice:2015rga}
\bibitem{Lattice:2015rga}
  J.~A.~Bailey {\it et al.} [MILC Collaboration],
  %``B→Dℓν form factors at nonzero recoil and |V$_{cb}$| from 2+1-flavor lattice QCD,''
  {\it Phys. Rev.} {\bf D92} (2015) 034506.
 % doi:10.1103/PhysRevD.92.034506
 % [arXiv:1503.07237 [hep-lat]].
  %%CITATION = doi:10.1103/PhysRevD.92.034506;%%
  %127 citations counted in INSPIRE as of 29 Apr 2018

%\cite{Bigi:2016mdz}
\bibitem{Bigi:2016mdz}
  D.~Bigi and P.~Gambino,
  %``Revisiting $B\to D \ell \nu$,''
  {\it Phys. Rev.} {\bf D94} (2016) 094008.
 % doi:10.1103/PhysRevD.94.094008
 % [arXiv:1606.08030 [hep-ph]].
  %%CITATION = doi:10.1103/PhysRevD.94.094008;%%
  %59 citations counted in INSPIRE as of 29 Apr 2018\\
  
%\cite{Bigi:2017njr}
\bibitem{Bigi:2017njr}
  D.~Bigi, P.~Gambino and S.~Schacht,
  %``A fresh look at the determination of $|V_{cb}|$ from $B\to D^{*} \ell \nu$,''
  {\it Phys. Lett.} {\bf B769} (2017) 441;
 % doi:10.1016/j.physletb.2017.04.022
 % [arXiv:1703.06124 [hep-ph]].
  %%CITATION = doi:10.1016/j.physletb.2017.04.022;%%
  %23 citations counted in INSPIRE as of 29 Apr 2018
%
%\cite{Bigi:2017jbd}
%\bibitem{Bigi:2017jbd}
%  D.~Bigi, P.~Gambino and S.~Schacht,
  %``$R(D^*)$, $|V_{cb}|$, and the Heavy Quark Symmetry relations between form factors,''
  {\it JHEP} {\bf 1711} (2017) 061.
 % doi:10.1007/JHEP11(2017)061
 % [arXiv:1707.09509 [hep-ph]].
  %%CITATION = doi:10.1007/JHEP11(2017)061;%%
  %26 citations counted in INSPIRE as of 29 Apr 2018

%\cite{Grinstein:2017nlq}
\bibitem{Grinstein:2017nlq}
  B.~Grinstein and A.~Kobach,
  %``Model-Independent Extraction of $|V_{cb}|$ from $\bar{B}\rightarrow D^* \ell \overline{\nu}$,''
  {\it Phys. Lett.} {\bf B771} (2017) 359.
 % doi:10.1016/j.physletb.2017.05.078
 % [arXiv:1703.08170 [hep-ph]].
  %%CITATION = doi:10.1016/j.physletb.2017.05.078;%%
  %20 citations counted in INSPIRE as of 29 Apr 2018

%\cite{Bernlochner:2017jka} 
\bibitem{Bernlochner:2017jka}
  F.~U.~Bernlochner, Z.~Ligeti, M.~Papucci and D.~J.~Robinson,
  %``Combined analysis of semileptonic $B$ decays to $D$ and $D^*$: $R(D^{(*)})$, $|V_{cb}|$, and new physics,''
  {\it Phys. Rev.} {\bf D95} (2017)115008
   [Err: {\bf D97} (2018) 059902];
 % doi:10.1103/PhysRevD.95.115008, 10.1103/PhysRevD.97.059902
 % [arXiv:1703.05330 [hep-ph]].
  %%CITATION = doi:10.1103/PhysRevD.95.115008, 10.1103/PhysRevD.97.059902;%%
  %48 citations counted in INSPIRE as of 29 Apr 2018
%
%\cite{Bernlochner:2017xyx}
%\bibitem{Bernlochner:2017xyx}
%  F.~U.~Bernlochner, Z.~Ligeti, M.~Papucci and D.~J.~Robinson,
  %``Tensions and correlations in $|V_{cb}|$ determinations,''
%  Phys.\ Rev.\ D 
{\bf D96} (2017) 091503.
%  doi:10.1103/PhysRevD.96.091503
%  [arXiv:1708.07134 [hep-ph]].
  %%CITATION = doi:10.1103/PhysRevD.96.091503;%%
  %11 citations counted in INSPIRE as of 29 Apr 2018

%\cite{Boyd:1997kz}
\bibitem{Boyd:1997kz}
  C.~G.~Boyd, B.~Grinstein and R.~F.~Lebed,
  %``Precision corrections to dispersive bounds on form-factors,''
  {\it Phys. Rev.} {\bf D56} (1997) 6895.
 % doi:10.1103/PhysRevD.56.6895
 % [hep-ph/9705252].
  %%CITATION = doi:10.1103/PhysRevD.56.6895;%%
  %151 citations counted in INSPIRE as of 29 Apr 2018

%\cite{Abdesselam:2017kjf}
\bibitem{Abdesselam:2017kjf} Belle Collaboration,
 % A.~Abdesselam {\it et al.} [Belle Collaboration],
  %``Precise determination of the CKM matrix element $\left| V_{cb}\right|$ with $\bar B^0 \to D^{*\,+} \, \ell^- \, \bar \nu_\ell$ decays with hadronic tagging at Belle,''
  arXiv:1702.01521 [hep-ex].
  %%CITATION = ARXIV:1702.01521;%%
  %22 citations counted in INSPIRE as of 29 Apr 2018

%\cite{Waheed:2018djm}
\bibitem{Waheed:2018djm}
E.~Waheed \textit{et al.} [Belle],
%``Measurement of the CKM matrix element $|V_{cb}|$ from $B^0\to D^{*-}\ell^ {+} \nu_\ell$ at Belle,''
{\it Phys. Rev.}  \textbf{D100} (2019) 052007.
%doi:10.1103/PhysRevD.100.052007
%[arXiv:1809.03290 [hep-ex]].
%41 citations counted in INSPIRE as of 15 Apr 2020

%\cite{Gambino:2019sif}
\bibitem{Gambino:2019sif}
P.~Gambino, M.~Jung and S.~Schacht,
%``The $V_{cb}$ puzzle: An update,''
{\it Phys. Lett.}  \textbf{B795} (2019) 386. %%%-390
%doi:10.1016/j.physletb.2019.06.039
%[arXiv:1905.08209 [hep-ph]].
%26 citations counted in INSPIRE as of 15 Apr 2020

%\cite{Aubert:2009ac}
\bibitem{Aubert:2009ac} BaBar Collaboration,
 % B.~Aubert {\it et al.} [BaBar Collaboration],
  %``Measurement of |V(cb)| and the Form-Factor Slope in anti-B ---> D l- anti-nu Decays in Events Tagged by a Fully Reconstructed B Meson,''
  {\it Phys. Rev. Lett.}  {\bf 104} (2010) 011802.
%  doi:10.1103/PhysRevLett.104.011802
%  [arXiv:0904.4063 [hep-ex]].
  %%CITATION = doi:10.1103/PhysRevLett.104.011802;%%
  %85 citations counted in INSPIRE as of 16 May 2018

%\cite{Glattauer:2015teq}
\bibitem{Glattauer:2015teq} Belle Collaboration,
 % R.~Glattauer {\it et al.} [Belle Collaboration],
  %``Measurement of the decay $B\to D\ell\nu_\ell$ in fully reconstructed events and determination of the Cabibbo-Kobayashi-Maskawa matrix element $|V_{cb}|$,''
  {\it Phys. Rev.} {\bf D93} (2016) 032006.
 % doi:10.1103/PhysRevD.93.032006
 % [arXiv:1510.03657 [hep-ex]].
  %%CITATION = doi:10.1103/PhysRevD.93.032006;%%
  %37 citations counted in INSPIRE as of 16 May 2018


\bibitem{BI:93} I.I.Y. Bigi et al., {\it Phys. Rev. Lett.} {\bf 71} (1993) 496;
% [hep-ph/9304225].
{\it Phys. Lett.} {\bf B323} (1994) 408.

\bibitem{MW:94} A.V. Manohar and M.B. Wise, {\it Phys. Rev.} {\bf D49} (1994) 1310.
% [hep-ph/9308246].

%\cite{Gremm:1996df}
\bibitem{Gremm:1996df}
  M.~Gremm and A.~Kapustin,
  %``Order 1/m(b)**3 corrections to B --> X(c) lepton anti-neutrino decay and their implication for the measurement of Lambda-bar and lambda(1),''
  {\it Phys. Rev.} {\bf D55} (1997) 6924.
 % doi:10.1103/PhysRevD.55.6924
 % [hep-ph/9603448].
  %%CITATION = doi:10.1103/PhysRevD.55.6924;%%
  %174 citations counted in INSPIRE as of 30 Apr 2018
  
\bibitem{BBMU:03} D. Benson, I.I. Bigi, T. Mannel and N. Uraltsev, {\it Nucl. Phys.}
   {\bf B665} (2003) 367.
% [hep-ph/0302262].

\bibitem{BLMT:04} C.W. Bauer, Z. Ligeti, M. Luke, A.V. Manohar and M. Trott,
{\it Phys. Rev.} {\bf D70} (2004) 094017.
% [hep-ph/0408002].

\bibitem{GU:04} P. Gambino and N. Uraltsev, {\it Eur. Phys. J.} {\bf C34} (2004) 181.
% [hep-ph/0401063].

\bibitem{BBU:05} D. Benson, I.I. Bigi and N. Uraltsev, {\it Nucl. Phys.} {\bf B710}
    (2005) 371.
% [hep-ph/0410080].

\bibitem{BF:06} O. Buchmuller and H. Flacher, {\it Phys. Rev.} {\bf D73} (2006) 073008.
% [hep-ph/0507253].


%\cite{Mannel:2010wj}
\bibitem{Mannel:2010wj}
  T.~Mannel, S.~Turczyk and N.~Uraltsev,
  %``Higher Order Power Corrections in Inclusive B Decays,''
  {\it JHEP} {\bf 1011} (2010) 109.
 % doi:10.1007/JHEP11(2010)109
 % [arXiv:1009.4622 [hep-ph]].
  %%CITATION = doi:10.1007/JHEP11(2010)109;%%
  %41 citations counted in INSPIRE as of 30 Apr 2018

%\cite{Gambino:2011cq}
\bibitem{Gambino:2011cq}
  P.~Gambino,
  %``B semileptonic moments at NNLO,''
  {\it JHEP} {\bf 1109} (2011) 055.
%  doi:10.1007/JHEP09(2011)055
%  [arXiv:1107.3100 [hep-ph]].
  %%CITATION = doi:10.1007/JHEP09(2011)055;%%
  %29 citations counted in INSPIRE as of 29 Apr 2018

%\cite{Alberti:2012dn}
\bibitem{Alberti:2012dn}
  A.~Alberti, T.~Ewerth, P.~Gambino and S.~Nandi,
  %``Kinetic operator effects in $\bar{B}\to X_c l \nu$ at O($\alpha_s$),''
  {\it Nucl. Phys.} {\bf B870} (2013) 16.
 % doi:10.1016/j.nuclphysb.2013.01.005
 % [arXiv:1212.5082 [hep-ph]].
  %%CITATION = doi:10.1016/j.nuclphysb.2013.01.005;%%
  %20 citations counted in INSPIRE as of 30 Apr 2018
  
%\cite{Alberti:2013kxa}
\bibitem{Alberti:2013kxa}
  A.~Alberti, P.~Gambino and S.~Nandi,
  %``Perturbative corrections to power suppressed effects in semileptonic B decays,''
  {\it JHEP} {\bf 1401} (2014) 147.
 % doi:10.1007/JHEP01(2014)147
 % [arXiv:1311.7381 [hep-ph]].
  %%CITATION = doi:10.1007/JHEP01(2014)147;%%
  %24 citations counted in INSPIRE as of 30 Apr 2018

%\cite{Mannel:2015jka}
\bibitem{Mannel:2015jka}
  T.~Mannel, A.~A.~Pivovarov and D.~Rosenthal,
  %``Inclusive weak decays of heavy hadrons with power suppressed terms at NLO,''
  {\it Phys. Rev.} {\bf D92} (2015) 054025.
 % doi:10.1103/PhysRevD.92.054025
 % [arXiv:1506.08167 [hep-ph]].
  %%CITATION = doi:10.1103/PhysRevD.92.054025;%%
  %7 citations counted in INSPIRE as of 30 Apr 2018

%\cite{Alberti:2014yda}
\bibitem{Alberti:2014yda}
  A.~Alberti, P.~Gambino, K.~J.~Healey and S.~Nandi,
  %``Precision Determination of the Cabibbo-Kobayashi-Maskawa Element $V_{cb}$,''
  {\it Phys. Rev. Lett.} {\bf 114} (2015) 061802.
%  doi:10.1103/PhysRevLett.114.061802
%  [arXiv:1411.6560 [hep-ph]].
  %%CITATION = doi:10.1103/PhysRevLett.114.061802;%%
  %57 citations counted in INSPIRE as of 30 Apr 2018
  
%\cite{Gambino:2016jkc}
\bibitem{Gambino:2016jkc}
  P.~Gambino, K.~J.~Healey and S.~Turczyk,
  %``Taming the higher power corrections in semileptonic B decays,''
  {\it Phys. Lett.} {\bf B763} (2016) 60.
 % doi:10.1016/j.physletb.2016.10.023
 % [arXiv:1606.06174 [hep-ph]].
  %%CITATION = doi:10.1016/j.physletb.2016.10.023;%%
  %32 citations counted in INSPIRE as of 29 Apr 2018


\bibitem{BZ:05} P. Ball and R. Zwicky, {\it Phys. Rev.} {\bf D71} (2005) 014015.
%%% [hep-ph/0406232].

\bibitem{DKMMO:08} G. Duplancic \etal,
% A. Khodjamirian, T. Mannel, B. Melic and N. Offen,
{\it JHEP} {\bf 0804} (2008) 014.    %%% [arXiv:0801.1796 [hep-ph]].

\bibitem{DMOW:11} A. Khodjamirian \etal,
% T. Mannel, N. Offen and Y. M.Wang,
{\it Phys. Rev.} {\bf D83} (2011) 094031.
%%% arXiv:1103.2655 [hep-ph]]

%\cite{Bharucha:2012wy}
\bibitem{Bharucha:2012wy}
  A.~Bharucha,
  %``Two-loop Corrections to the $B to \pi$ Form Factor from QCD Sum Rules on the Light-Cone and $|V_{ub}|$,''
  {\it JHEP} {\bf 1205} (2012) 092.
 % doi:10.1007/JHEP05(2012)092
 % [arXiv:1203.1359 [hep-ph]].
  %%CITATION = doi:10.1007/JHEP05(2012)092;%%
  %45 citations counted in INSPIRE as of 30 Apr 2018

%\cite{Lattice:2015tia}
\bibitem{Lattice:2015tia}
  J.~A.~Bailey {\it et al.} [Fermilab Lattice and MILC Collaborations],
  %``$|V_{ub}|$ from $B\to\pi\ell\nu$ decays and (2+1)-flavor lattice QCD,''
  {\it Phys. Rev.} {\bf D92} (2015) 014024.
 % doi:10.1103/PhysRevD.92.014024
 % [arXiv:1503.07839 [hep-lat]].
  %%CITATION = doi:10.1103/PhysRevD.92.014024;%%
  %86 citations counted in INSPIRE as of 30 Apr 2018

%\cite{Flynn:2015mha}
\bibitem{Flynn:2015mha} J.~M.~Flynn \etal,
 % J.~M.~Flynn, T.~Izubuchi, T.~Kawanai, C.~Lehner, A.~Soni, R.~S.~Van de Water and O.~Witzel,
  %``$B \to \pi \ell \nu$ and $B_s \to K \ell \nu$ form factors and $|V_{ub}|$ from 2+1-flavor lattice QCD with domain-wall light quarks and relativistic heavy quarks,''
  {\it Phys. Rev.} {\bf D91} (2015) 074510.
%  doi:10.1103/PhysRevD.91.074510
%  [arXiv:1501.05373 [hep-lat]].
  %%CITATION = doi:10.1103/PhysRevD.91.074510;%%
  %65 citations counted in INSPIRE as of 30 Apr 2018

%\cite{Antonelli:2009ws}
\bibitem{Antonelli:2009ws}
  M.~Antonelli {\it et al.},
  %``Flavor Physics in the Quark Sector,''
  {\it Phys. Rept.}  {\bf 494} (2010) 197.
 % doi:10.1016/j.physrep.2010.05.003
 % [arXiv:0907.5386 [hep-ph]].
  %%CITATION = doi:10.1016/j.physrep.2010.05.003;%%
  %287 citations counted in INSPIRE as of 30 Apr 2018
  
\bibitem{BLNP:05} B.O. Lange \etal,
% B. O. Lange, M. Neubert, G. Paz,
{\it Phys. Rev.} {\bf D72} (2005) 073006;   %%% [hep-ph/0504071].
% S.W. Bosch, B. O. Lange, M. Neubert, G. Paz,
{\it Nucl. Phys.} {\bf B699} (2004) 335. %%% -386. [hep-ph/0402094];

\bibitem{AG:06} J.R. Andersen and E. Gardi, {\it JHEP} {\bf 0601} (2006) 097.
%%%   [hep-ph/0509360].

\bibitem{GGOU:07} P. Gambino, P. Giordano, G. Ossola and N. Uraltsev,
{\it JHEP} {\bf 0710} (2007) 058. %%% [arXiv:0707.2493 [hep-ph]].

\bibitem{ALFR:09} U. Aglietti, F. Di Lodovico, G. Ferrera and G. Ricciardi,
{\it Eur. Phys. J.} {\bf C59} (2009) 831. %%% [arXiv:0711.0860 [hep-ph]].

\bibitem{GNP:10} C. Greub, M. Neubert and B. D. Pecjak, {\it Eur. Phys. J.} {\bf C65} (2010)  501.
% [arXiv:0909.1609 [hep-ph]].

\bibitem{PA:09} G. Paz, {\it JHEP} {\bf 0906} (2009) 083. % [arXiv:0903.3377 [hep-ph]].

\bibitem{LLM:10} Z. Ligeti \etal,
% Z. Ligeti, M. Luke, A. V. Manohar,
{\it Phys. Rev.} {\bf D82} (2010) 033003;    %%% [arXiv:1003.1351 [hep-ph]].
% Z. Ligeti, I. W. Stewart, F. J. Tackmann, Phys. Rev.
{\bf D78} (2008) 114014; %%% [arXiv:0807.1926 [hep-ph]].
% F. U. Bernlochner, H. Lacker, Z. Ligeti, I. W. Stewart, F. J. Tackmann and K. Tackmann,
arXiv:1101.3310 [hep-ph].

\bibitem{GK:10} P. Gambino and J.F. Kamenik, {\it Nucl. Phys.} {\bf B840} (2010) 424.
%%%   [arXiv:1004.0114 [hep-ph]].

%\cite{Aaij:2015bfa}
\bibitem{Aaij:2015bfa} LHCb Collaboration,
  % R.~Aaij {\it et al.} [LHCb Collaboration],
  %``Determination of the quark coupling strength $|V_{ub}|$ using baryonic decays,''
  {\it Nature Phys.}  {\bf 11} (2015) 743.
 % doi:10.1038/nphys3415
 % [arXiv:1504.01568 [hep-ex]].
  %%CITATION = doi:10.1038/nphys3415;%%
  %96 citations counted in INSPIRE as of 30 Apr 2018

%\cite{Aubert:2007xj}
\bibitem{Aubert:2007xj} BaBar Collaboration,
 % B.~Aubert {\it et al.} [BaBar Collaboration],
  %``A Search for $B^{+} \to \tau^{+} \nu$ with Hadronic $B$ tags,''
  {\it Phys. Rev.} {\bf D77} (2008) 011107;
 % doi:10.1103/PhysRevD.77.011107
 % [arXiv:0708.2260 [hep-ex]].
  %%CITATION = doi:10.1103/PhysRevD.77.011107;%%
  %149 citations counted in INSPIRE as of 08 May 2018
%
%\cite{Aubert:2009wt}
%\bibitem{Aubert:2009wt}
%  B.~Aubert {\it et al.} [BaBar Collaboration],
%  %``A Search for $B^+ \to \ell^+ \nu_{\ell}$ Recoiling Against $B^{-}\to D^{0} \ell^{-}\bar{\nu} X$,''
%  Phys.\ Rev.\ D 
{\bf D81} (2010) 051101;
%  doi:10.1103/PhysRevD.81.051101
%  [arXiv:0912.2453 [hep-ex]].
%  %%CITATION = doi:10.1103/PhysRevD.81.051101;%%
%  %157 citations counted in INSPIRE as of 08 May 2018
%
%\cite{Lees:2012ju}
%\bibitem{Lees:2012ju}
%  J.~P.~Lees {\it et al.} [BaBar Collaboration],
%  %``Evidence of $B^+ \to \tau^+\nu$ decays with hadronic B tags,''
%  Phys.\ Rev.\ D 
{\bf D88} (2013) 031102.
%  doi:10.1103/PhysRevD.88.031102
%  [arXiv:1207.0698 [hep-ex]].
%  %%CITATION = doi:10.1103/PhysRevD.88.031102;%%
%  %162 citations counted in INSPIRE as of 08 May 2018

%\cite{Adachi:2012mm}
\bibitem{Adachi:2012mm} Belle Collaboration,
 % I.~Adachi {\it et al.} [Belle Collaboration],
  %``Evidence for $B^- \to \tau^- \bar{\nu}_\tau$  with a Hadronic Tagging Method Using the Full Data Sample of Belle,''
  {\it Phys. Rev. Lett.}  {\bf 110} (2013) 131801.
 % doi:10.1103/PhysRevLett.110.131801
 % [arXiv:1208.4678 [hep-ex]].
  %%CITATION = doi:10.1103/PhysRevLett.110.131801;%%
  %216 citations counted in INSPIRE as of 08 May 2018
%
%\cite{Kronenbitter:2015kls}
%\bibitem{Kronenbitter:2015kls}
%  B.~Kronenbitter {\it et al.} [Belle Collaboration],
%  %``Measurement of the branching fraction of B^+ -> tau^+ nu_tau decays with the semileptonic tagging method,''
 {\it Phys. Rev.} {\bf D92} (2015) 051102.
%  doi:10.1103/PhysRevD.92.051102
%  [arXiv:1503.05613 [hep-ex]].
%  %%CITATION = doi:10.1103/PhysRevD.92.051102;%%
%  %57 citations counted in INSPIRE as of 08 May 2018

%\cite{Khachatryan:2014nda}
\bibitem{Khachatryan:2014nda} CMS Collaboration,
 % V.~Khachatryan {\it et al.} [CMS Collaboration],
  %``Measurement of the ratio $\mathcal B(t \to Wb)/\mathcal B(t \to Wq)$ in pp collisions at $\sqrt{s}$ = 8 TeV,''
  {\it Phys. Lett.} {\bf B736} (2014) 33.
 % doi:10.1016/j.physletb.2014.06.076
 % [arXiv:1404.2292 [hep-ex]].
  %%CITATION = doi:10.1016/j.physletb.2014.06.076;%%
  %69 citations counted in INSPIRE as of 30 Apr 2018

\bibitem{LEPEWWG}
The ALEPH, CDF, D0, DELPHI, L3, OPAL, SLD Collaborations, the LEP
Electroweak Working Group, the Tevatron Electroweak Working Group and
the SLD Electroweak and Heavy Flavour Groups,
%%% Precision Electroweak Measurements and Constraints on the Standard Model
arXiv:1012.2367 [hep-ex];
%%% arXiv:hep-ex/0612034;
%%% arXiv:hep-ex/0412015;
%%% arXiv:hep-ex/0312023;  %%%CERN-EP/2003-091;
%%% http://www.cern.ch/LEPEWWG.
http://www.cern.ch/LEPEWWG/.

\bibitem{LEPEWWG_SLD:06}
The ALEPH, DELPHI, L3, OPAL and SLD Collaborations, the LEP
Electroweak Working Group and the SLD Electroweak and Heavy Flavour
Groups, {\it Phys. Rept.} {\bf 427} (2006) 257. %%%arXiv:hep-ex/0509008

\bibitem{WO:83} L. Wolfenstein, {\it Phys. Rev. Lett.} {\bf 51} (1983) 1945.

\bibitem{BLO:94} A.J. Buras, M.E. Lautenbacher and G. Ostermaier, {\it Phys. Rev.}
{\bf D50} (1994) 3433.

%%%%%%%%%%%%%%%%%%%%%%%%%%%%%%%%%%

\bibitem{IL:81} T. Inami and C.S. Lim, {\it Progr. Theor. Phys.} {\bf 65} (1981) 297.

\bibitem{GL:74} M.K. Gaillard and B.W. Lee, {\it Phys. Rev.} {\bf D10} (1974) 897.

\bibitem{ARGUS:87} ARGUS Collaboration, {\it Phys. Lett.} {\bf B192} (1987) 245.

\bibitem{CDF:06} CDF Collaboration, {\it Phys. Rev. Lett.} {\bf 97} (2006) 242003.

\bibitem{BJW:90} A.J. Buras, M. Jamin and P.H. Weisz, {\it Nucl. Phys.} {\bf B347} (1990) 491.

\bibitem{HN:94} S. Herrlich and U. Nierste, {\it Nucl. Phys.} {\bf B419} (1994) 292;
{\bf B476} (1996) 27.

\bibitem{PP:95} A. Pich and J. Prades, {\it Phys. Lett.} {\bf B346} (1995) 342.



%%%%%%%%%%%%%%%%%%%%%%%%%%%%%%%%%%

\bibitem{JA:85} C. Jarlskog, {\it Phys. Rev. Lett.} {\bf 55} (1985) 1039;
   {\it Z. Phys.} {\bf C29} (1985) 491.

\bibitem{CKMfitter} CKMfitter Group, %%% (J. Charles et al.),
{\it Eur. Phys. J.} {\bf C41} (2005) 1;    %%%[hep-ph/0406184]
and 2019 update at http://ckmfitter.in2p3.fr/.

\bibitem{NA48} NA48 Collaboration, {\it Phys. Lett.} {\bf B544} (2002) 97;
{\bf B465} (1999) 335; {\it Eur. Phys. J.} {\bf C22} (2001) 231.

\bibitem{KTeV} KTeV Collaboration, {\it Phys. Rev.} {\bf D83} (2011) 092001;
{\bf D67} (2003) 012005; {\it Phys. Rev. Lett.} {\bf 83} (1999) 22.

\bibitem{NA31} NA31 Collaboration, {\it Phys. Lett.} {\bf B317} (1993) 233;
 {\bf B206} (1988) 169.

\bibitem{E731} E731 Collaboration, {\it Phys. Rev. Lett.} {\bf 70} (1993) 1203.
% Gibbons, L. K., et al.,

\bibitem{BJL:93} A.J.~Buras, M.~Jamin and  M.E.~Lautenbacher,
   {\it Nucl. Phys.} {\bf B408} (1993) 209; {\it Phys. Lett.} {\bf B389} (1996) 749.

\bibitem{ciuc:93} M.~Ciuchini  \etal, {\it Phys. Lett.} {\bf B301} (1993) 263;
    {\it Z. Phys.} {\bf C68} (1995) 239.

\bibitem{PP:00} E. Pallante and A. Pich, {\it Phys. Rev. Lett.} {\bf 84} (2000) 2568;
{\it Nucl. Phys.} {\bf B592} 294.

\bibitem{PPS:01}
E. Pallante, A. Pich and I. Scimemi, {\it Nucl. Phys.} {\bf B617} (2001) 441.

\bibitem{CPEN:03}
V. Cirigliano, A. Pich, G. Ecker and H. Neufeld, {\it Phys. Rev. Lett.} {\bf 91} (2003) 162001;
{\it Eur. Phys. J.} {\bf C33} (2004) 369

%\cite{Gisbert:2017vvj}
\bibitem{Gisbert:2017vvj}
H.~Gisbert and A.~Pich,
%``Direct CP violation in $K^0\to\pi\pi$: Standard Model Status,''
{\it Rept. Prog. Phys.} \textbf{81} (2018) 076201.
%doi:10.1088/1361-6633/aac18e
%[arXiv:1712.06147 [hep-ph]].
%43 citations counted in INSPIRE as of 17 Apr 2020

%\cite{Cirigliano:2019cpi}
\bibitem{Cirigliano:2019cpi}
V.~Cirigliano, H.~Gisbert, A.~Pich and A.~Rodr\'{\i}guez-S\'anchez,
%``Isospin-violating contributions to  $\epsilon'/\epsilon$,''
{\it JHEP} \textbf{02} (2020) 032.
%doi:10.1007/JHEP02(2020)032
%[arXiv:1911.01359 [hep-ph]].
%9 citations counted in INSPIRE as of 17 Apr 2020

%\cite{Abbott:2020hxn}
\bibitem{Abbott:2020hxn} R.~Abbott {\it et al.} [RBC and UKQCD Collaborations],
%R.~Abbott, T.~Blum, P.~A.~Boyle, M.~Bruno, N.~H.~Christ, D.~Hoying, C.~Jung, C.~Kelly, C.~Lehner, R.~D.~Mawhinney, D.~J.~Murphy, C.~T.~Sachrajda, A.~Soni, M.~Tomii and T.~Wang,
%``Direct CP violation and the $\Delta I=1/2$ rule in $K\to\pi\pi$ decay from the Standard Model,''
arXiv:2004.09440 [hep-lat].
%0 citations counted in INSPIRE as of 24 Apr 2020\\

%\cite{Buchalla:1995vs}
\bibitem{Buchalla:1995vs}
  G.~Buchalla, A.~J.~Buras and M.~E.~Lautenbacher,
  %``Weak decays beyond leading logarithms,''
 {\it Rev. Mod. Phys.} {\bf 68} (1996) 1125.
 % doi:10.1103/RevModPhys.68.1125
 % [hep-ph/9512380].
  %%CITATION = doi:10.1103/RevModPhys.68.1125;%%
  %2267 citations counted in INSPIRE as of 04 May 2018  
  
\bibitem{BG:10} J. Brod and M. Gorbahn, {\it Phys. Rev.} {\bf D82} (2010) 094026.

%%%%%%%%%%%%%%%%%%%%%%%%%%%%%%%%%%%%%%%%%%%%%%%%%

\bibitem{CS:80}
  A.B.~Carter and A.I. Sanda, {\it Phys. Rev. Lett.} {\bf 45} (1980) 952;
   {\it Phys. Rev.} {\bf D23} (1981) 1567.

\bibitem{BS:81}
  I.I. Bigi and A.I. Sanda, {\it Nucl. Phys.} {\bf B193} (1981) 85.

\bibitem{KLPS:88} P. Krawczyk \etal, {\it Nucl. Phys.} {\bf B307} (1988) 19.
% P. Krawczyk?, D. London, R.D. Peccei, H. Steger

%\bibitem{BI:07} I.I. Bigi, arXiv:hep-ph/0701273.

%\cite{Adachi:2012et}
\bibitem{Adachi:2012et} Belle collaboration,
  %I.~Adachi {\it et al.},
  %``Precise measurement of the CP violation parameter sin2phi_1 in B0-->(c\bar c)K0 decays,''
  {\it Phys. Rev. Lett.}  {\bf 108} (2012) 171802.
%  doi:10.1103/PhysRevLett.108.171802
%  [arXiv:1201.4643 [hep-ex]].
  %%CITATION = doi:10.1103/PhysRevLett.108.171802;%%
  %118 citations counted in INSPIRE as of 03 May 2018

%\cite{Adachi:2018itz}
\bibitem{Adachi:2018itz} BaBar and Belle Collaborations,
%I.~Adachi \textit{et al.} [BaBar and Belle],
%``First Evidence for cos2β>0 and Resolution of the Cabibbo-Kobayashi-Maskawa Quark-Mixing Unitarity Triangle Ambiguity,''
{\it Phys. Rev. Lett.} \textbf{121} (2018) 261801;
%doi:10.1103/PhysRevLett.121.261801
%[arXiv:1804.06152 [hep-ex]].
%8 citations counted in INSPIRE as of 17 Apr 2020
%
%\cite{Adachi:2018jqe}
%\bibitem{Adachi:2018jqe}
%I.~Adachi \textit{et al.} [BaBar and Belle],
%``Measurement of $\cos{2\beta}$ in $B^{0} \to D^{(*)} h^{0}$ with $D \to K_{S}^{0} \pi^{+} \pi^{-}$ decays by a combined time-dependent Dalitz plot analysis of BaBar and Belle data,''
{\it Phys. Rev.}  \textbf{D98} (2018) 112012.
%doi:10.1103/PhysRevD.98.112012
%[arXiv:1804.06153 [hep-ex]].
%7 citations counted in INSPIRE as of 17 Apr 2020

  
%\bibitem{Babar:05} BABAR Collaboration,  {\it Phys. Rev.} {\bf D71} (2005) 032005;
%{\it Phys. Rev. Lett.} {\bf 99} (2007) 231802.
%
%\bibitem{Belle:05} BELLE Collaboration, {\it Phys. Rev. Lett.} {\bf 95} (2005) 091601; {\bf 97} (2006) 081801.

\bibitem{GL:90} M. Gronau and D. London, {\it Phys. Rev. Lett.} {\bf 65} (1990) 3381.

\bibitem{GL:91} M. Gronau and D. London, {\it Phys. Lett.} {\bf B253} (1991) 483.

\bibitem{GW:91}  M. Gronau and D. Wyler, {\it Phys. Lett.} {\bf B265} (1991) 172.

\bibitem{AT:97} D. Atwood \etal, {\it Phys. Rev. Lett} {\bf 78} (1997) 3257;
{\it Phys. Rev.} {\bf D63} (2001) 036005.

%\cite{Lenz:2006hd}
\bibitem{Lenz:2006hd}
  A.~Lenz and U.~Nierste,
  %``Theoretical update of $B_s - \bar{B}_s$ mixing,''
  {\it JHEP} {\bf 0706} (2007) 072.
%  doi:10.1088/1126-6708/2007/06/072
%  [hep-ph/0612167].
  %%CITATION = doi:10.1088/1126-6708/2007/06/072;%%
  %587 citations counted in INSPIRE as of 04 May 2018
  
%\cite{Artuso:2015swg}
\bibitem{Artuso:2015swg}
  M.~Artuso, G.~Borissov and A.~Lenz,
  %``CP violation in the $B_s^0$ system,''
  {\it Rev. Mod. Phys.} {\bf 88} (2016) 045002.
 % doi:10.1103/RevModPhys.88.045002
 % [arXiv:1511.09466 [hep-ph]].
  %%CITATION = doi:10.1103/RevModPhys.88.045002;%%
  %48 citations counted in INSPIRE as of 04 May 2018\\
  
%\cite{Jubb:2016mvq}
\bibitem{Jubb:2016mvq}
  T.~Jubb, M.~Kirk, A.~Lenz and G.~Tetlalmatzi-Xolocotzi,
  %``On the ultimate precision of meson mixing observables,''
  {\it Nucl. Phys.} {\bf B915} (2017) 431.
  %doi:10.1016/j.nuclphysb.2016.12.020
  %[arXiv:1603.07770 [hep-ph]].
  %%CITATION = doi:10.1016/j.nuclphysb.2016.12.020;%%
  %12 citations counted in INSPIRE as of 04 May 2018

\bibitem{UTfit} UTfit Collaboration,  %%%M. Bona et al.
{\it JHEP} {\bf 0507} (2005) 028;  %  [hep-ph/0501199]
% {\it JHEP} {\bf 0610} (2006) 081; %%%arXiv:hep-ph/0606167;
and 2018 update at http://www.utfit.org/UTfit/.


%\cite{Aaij:2019kcg}
\bibitem{Aaij:2019kcg} LHCb Collaboration,
% R.~Aaij \textit{et al.} [LHCb],
%``Observation of CP Violation in Charm Decays,''
{\it Phys. Rev. Lett.} \textbf{122} (2019) 211803.
%doi:10.1103/PhysRevLett.122.211803
%[arXiv:1903.08726 [hep-ex]].
%75 citations counted in INSPIRE as of 18 Apr 2020

%%%%%%%%%%%%%%%%%%%%%%%%%%%%%%%%%%%%%%%%%%%%%%

%\cite{Aaij:2020sbt}
\bibitem{Aaij:2020sbt} LHCb Collaboration,
%R.~Aaij \textit{et al.} [LHCb],
%``Strong constraints on the $K^0_{\mathrm{S}}\rightarrow\mu^{+}\mu^{-}$ branching fraction,''
arXiv:2001.10354 [hep-ex].
%1 citations counted in INSPIRE as of 18 Apr 2020

%%\cite{Aaij:2017tia}
%\bibitem{Aaij:2017tia} LHCb Collaboration,
% % R.~Aaij {\it et al.} [LHCb Collaboration],
%  %``Improved limit on the branching fraction of the rare decay ${{K} ^0_{\mathrm { \scriptscriptstyle S}}} \rightarrow \mu ^+\mu ^-$,''
%  {\it Eur. Phys. J.} {\bf C77} (2017) 678.
%  % doi:10.1140/epjc/s10052-017-5230-x
%  % [arXiv:1706.00758 [hep-ex]].
%  %%CITATION = doi:10.1140/epjc/s10052-017-5230-x;%%
%  %6 citations counted in INSPIRE as of 04 May 2018

%\cite{Gorbahn:2006bm}
\bibitem{Gorbahn:2006bm}
  M.~Gorbahn and U.~Haisch,
  %``Charm Quark Contribution to K(L) ---> mu+ mu- at Next-to-Next-to-Leading,''
  {\it Phys. Rev. Lett.} {\bf 97} (2006) 122002.
 % doi:10.1103/PhysRevLett.97.122002
 % [hep-ph/0605203].
  %%CITATION = doi:10.1103/PhysRevLett.97.122002;%%
  %46 citations counted in INSPIRE as of 04 May 2018
  
%\cite{GomezDumm:1998gw}
\bibitem{GomezDumm:1998gw}
  D.~Gomez Dumm and A.~Pich,
  %``Long distance contributions to the K(L) ---> mu+ mu- decay width,''
  {\it Phys. Rev. Lett.} {\bf 80} (1998) 4633.
 % doi:10.1103/PhysRevLett.80.4633
 % [hep-ph/9801298].
  %%CITATION = doi:10.1103/PhysRevLett.80.4633;%%
  %84 citations counted in INSPIRE as of 04 May 2018

%\cite{Ecker:1991ru}
\bibitem{Ecker:1991ru}
  G.~Ecker and A.~Pich,
  %``The Longitudinal muon polarization in K(L) ---> mu+ mu-,''
  {\it Nucl. Phys.} {\bf B366} (1991) 189.
 % doi:10.1016/0550-3213(91)90056-4
  %%CITATION = doi:10.1016/0550-3213(91)90056-4;%%
  %90 citations counted in INSPIRE as of 04 May 2018

%\cite{Ecker:1987hd}
\bibitem{Ecker:1987hd}
  G.~Ecker, A.~Pich and E.~de Rafael,
  %``Radiative Kaon Decays and CP Violation in Chiral Perturbation Theory,''
  {\it Nucl. Phys.} {\bf B303} (1988) 665;
%  doi:10.1016/0550-3213(88)90425-7
  %%CITATION = doi:10.1016/0550-3213(88)90425-7;%%
  %339 citations counted in INSPIRE as of 04 May 2018
  %
  %\cite{Ecker:1987qi}
%\bibitem{Ecker:1987qi}
%  G.~Ecker, A.~Pich and E.~de Rafael,
  %``K ---> pi Lepton+ Lepton- Decays in the Effective Chiral Lagrangian of the Standard Model,''
 % Nucl.\ Phys.\ B 
 {\bf B291} (1987) 692;
 % doi:10.1016/0550-3213(87)90491-3
  %%CITATION = doi:10.1016/0550-3213(87)90491-3;%%
  %298 citations counted in INSPIRE as of 04 May 2018
%
%\cite{Ecker:1987fm}
%\bibitem{Ecker:1987fm}
%  G.~Ecker, A.~Pich and E.~de Rafael,
  %``K0 ---> pi0 gamma Gamma Decays in Chiral Perturbation Theory,''
  {\it Phys. Lett.} {\bf B189} (1987) 363.
 % doi:10.1016/0370-2693(87)91448-1
  %%CITATION = doi:10.1016/0370-2693(87)91448-1;%%
  %205 citations counted in INSPIRE as of 04 May 2018

%\cite{Buras:1994qa}
\bibitem{Buras:1994qa}
  A.~J.~Buras, M.~E.~Lautenbacher, M.~Misiak and M.~Munz,
  %``Direct CP violation in K(L) ---> pi0 e+ e- beyond leading logarithms,''
  {\it Nucl. Phys.} {\bf B423} (1994) 349.
 % doi:10.1016/0550-3213(94)90138-4
 % [hep-ph/9402347].
  %%CITATION = doi:10.1016/0550-3213(94)90138-4;%%
  %110 citations counted in INSPIRE as of 04 May 2018

%\cite{Buchalla:2003sj}
\bibitem{Buchalla:2003sj}
  G.~Buchalla, G.~D'Ambrosio and G.~Isidori,
  %``Extracting short distance physics from K(L,S) ---> pi0 e+ e- decays,''
  {\it Nucl. Phys.} {\bf B672} (2003) 387.
 % doi:10.1016/j.nuclphysb.2003.09.010
 % [hep-ph/0308008].
  %%CITATION = doi:10.1016/j.nuclphysb.2003.09.010;%%
  %134 citations counted in INSPIRE as of 04 May 2018

%\cite{AlaviHarati:2003mr}
\bibitem{AlaviHarati:2003mr} KTeV Collaboration,
  % A.~Alavi-Harati {\it et al.} [KTeV Collaboration],
  %``Search for the rare decay K(L) ---> pi0 e+ e-,''
  {\it Phys. Rev. Lett.} {\bf 93} (2004) 021805.
 % doi:10.1103/PhysRevLett.93.021805
 % [hep-ex/0309072].
  %%CITATION = doi:10.1103/PhysRevLett.93.021805;%%
  %95 citations counted in INSPIRE as of 04 May 2018

%\cite{Buras:2005gr}
\bibitem{Buras:2005gr}
  A.~J.~Buras, M.~Gorbahn, U.~Haisch and U.~Nierste,
  %``The Rare decay K+ ---> pi+ nu anti-nu at the next-to-next-to-leading order in QCD,''
  {\it Phys. Rev. Lett.} {\bf 95} (2005) 261805;
 % doi:10.1103/PhysRevLett.95.261805
 % [hep-ph/0508165].
  %%CITATION = doi:10.1103/PhysRevLett.95.261805;%%
  %150 citations counted in INSPIRE as of 04 May 2018
%
%\cite{Buras:2006gb}
%\bibitem{Buras:2006gb}
%  A.~J.~Buras, M.~Gorbahn, U.~Haisch and U.~Nierste,
  %``Charm quark contribution to K+ ---> pi+ nu anti-nu at next-to-next-to-leading order,''
  {\it JHEP} {\bf 0611} (2006) 002
   [Err: {\bf 1211} (2012) 167].
%  doi:10.1007/JHEP11(2012)167, 10.1088/1126-6708/2006/11/002
%  [hep-ph/0603079].
  %%CITATION = doi:10.1007/JHEP11(2012)167, 10.1088/1126-6708/2006/11/002;%%
  %166 citations counted in INSPIRE as of 04 May 2018

%\cite{Brod:2010hi}
\bibitem{Brod:2010hi}
  J.~Brod, M.~Gorbahn and E.~Stamou,
  %``Two-Loop Electroweak Corrections for the $K \to \pi \nu \bar{\nu}$ Decays,''
  {\it Phys. Rev.} {\bf D83} (2011) 034030.
 % doi:10.1103/PhysRevD.83.034030
 % [arXiv:1009.0947 [hep-ph]].
  %%CITATION = doi:10.1103/PhysRevD.83.034030;%%
  %238 citations counted in INSPIRE as of 04 May 2018\\
  
%%\cite{Buras:2015qea}
%\bibitem{Buras:2015qea}
%  A.~J.~Buras, D.~Buttazzo, J.~Girrbach-Noe and R.~Knegjens,
%  %``$ {K}^{+}\to {\pi}^{+}\nu \overline{\nu} $ and $ {K}_L\to {\pi}^0\nu \overline{\nu} $ in the Standard Model: status and perspectives,''
%  {\it JHEP} {\bf 1511} (2015) 033.
%  % doi:10.1007/JHEP11(2015)033
%  % [arXiv:1503.02693 [hep-ph]].
%  %%CITATION = doi:10.1007/JHEP11(2015)033;%%
%  %129 citations counted in INSPIRE as of 04 May 2018

\bibitem{Gorbahn:KAON2019}
M. Gorbahn, 
%{\it Latest results of the $K\to\pi\nu\bar\nu$ branching ratio calculations}, 
talk at Kaon 2019, Perugia (Italy), September 10th, 2019.

\bibitem{Ruggiero:KAON2019} NA62 Collaboration, G. Ruggiero,
%{\it New result on $K^+\to\pi^+\nu\bar\nu$ from the NA62 experiment}, 
talk at Kaon 2019, Perugia (Italy), September 10th, 2019.


%\cite{Ahn:2018mvc}
\bibitem{Ahn:2018mvc} KOTO Collaboration,
%J.~Ahn \textit{et al.} [KOTO],
%``Search for the $K_L \!\to\! \pi^0 \nu \overline{\nu}$ and $K_L \!\to\! \pi^0 X^0$ decays at the J-PARC KOTO experiment,''
{\it Phys. Rev. Lett.} \textbf{122} (2019) 021802.
%doi:10.1103/PhysRevLett.122.021802
%[arXiv:1810.09655 [hep-ex]].
%44 citations counted in INSPIRE as of 19 Apr 2020

%\bibitem{E949:08} E949 Collaboration, % Artamonov, A. V., et al. (E949), 2008,
%{\it Phys. Rev. Lett.} {\bf 101} (2008) 191802.
%
%\bibitem{E391a:08} E391a Collaboration, % Ahn et al.
%{\it Phys. Rev. Lett.} {\bf 100} (2008) 201802;
%{\it Phys. Rev.} {\bf D81} (2010) 072004.

%%\cite{Anelli:2005ju}
%\bibitem{Anelli:2005ju}
%  G.~Anelli {\it et al.},
%  %``Proposal to measure the rare decay K+ ---> pi+ nu anti-nu at the CERN SPS,''
%  CERN-SPSC-2005-013, CERN-SPSC-P-326 (2005).
%  %%CITATION = CERN-SPSC-2005-013, CERN-SPSC-P-326;%%
%  %155 citations counted in INSPIRE as of 04 May 2018
%
%%\cite{Beckford:2017gsf}
%\bibitem{Beckford:2017gsf}
%  B.~Beckford [KOTO Collaboration],
%  %``Present status of the search for the K$^{0}_{L} \rightarrow \pi^{0}\nu\bar{\nu}$ decay with the KOTO detector at J-PARC,''
%  arXiv:1710.01412 [hep-ex].
%  %%CITATION = ARXIV:1710.01412;%%
%  %2 citations counted in INSPIRE as of 04 May 2018

%\cite{Misiak:2015xwa}
\bibitem{Misiak:2015xwa}
  M.~Misiak {\it et al.},
  %``Updated NNLO QCD predictions for the weak radiative B-meson decays,''
  {\it Phys. Rev. Lett.}  {\bf 114} (2015)  221801.
 % doi:10.1103/PhysRevLett.114.221801
 % [arXiv:1503.01789 [hep-ph]].
  %%CITATION = doi:10.1103/PhysRevLett.114.221801;%%
  %214 citations counted in INSPIRE as of 04 May 2018

%\cite{Czakon:2015exa}
\bibitem{Czakon:2015exa} M.~Czakon {\it et al.},
 % M.~Czakon, P.~Fiedler, T.~Huber, M.~Misiak, T.~Schutzmeier and M.~Steinhauser,
  %``The $(Q_{7}, Q_{1,2})$ contribution to $ \overline{B}\to {X}_s\gamma $ at $ \mathcal{O}\left({\alpha}_{\mathrm{s}}^2\right) $,''
 {\it JHEP} {\bf 1504} (2015) 168.
 % doi:10.1007/JHEP04(2015)168
 % [arXiv:1503.01791 [hep-ph]].
  %%CITATION = doi:10.1007/JHEP04(2015)168;%%
  %55 citations counted in INSPIRE as of 04 May 2018

%\cite{Misiak:2020vlo}
\bibitem{Misiak:2020vlo}
M.~Misiak, A.~Rehman and M.~Steinhauser,
%``Towards $\bar{B}\rightarrow X_s\gamma$ at the NNLO in QCD without interpolation in $m_c$,''
arXiv:2002.01548 [hep-ph].
%2 citations counted in INSPIRE as of 19 Apr 2020

%%\cite{Aaij:2013cza}
%\bibitem{Aaij:2013cza} LHCb Collaboration,
%%R.~Aaij \textit{et al.} [LHCb],
%%``Search for the rare decay $D^0 \to \mu^+ \mu^-$,''
%{\it Phys. Lett.}  \textbf{B725} (2013) 15.
%%doi:10.1016/j.physletb.2013.06.037
%%[arXiv:1305.5059 [hep-ex]].
%%62 citations counted in INSPIRE as of 19 Apr 2020


%\cite{CMS:2014xfa}
\bibitem{CMS:2014xfa} CMS and LHCb Collaborations,
 % V.~Khachatryan {\it et al.} [CMS and LHCb Collaborations],
  %``Observation of the rare $B^0_s\to\mu^+\mu^-$ decay from the combined analysis of CMS and LHCb data,''
  {\it Nature} {\bf 522} (2015) 68.
 % doi:10.1038/nature14474
 % [arXiv:1411.4413 [hep-ex]].
  %%CITATION = doi:10.1038/nature14474;%%
  %380 citations counted in INSPIRE as of 04 May 2018
  
%\cite{Aaij:2017vad}
\bibitem{Aaij:2017vad} LHCb Collaboration,
 % R.~Aaij {\it et al.} [LHCb Collaboration],
  %``Measurement of the $B^0_s\to\mu^+\mu^-$ branching fraction and effective lifetime and search for $B^0\to\mu^+\mu^-$ decays,''
  {\it Phys. Rev. Lett.} {\bf 118} (2017) 191801.
 % doi:10.1103/PhysRevLett.118.191801
 % [arXiv:1703.05747 [hep-ex]].
  %%CITATION = doi:10.1103/PhysRevLett.118.191801;%%
  %88 citations counted in INSPIRE as of 04 May 2018

%\cite{Aaboud:2018mst}
\bibitem{Aaboud:2018mst} ATLAS Collaboration,
% M.~Aaboud \textit{et al.} [ATLAS],
%``Study of the rare decays of $B^0_s$ and $B^0$ mesons into muon pairs using data collected during 2015 and 2016 with the ATLAS detector,''
{\it JHEP} \textbf{04} (2019) 098.
%doi:10.1007/JHEP04(2019)098
%[arXiv:1812.03017 [hep-ex]].
%51 citations counted in INSPIRE as of 19 Apr 2020




%\cite{DeBruyn:2012wk}
\bibitem{DeBruyn:2012wk} K.~De Bruyn {\it et al.},
 % K.~De Bruyn, R.~Fleischer, R.~Knegjens, P.~Koppenburg, M.~Merk, A.~Pellegrino and N.~Tuning,
  %``Probing New Physics via the $B^0_s\to \mu^+\mu^-$ Effective Lifetime,''
  {\it Phys. Rev. Lett.} {\bf 109} (2012) 041801.
 % doi:10.1103/PhysRevLett.109.041801
 % [arXiv:1204.1737 [hep-ph]].
  %%CITATION = doi:10.1103/PhysRevLett.109.041801;%%
  %248 citations counted in INSPIRE as of 04 May 2018

%\cite{Bobeth:2013uxa}
\bibitem{Bobeth:2013uxa} C.~Bobeth {\it et al.},
  % C.~Bobeth, M.~Gorbahn, T.~Hermann, M.~Misiak, E.~Stamou and M.~Steinhauser,
  %``$B_{s,d} \to l^+ l^-$ in the Standard Model with Reduced Theoretical Uncertainty,''
  {\it Phys. Rev. Lett.}  {\bf 112} (2014) 101801.
 % doi:10.1103/PhysRevLett.112.101801
 % [arXiv:1311.0903 [hep-ph]].
  %%CITATION = doi:10.1103/PhysRevLett.112.101801;%%
  %297 citations counted in INSPIRE as of 04 May 2018

%%%%%%%%%%%%%%%%%%%%%%%%%%%%%%%%%%%%%%%%%%%%%%%%%%%%%%

%\cite{Weinberg:1979sa}
\bibitem{Weinberg:1979sa}
  S.~Weinberg,
  %``Baryon and Lepton Nonconserving Processes,''
  {\it Phys. Rev. Lett.} {\bf 43} (1979) 1566.
 % doi:10.1103/PhysRevLett.43.1566
  %%CITATION = doi:10.1103/PhysRevLett.43.1566;%%
  %1404 citations counted in INSPIRE as of 05 May 2018

%\cite{Isidori:2010kg}
\bibitem{Isidori:2010kg}
  G.~Isidori, Y.~Nir and G.~Perez,
  %``Flavor Physics Constraints for Physics Beyond the Standard Model,''
  {\it Ann. Rev. Nucl. Part. Sci.}  {\bf 60} (2010) 355.
 % doi:10.1146/annurev.nucl.012809.104534
 % [arXiv:1002.0900 [hep-ph]].
  %%CITATION = doi:10.1146/annurev.nucl.012809.104534;%%
  %318 citations counted in INSPIRE as of 05 May 2018


%%\cite{Isidori:2014rba}
%\bibitem{Isidori:2014rba}
%  G.~Isidori and F.~Teubert,
%  %``Status of indirect searches for New Physics with heavy flavour decays after the initial LHC run,''
%  {\it Eur. Phys. J. Plus} {\bf 129} (2014) 40.
% % doi:10.1140/epjp/i2014-14040-4
% % [arXiv:1402.2844 [hep-ph]].
%  %%CITATION = doi:10.1140/epjp/i2014-14040-4;%%
%  %16 citations counted in INSPIRE as of 05 May 2018

%\cite{Hall:1990ac}
\bibitem{Hall:1990ac}
  L.~J.~Hall and L.~Randall,
  %``Weak scale effective supersymmetry,''
  {\it Phys. Rev. Lett.} {\bf 65} (1990) 2939.
  % doi:10.1103/PhysRevLett.65.2939
  %%CITATION = doi:10.1103/PhysRevLett.65.2939;%%
  %430 citations counted in INSPIRE as of 05 May 2018

%\cite{Chivukula:1987py}
\bibitem{Chivukula:1987py}
  R.~S.~Chivukula and H.~Georgi,
  %``Composite Technicolor Standard Model,''
  {\it Phys. Lett.} {\bf B188} (1987) 99.
 % doi:10.1016/0370-2693(87)90713-1
  %%CITATION = doi:10.1016/0370-2693(87)90713-1;%%
  %541 citations counted in INSPIRE as of 05 May 2018

%\cite{DAmbrosio:2002vsn}
\bibitem{DAmbrosio:2002vsn}
  G.~D'Ambrosio, G.~F.~Giudice, G.~Isidori and A.~Strumia,
  %``Minimal flavor violation: An Effective field theory approach,''
  {\it Nucl. Phys.} {\bf B645} (2002) 155.
 % doi:10.1016/S0550-3213(02)00836-2
 % [hep-ph/0207036].
  %%CITATION = doi:10.1016/S0550-3213(02)00836-2;%%
  %1306 citations counted in INSPIRE as of 05 May 2018
  
%\cite{Pich:2009sp}
\bibitem{Pich:2009sp}
  A.~Pich and P.~Tuz\'on,
  %``Yukawa Alignment in the Two-Higgs-Doublet Model,''
  {\it Phys. Rev.} {\bf D80} (2009) 091702.
 % doi:10.1103/PhysRevD.80.091702
 % [arXiv:0908.1554 [hep-ph]].
  %%CITATION = doi:10.1103/PhysRevD.80.091702;%%
  %236 citations counted in INSPIRE as of 05 May 2018

%\cite{Penuelas:2017ikk}
\bibitem{Penuelas:2017ikk}
  A.~Pe\~nuelas and A.~Pich,
  %``Flavour alignment in multi-Higgs-doublet models,''
  {\it JHEP} {\bf 84} (2017).
 % doi:10.1007/JHEP12(2017)084
 % [arXiv:1710.02040 [hep-ph]].
  %%CITATION = doi:10.1007/JHEP12(2017)084;%%
  %2 citations counted in INSPIRE as of 05 May 2018

%\cite{Jung:2010ik}
\bibitem{Jung:2010ik}
  M.~Jung, A.~Pich and P.~Tuz\'on,
  %``Charged-Higgs phenomenology in the Aligned two-Higgs-doublet model,''
  {\it JHEP} {\bf 1011} (2010) 003;
 % doi:10.1007/JHEP11(2010)003
 % [arXiv:1006.0470 [hep-ph]].
  %%CITATION = doi:10.1007/JHEP11(2010)003;%%
  %130 citations counted in INSPIRE as of 05 May 2018
%
%\cite{Jung:2010ab}
%\bibitem{Jung:2010ab}
 % M.~Jung, A.~Pich and P.~Tuzon,
  %``The B -> Xs gamma Rate and CP Asymmetry within the Aligned Two-Higgs-Doublet Model,''
  {\it Phys.\ Rev.} {\bf D83} (2011) 074011.
%  doi:10.1103/PhysRevD.83.074011
%  [arXiv:1011.5154 [hep-ph]].
  %%CITATION = doi:10.1103/PhysRevD.83.074011;%%
  %33 citations counted in INSPIRE as of 05 May 2018


%\cite{Jung:2012vu}
\bibitem{Jung:2012vu}
  M.~Jung, X.~Q.~Li and A.~Pich,
  %``Exclusive radiative B-meson decays within the aligned two-Higgs-doublet model,''
  {\it JHEP} {\bf 1210} (2012) 063.
 % doi:10.1007/JHEP10(2012)063
 % [arXiv:1208.1251 [hep-ph]].
  %%CITATION = doi:10.1007/JHEP10(2012)063;%%
  %37 citations counted in INSPIRE as of 05 May 2018

%\cite{Jung:2013hka}
\bibitem{Jung:2013hka}
  M.~Jung and A.~Pich,
  %``Electric Dipole Moments in Two-Higgs-Doublet Models,''
  {\it JHEP} {\bf 1404} (2014) 076.
 % doi:10.1007/JHEP04(2014)076
 % [arXiv:1308.6283 [hep-ph]].
  %%CITATION = doi:10.1007/JHEP04(2014)076;%%
  %82 citations counted in INSPIRE as of 06 May 2018

%\cite{Li:2014fea}
\bibitem{Li:2014fea}
  X.~Q.~Li, J.~Lu and A.~Pich,
  %``$B_{s,d}^0 \to \ell^+\ell^-$ Decays in the Aligned Two-Higgs-Doublet Model,''
  {\it JHEP} {\bf 1406} (2014) 022.
 % doi:10.1007/JHEP06(2014)022
 % [arXiv:1404.5865 [hep-ph]].
  %%CITATION = doi:10.1007/JHEP06(2014)022;%%
  %44 citations counted in INSPIRE as of 06 May 2018
  
%\cite{Ilisie:2015tra}
\bibitem{Ilisie:2015tra}
  V.~Ilisie,
  %``New Barr-Zee contributions to $\mathbf{(g-2)_\mu}$ in two-Higgs-doublet models,''
  {\it JHEP} {\bf 1504} (2015) 077.
 % doi:10.1007/JHEP04(2015)077
 % [arXiv:1502.04199 [hep-ph]].
  %%CITATION = doi:10.1007/JHEP04(2015)077;%%
  %28 citations counted in INSPIRE as of 17 May 2018

%\cite{Cherchiglia:2016eui}
\bibitem{Cherchiglia:2016eui}
  A.~Cherchiglia, P.~Kneschke, D.~St\"ockinger and H.~St\"ockinger-Kim,
  %``The muon magnetic moment in the 2HDM: complete two-loop result,''
 {\it JHEP} {\bf 1701} (2017) 007.
%  doi:10.1007/JHEP01(2017)007
%  [arXiv:1607.06292 [hep-ph]].
  %%CITATION = doi:10.1007/JHEP01(2017)007;%%
  %13 citations counted in INSPIRE as of 17 May 2018

%\cite{Chang:2015rva}
\bibitem{Chang:2015rva}
  Q.~Chang, P.~F.~Li and X.~Q.~Li,
  %``${B_s^0}$ – ${\bar{B}}_s^0$ mixing within minimal flavor-violating two-Higgs-doublet models,''
  {\it Eur. Phys. J.} {\bf C75} (2015) 594.
 % doi:10.1140/epjc/s10052-015-3813-y
 % [arXiv:1505.03650 [hep-ph]].
  %%CITATION = doi:10.1140/epjc/s10052-015-3813-y;%%
  %6 citations counted in INSPIRE as of 17 May 2018

%\cite{Hu:2016gpe}
\bibitem{Hu:2016gpe}
  Q.~Y.~Hu, X.~Q.~Li and Y.~D.~Yang,
  %``$B^0\to K^{\ast 0}\mu^+\mu^-$ decay in the Aligned Two-Higgs-Doublet Model,''
  {\it Eur. Phys. J.} {\bf C77} (2017) 190;
%  doi:10.1140/epjc/s10052-017-4748-2
%  [arXiv:1612.08867 [hep-ph]].
  %%CITATION = doi:10.1140/epjc/s10052-017-4748-2;%%
  %11 citations counted in INSPIRE as of 17 May 2018
%
%%\cite{Hu:2017qxj}
%\bibitem{Hu:2017qxj}
%  Q.~Y.~Hu, X.~Q.~Li and Y.~D.~Yang,
%  %``The $\varvec{\Lambda _b\rightarrow \Lambda (\rightarrow p\pi ^-)\mu ^+\mu ^-}$ decay in the aligned two-Higgs-doublet model,''
%  Eur.\ Phys.\ J.\ C 
{\bf C77} (2017) 228.
%  doi:10.1140/epjc/s10052-017-4794-9
%  [arXiv:1701.04029 [hep-ph]].
%  %%CITATION = doi:10.1140/epjc/s10052-017-4794-9;%%
%  %4 citations counted in INSPIRE as of 17 May 2018

%\cite{Cho:2017jym}
\bibitem{Cho:2017jym}
  N.~Cho, X.~Q.~Li, F.~Su and X.~Zhang,
  %``$K^0-\overline{K}^0$ mixing in the minimal flavor-violating two-Higgs-doublet models,''
  {\it Adv. High Energy Phys.} {\bf 2017} (2017) 2863647.
 % doi:10.1155/2017/2863647
 % [arXiv:1705.07638 [hep-ph]].
  %%CITATION = doi:10.1155/2017/2863647;%%
  %2 citations counted in INSPIRE as of 17 May 2018

%\cite{Han:2015yys}
\bibitem{Han:2015yys}
  T.~Han, S.~K.~Kang and J.~Sayre,
  %``Muon $g-2$ in the aligned two Higgs doublet model,''
  {\it JHEP} {\bf 1602} (2016) 097.
%  doi:10.1007/JHEP02(2016)097
%  [arXiv:1511.05162 [hep-ph]].
  %%CITATION = doi:10.1007/JHEP02(2016)097;%%
  %13 citations counted in INSPIRE as of 17 May 2018

%\cite{Enomoto:2015wbn}
\bibitem{Enomoto:2015wbn}
  T.~Enomoto and R.~Watanabe,
  %``Flavor constraints on the Two Higgs Doublet Models of Z$_{2}$ symmetric and aligned types,''
  {\it JHEP} {\bf 1605} (2016) 002.
%  doi:10.1007/JHEP05(2016)002
%  [arXiv:1511.05066 [hep-ph]].
  %%CITATION = doi:10.1007/JHEP05(2016)002;%%
  %39 citations counted in INSPIRE as of 17 May 2018
  
%%% 


%\cite{Celis:2013rcs}
\bibitem{Celis:2013rcs}
  A.~Celis, V.~Ilisie and A.~Pich,
  %``LHC constraints on two-Higgs doublet models,''
  {\it JHEP} {\bf 1307} (2013) 053;
 % doi:10.1007/JHEP07(2013)053
 % [arXiv:1302.4022 [hep-ph]].
  %%CITATION = doi:10.1007/JHEP07(2013)053;%%
  %134 citations counted in INSPIRE as of 05 May 2018  
%
%\cite{Celis:2013ixa}
%\bibitem{Celis:2013ixa}
 % A.~Celis, V.~Ilisie and A.~Pich,
  %``Towards a general analysis of LHC data within two-Higgs-doublet models,''
 % JHEP 
 {\bf 1312} (2013) 095.
 % doi:10.1007/JHEP12(2013)095
 % [arXiv:1310.7941 [hep-ph]].
  %%CITATION = doi:10.1007/JHEP12(2013)095;%%
  %75 citations counted in INSPIRE as of 05 May 2018

%\cite{Abbas:2015cua}
\bibitem{Abbas:2015cua}
  G.~Abbas, A.~Celis, X.~Q.~Li, J.~Lu and A.~Pich,
  %``Flavour-changing top decays in the aligned two-Higgs-doublet model,''
  {\it JHEP} {\bf 1506} (2015) 005.
 % doi:10.1007/JHEP06(2015)005
 % [arXiv:1503.06423 [hep-ph]].
  %%CITATION = doi:10.1007/JHEP06(2015)005;%%
  %19 citations counted in INSPIRE as of 05 May 2018

%\cite{Ilisie:2014hea}
\bibitem{Ilisie:2014hea}
  V.~Ilisie and A.~Pich,
  %``Low-mass fermiophobic charged Higgs phenomenology in two-Higgs-doublet models,''
  {\it JHEP} {\bf 1409} (2014) 089.
 % doi:10.1007/JHEP09(2014)089
 % [arXiv:1405.6639 [hep-ph]].
  %%CITATION = doi:10.1007/JHEP09(2014)089;%%
  %12 citations counted in INSPIRE as of 17 May 2018

%\cite{Altmannshofer:2012ar}
\bibitem{Altmannshofer:2012ar}
  W.~Altmannshofer, S.~Gori and G.~D.~Kribs,
  %``A Minimal Flavor Violating 2HDM at the LHC,''
  {\it Phys. Rev.} {\bf D86} (2012) 115009.
 % doi:10.1103/PhysRevD.86.115009
 % [arXiv:1210.2465 [hep-ph]].
  %%CITATION = doi:10.1103/PhysRevD.86.115009;%%
  %105 citations counted in INSPIRE as of 17 May 2018
  
%\cite{Bai:2012ex}
\bibitem{Bai:2012ex}
  Y.~Bai, V.~Barger, L.~L.~Everett and G.~Shaughnessy,
  %``General two Higgs doublet model (2HDM-G) and Large Hadron Collider data,''
  {\it Phys. Rev.} {\bf D87} (2013) 115013.
%  doi:10.1103/PhysRevD.87.115013
%  [arXiv:1210.4922 [hep-ph]].
  %%CITATION = doi:10.1103/PhysRevD.87.115013;%%
  %60 citations counted in INSPIRE as of 17 May 2018
  
%\cite{Duarte:2013zfa}
\bibitem{Duarte:2013zfa}
  L.~Duarte, G.~A.~Gonz\'alez-Sprinberg and J.~Vidal,
  %``Top quark anomalous tensor couplings in the two-Higgs-doublet models,''
  {\it JHEP} {\bf 1311} (2013) 114.
 % doi:10.1007/JHEP11(2013)114
 % [arXiv:1308.3652 [hep-ph]].
  %%CITATION = doi:10.1007/JHEP11(2013)114;%%
  %14 citations counted in INSPIRE as of 17 May 2018

%\cite{Ayala:2016djv}
\bibitem{Ayala:2016djv}
  C.~Ayala, G.~A.~Gonz\'alez-Sprinberg, R.~Martinez and J.~Vidal,
  %``The top right coupling in the aligned two-Higgs-doublet model,''
  {\it JHEP} {\bf 1703} (2017) 128.
%  doi:10.1007/JHEP03(2017)128
%  [arXiv:1611.07756 [hep-ph]].
  %%CITATION = doi:10.1007/JHEP03(2017)128;%%
  %1 citations counted in INSPIRE as of 17 May 2018

%\cite{Wang:2013sha}
\bibitem{Wang:2013sha}
  L.~Wang and X.~F.~Han,
  %``Status of the aligned two-Higgs-doublet model confronted with the Higgs data,''
  {\it JHEP} {\bf 1404} (2014) 128.
 % doi:10.1007/JHEP04(2014)128
 % [arXiv:1312.4759 [hep-ph]].
  %%CITATION = doi:10.1007/JHEP04(2014)128;%%
  %34 citations counted in INSPIRE as of 17 May 2018
    
%\cite{Pich:2019pzg}
\bibitem{Pich:2019pzg}
A.~Pich,
%``Flavour Anomalies,''
{\it PoS} \textbf{LHCP2019} (2019) 078.
%doi:10.22323/1.350.0078
%[arXiv:1911.06211 [hep-ph]].
%3 citations counted in INSPIRE as of 23 Apr 2020
  
%%%%%%%%%%%%%%%%%%%%%%%%%%%%%%%%%%%%%%%%%%%%%%%%%%%%%%

%\cite{Filipuzzi:2012mg}
\bibitem{Filipuzzi:2012mg}
  A.~Filipuzzi, J.~Portol\'es and M.~Gonz\'alez-Alonso,
  %``U(2)$^5$ flavor symmetry and lepton universality violation in $W \to \tau \nu_\tau$,''
  {\it Phys. Rev.} {\bf D85} (2012) 116010.
 % doi:10.1103/PhysRevD.85.116010
 % [arXiv:1203.2092 [hep-ph]].
  %%CITATION = doi:10.1103/PhysRevD.85.116010;%%
  %18 citations counted in INSPIRE as of 08 May 2018

%\cite{BABAR:2011aa}
\bibitem{BABAR:2011aa} BaBar Collaboration,
 % J.~P.~Lees {\it et al.} [BaBar Collaboration],
  %``Search for CP Violation in the Decay $\tau^- -> \pi^- K^0_S (>= 0 \pi^0) \nu_tau$,''
  {\it Phys. Rev.} {\bf D85} (2012) 031102
   [Err: {\bf D85} (2012) 099904].
%  doi:10.1103/PhysRevD.85.099904, 10.1103/PhysRevD.85.031102
%  [arXiv:1109.1527 [hep-ex]].
  %%CITATION = doi:10.1103/PhysRevD.85.099904, 10.1103/PhysRevD.85.031102;%%
  %48 citations counted in INSPIRE as of 08 May 2018

%\cite{Bigi:2005ts}
\bibitem{Bigi:2005ts}
  I.~I.~Bigi and A.~I.~Sanda,
  %``A 'Known' CP asymmetry in tau decays,''
  {\it Phys. Lett.} {\bf B625} (2005) 47.
 % doi:10.1016/j.physletb.2005.08.033
 % [hep-ph/0506037].
  %%CITATION = doi:10.1016/j.physletb.2005.08.033;%%
  %67 citations counted in INSPIRE as of 08 May 2018
  
%\cite{Grossman:2011zk}
\bibitem{Grossman:2011zk}
  Y.~Grossman and Y.~Nir,
  %``CP Violation in \tau ->\nu\pi K_S and D->\pi K_S: The Importance of K_S-K_L Interference,''
  {\it JHEP} {\bf 1204} (2012) 002.
 % doi:10.1007/JHEP04(2012)002
 % [arXiv:1110.3790 [hep-ph]].
  %%CITATION = doi:10.1007/JHEP04(2012)002;%%
  %63 citations counted in INSPIRE as of 08 May 2018  
  
  
  
 
%\cite{Cirigliano:2017tqn}
\bibitem{Cirigliano:2017tqn}
  V.~Cirigliano, A.~Crivellin and M.~Hoferichter,
  %``No-go theorem for nonstandard explanations of the $\tau\to K_S\pi\nu_\tau$ CP asymmetry,''
  {\it Phys. Rev. Lett.} {\bf 120} (2018) 141803.
 % doi:10.1103/PhysRevLett.120.141803
 % [arXiv:1712.06595 [hep-ph]].
  %%CITATION = doi:10.1103/PhysRevLett.120.141803;%%
  %1 citations counted in INSPIRE as of 08 May 2018

%\cite{Ikado:2006un}
\bibitem{Ikado:2006un} Belle Collaboration,
 % K.~Ikado {\it et al.} [Belle Collaboration],
  %``Evidence of the Purely Leptonic Decay B- ---> tau- anti-nu(tau),''
  {\it Phys. Rev. Lett.}  {\bf 97} (2006) 251802;
 % doi:10.1103/PhysRevLett.97.251802
 % [hep-ex/0604018].
  %%CITATION = doi:10.1103/PhysRevLett.97.251802;%%
  %345 citations counted in INSPIRE as of 08 May 2018
%
%\cite{Hara:2010dk}
%\bibitem{Hara:2010dk}
%  K.~Hara {\it et al.} [Belle Collaboration],
%  %``Evidence for $B^- -> \tau^- \bar{\nu}$ with a Semileptonic Tagging Method,''
  {\it Phys. Rev.} {\bf D82} (2010) 071101.
%  doi:10.1103/PhysRevD.82.071101
%  [arXiv:1006.4201 [hep-ex]].
%  %%CITATION = doi:10.1103/PhysRevD.82.071101;%%
%  %137 citations counted in INSPIRE as of 08 May 2018  
%  






%%%%%%%%%%%%%%%%%%%%%%%%%%%%%%%%%%%%%%%%%%%%%%%%%%%  

%\cite{Lees:2012xj}
\bibitem{Lees:2012xj} BaBar Collaboration,
 % J.~P.~Lees {\it et al.} [BaBar Collaboration],
  %``Evidence for an excess of $\bar{B} \to D^{(*)} \tau^-\bar{\nu}_\tau$ decays,''
 {\it Phys. Rev. Lett.} {\bf 109} (2012) 101802;
 % doi:10.1103/PhysRevLett.109.101802
 % [arXiv:1205.5442 [hep-ex]].
  %%CITATION = doi:10.1103/PhysRevLett.109.101802;%%
  %524 citations counted in INSPIRE as of 06 May 2018
%
%\cite{Lees:2013uzd}
%\bibitem{Lees:2013uzd}
%J.~Lees \textit{et al.} [BaBar],
%``Measurement of an Excess of $\bar{B} \to D^{(*)}\tau^- \bar{\nu}_\tau$ Decays and Implications for Charged Higgs Bosons,''
{\it Phys. Rev.} \textbf{D88} (2013) 072012.
%doi:10.1103/PhysRevD.88.072012
%[arXiv:1303.0571 [hep-ex]].
%651 citations counted in INSPIRE as of 19 Apr 2020

%\cite{Fajfer:2012vx}
\bibitem{Fajfer:2012vx}
  S.~Fajfer, J.~F.~Kamenik and I.~Nisandzic,
  %``On the $B \to D^* \tau \bar \nu_{\tau}$ Sensitivity to New Physics,''
  {\it Phys\ Rev.} {\bf D85} (2012) 094025.
 % doi:10.1103/PhysRevD.85.094025
 % [arXiv:1203.2654 [hep-ph]].
  %%CITATION = doi:10.1103/PhysRevD.85.094025;%%
  %348 citations counted in INSPIRE as of 06 May 2018  

%\cite{Aaij:2015yra}
\bibitem{Aaij:2015yra} LHCb Collaboration,
 % R.~Aaij {\it et al.} [LHCb Collaboration],
  %``Measurement of the ratio of branching fractions $\mathcal{B}(\bar{B}^0 \to D^{*+}\tau^{-}\bar{\nu}_{\tau})/\mathcal{B}(\bar{B}^0 \to D^{*+}\mu^{-}\bar{\nu}_{\mu})$,''
 {\it Phys. Rev. Lett.} {\bf 115} (2015) 111803
   [Err: {\bf 115} (2015) 159901];
%  doi:10.1103/PhysRevLett.115.159901, 10.1103/PhysRevLett.115.111803
%  [arXiv:1506.08614 [hep-ex]].
  %%CITATION = doi:10.1103/PhysRevLett.115.159901, 10.1103/PhysRevLett.115.111803;%%
  %383 citations counted in INSPIRE as of 06 May 2018
%
%\cite{Aaij:2017uff}
%\bibitem{Aaij:2017uff}
%  R.~Aaij {\it et al.} [LHCb Collaboration],
  %``Measurement of the ratio of the $B^0 \to D^{*-} \tau^+ \nu_{\tau}$ and $B^0 \to D^{*-} \mu^+ \nu_{\mu}$ branching fractions using three-prong $\tau$-lepton decays,''
 % Phys.\ Rev.\ Lett.\  
 {\bf 120} (2018) 171802;
 % doi:10.1103/PhysRevLett.120.171802
 % [arXiv:1708.08856 [hep-ex]].
  %%CITATION = doi:10.1103/PhysRevLett.120.171802;%%
  %31 citations counted in INSPIRE as of 06 May 2018
%
%\cite{Aaij:2017deq}
%\bibitem{Aaij:2017deq}
 % R.~Aaij {\it et al.} [LHCb Collaboration],
  %``Test of Lepton Flavor Universality by the measurement of the $B^0 \to D^{*-} \tau^+ \nu_{\tau}$ branching fraction using three-prong $\tau$ decays,''
 {\it Phys. Rev.} {\bf D97} (2018) 072013.
 % doi:10.1103/PhysRevD.97.072013
 % [arXiv:1711.02505 [hep-ex]].
  %%CITATION = doi:10.1103/PhysRevD.97.072013;%%
  %14 citations counted in INSPIRE as of 06 May 2018

%\cite{Huschle:2015rga}
\bibitem{Huschle:2015rga} Belle Collaboration,
  % M.~Huschle {\it et al.} [Belle Collaboration],
  %``Measurement of the branching ratio of $\bar{B} \to D^{(\ast)} \tau^- \bar{\nu}_\tau$ relative to $\bar{B} \to D^{(\ast)} \ell^- \bar{\nu}_\ell$ decays with hadronic tagging at Belle,''
  {\it Phys. Rev.} {\bf D92} (2015) 072014;
 % doi:10.1103/PhysRevD.92.072014
 % [arXiv:1507.03233 [hep-ex]].
  %%CITATION = doi:10.1103/PhysRevD.92.072014;%%
  %314 citations counted in INSPIRE as of 06 May 2018
%
%\cite{Sato:2016svk}
%\bibitem{Sato:2016svk}
%  Y.~Sato {\it et al.} [Belle Collaboration],
  %``Measurement of the branching ratio of $\bar{B}^0 \rightarrow D^{*+} \tau^- \bar{\nu}_{\tau}$ relative to $\bar{B}^0 \rightarrow D^{*+} \ell^- \bar{\nu}_{\ell}$ decays with a semileptonic tagging method,''
 % Phys.\ Rev.\ D 
 {\bf D94} (2016)  072007;
 % doi:10.1103/PhysRevD.94.072007
 % [arXiv:1607.07923 [hep-ex]].
  %%CITATION = doi:10.1103/PhysRevD.94.072007;%%
  %106 citations counted in INSPIRE as of 06 May 2018
%
%\cite{Hirose:2017dxl}
%\bibitem{Hirose:2017dxl}
%  S.~Hirose {\it et al.} [Belle Collaboration],
  %``Measurement of the $\tau$ lepton polarization and $R(D^*)$ in the decay $\bar{B} \rightarrow D^* \tau^- \bar{\nu}_\tau$ with one-prong hadronic $\tau$ decays at Belle,''
  % Phys.\ Rev.\ D 
  {\bf D97} (2018) 012004;
%  doi:10.1103/PhysRevD.97.012004
%  [arXiv:1709.00129 [hep-ex]].
  %%CITATION = doi:10.1103/PhysRevD.97.012004;%%
  %15 citations counted in INSPIRE as of 07 May 2018
%  
%\cite{Hirose:2016wfn}
%\bibitem{Hirose:2016wfn}
%  S.~Hirose {\it et al.} [Belle Collaboration],
  %``Measurement of the $\tau$ lepton polarization and $R(D^*)$ in the decay $\bar{B} \to D^* \tau^- \bar{\nu}_\tau$,''
  {\it Phys. Rev. Lett.} {\bf 118} (2017) 211801;
%  doi:10.1103/PhysRevLett.118.211801
%  [arXiv:1612.00529 [hep-ex]].
  %%CITATION = doi:10.1103/PhysRevLett.118.211801;%%
  %114 citations counted in INSPIRE as of 06 May 2018
%
%\cite{Abdesselam:2019dgh}
%\bibitem{Abdesselam:2019dgh}
%A.~Abdesselam \textit{et al.} [Belle],
%``Measurement of $\mathcal{R}(D)$ and $\mathcal{R}(D^{\ast})$ with a semileptonic tagging method,''
arXiv:1904.08794 [hep-ex].
%92 citations counted in INSPIRE as of 19 Apr 2020

%%\cite{Na:2015kha}
%\bibitem{Na:2015kha}
%  H.~Na {\it et al.} [HPQCD Collaboration],
%  %``$B \rightarrow D l \nu$ form factors at nonzero recoil and extraction of $|V_{cb}|$,''
%  {\it Phys. Rev.} {\bf D92} (2015) 054510
%   [Err: {\bf D93} (2016) 119906].
% % doi:10.1103/PhysRevD.93.119906, 10.1103/PhysRevD.92.054510
% % [arXiv:1505.03925 [hep-lat]].
%  %%CITATION = doi:10.1103/PhysRevD.93.119906, 10.1103/PhysRevD.92.054510;%%
%  %144 citations counted in INSPIRE as of 06 May 2018
  
%%\cite{Bigi:2016mdz}
%\bibitem{Bigi:2016mdz}
%D.~Bigi and P.~Gambino,
%%``Revisiting $B\to D \ell \nu$,''
%{\it Phys. Rev.}  \textbf{D94} (2016) 094008.
%%doi:10.1103/PhysRevD.94.094008
%%[arXiv:1606.08030 [hep-ph]].
%%173 citations counted in INSPIRE as of 19 Apr 2020

%\cite{Bigi:2017jbd}
\bibitem{Bigi:2017jbd}
D.~Bigi, P.~Gambino and S.~Schacht,
%``$R(D^*)$, $|V_{cb}|$, and the Heavy Quark Symmetry relations between form factors,''
{\it JHEP} \textbf{11} (2017) 061.
%doi:10.1007/JHEP11(2017)061
%[arXiv:1707.09509 [hep-ph]].
%161 citations counted in INSPIRE as of 19 Apr 2020

%\cite{Jaiswal:2017rve}
\bibitem{Jaiswal:2017rve}
S.~Jaiswal, S.~Nandi and S.~K.~Patra,
%``Extraction of $|V_{cb}|$ from $B\to D^{(*)}\ell\nu_\ell$ and the Standard Model predictions of $R(D^{(*)})$,''
{\it JHEP} \textbf{12} (2017) 060.
%doi:10.1007/JHEP12(2017)060
%[arXiv:1707.09977 [hep-ph]].
%134 citations counted in INSPIRE as of 19 Apr 2020

%\cite{Jung:2018lfu}
\bibitem{Jung:2018lfu}
M.~Jung and D.~M.~Straub,
%``Constraining new physics in $b\to c\ell\nu$ transitions,''
{\it JHEP} \textbf{01} (2019) 009.
%doi:10.1007/JHEP01(2019)009
%[arXiv:1801.01112 [hep-ph]].
%66 citations counted in INSPIRE as of 19 Apr 2020

%\cite{Murgui:2019czp}
\bibitem{Murgui:2019czp}
C.~Murgui, A.~Peñuelas, M.~Jung and A.~Pich,
%``Global fit to $b \to c \tau \nu$ transitions,''
{\it JHEP} \textbf{09} (2019) 103.
%doi:10.1007/JHEP09(2019)103
%[arXiv:1904.09311 [hep-ph]].
%42 citations counted in INSPIRE as of 19 Apr 2020

%\cite{Mandal:2020htr}
\bibitem{Mandal:2020htr}
R.~Mandal, C.~Murgui, A.~Peñuelas and A.~Pich,
%``The role of right-handed neutrinos in $b \to c \tau \bar{\nu}$ anomalies,''
arXiv:2004.06726 [hep-ph].
%0 citations counted in INSPIRE as of 19 Apr 2020


%\cite{Celis:2016azn}
\bibitem{Celis:2016azn}
A.~Celis, M.~Jung, X.~Li and A.~Pich,
%``Scalar contributions to $b\to c (u) \tau \nu$ transitions,''
{\it Phys. Lett.}  \textbf{B771} (2017) 168;
%doi:10.1016/j.physletb.2017.05.037
%[arXiv:1612.07757 [hep-ph]].
%108 citations counted in INSPIRE as of 19 Apr 2020
%
%\bibitem{Celis:2012dk}
%  A.~Celis, M.~Jung, X.~Q.~Li and A.~Pich,
  %``Sensitivity to charged scalars in $\boldsymbol{B\to D^{(*)}\tau\nu_\tau}$ and $\boldsymbol{B\to\tau\nu_\tau}$ decays,''
  {\it JHEP} {\bf 1301} (2013) 054.
 % doi:10.1007/JHEP01(2013)054
 % [arXiv:1210.8443 [hep-ph]].
  %%CITATION = doi:10.1007/JHEP01(2013)054;%%
  %170 citations counted in INSPIRE as of 06 May 2018

%\cite{Alonso:2016oyd}
\bibitem{Alonso:2016oyd}
R.~Alonso, B.~Grinstein and J.~Martin Camalich,
%``Lifetime of $B_c^-$ Constrains Explanations for Anomalies in  $B\to D^{(*)}\tau\nu$,''
{\it Phys. Rev. Lett.} \textbf{118} (2017) 081802.
%doi:10.1103/PhysRevLett.118.081802
%[arXiv:1611.06676 [hep-ph]].
%147 citations counted in INSPIRE as of 19 Apr 2020

%\cite{Akeroyd:2017mhr}
\bibitem{Akeroyd:2017mhr}
A.~Akeroyd and C.~Chen,
%``Constraint on the branching ratio of $B_c \to \tau \bar{\nu}$ from LEP1 and consequences for $R(D^{(*)})$ anomaly,''
{\it Phys. Rev.}  \textbf{D96} (2017) 075011.
%doi:10.1103/PhysRevD.96.075011
%[arXiv:1708.04072 [hep-ph]].
%86 citations counted in INSPIRE as of 19 Apr 2020


%\cite{Aaij:2017tyk}
\bibitem{Aaij:2017tyk} LHCb Collaboration,
 % R.~Aaij {\it et al.} [LHCb Collaboration],
  %``Measurement of the ratio of branching fractions $\mathcal{B}(B_c^+\,\to\,J/\psi\tau^+\nu_\tau)$/$\mathcal{B}(B_c^+\,\to\,J/\psi\mu^+\nu_\mu)$,''
  {\it Phys. Rev. Lett.} {\bf 120} (2018) 121801.
 % doi:10.1103/PhysRevLett.120.121801
 % [arXiv:1711.05623 [hep-ex]].
  %%CITATION = doi:10.1103/PhysRevLett.120.121801;%%
  %33 citations counted in INSPIRE as of 06 May 2018

%\cite{Anisimov:1998uk}
\bibitem{Anisimov:1998uk}
  A.~Y.~Anisimov, I.~M.~Narodetsky, C.~Semay and B.~Silvestre-Brac,
  %``The $B_c$ meson lifetime in the light front constituent quark model,''
  {\it Phys. Lett.} {\bf B452} (1999) 129.
 % doi:10.1016/S0370-2693(99)00273-7
 % [hep-ph/9812514].
  %%CITATION = doi:10.1016/S0370-2693(99)00273-7;%%
  %64 citations counted in INSPIRE as of 06 May 2018

%\cite{Kiselev:2002vz}
\bibitem{Kiselev:2002vz}
  V.~V.~Kiselev,
  %``Exclusive decays and lifetime of $B_c$ meson in QCD sum rules,''
  hep-ph/0211021.
  %%CITATION = HEP-PH/0211021;%%
  %88 citations counted in INSPIRE as of 06 May 2018

%\cite{Ivanov:2006ni}
\bibitem{Ivanov:2006ni}
  M.~A.~Ivanov, J.~G.~Korner and P.~Santorelli,
  %``Exclusive semileptonic and nonleptonic decays of the $B_c$ meson,''
  {\it Phys. Rev.} {\bf D73} (2006) 054024.
%  doi:10.1103/PhysRevD.73.054024
%  [hep-ph/0602050].
  %%CITATION = doi:10.1103/PhysRevD.73.054024;%%
  %122 citations counted in INSPIRE as of 06 May 2018

%\cite{Hernandez:2006gt}
\bibitem{Hernandez:2006gt}
  E.~Hernandez, J.~Nieves and J.~M.~Verde-Velasco,
  %``Study of exclusive semileptonic and non-leptonic decays of $B_c$ - in a nonrelativistic quark model,''
  {\it Phys. Rev.} {\bf D74} (2006) 074008.
 % doi:10.1103/PhysRevD.74.074008
 % [hep-ph/0607150].
  %%CITATION = doi:10.1103/PhysRevD.74.074008;%%
  %72 citations counted in INSPIRE as of 06 May 2018

%\cite{Alonso:2015sja}
\bibitem{Alonso:2015sja}
  R.~Alonso, B.~Grinstein and J.~Martin Camalich,
  %``Lepton universality violation and lepton flavor conservation in $B$-meson decays,''
  {\it JHEP} {\bf 1510} (2015) 184.
 % doi:10.1007/JHEP10(2015)184
 % [arXiv:1505.05164 [hep-ph]].
  %%CITATION = doi:10.1007/JHEP10(2015)184;%%
  %153 citations counted in INSPIRE as of 06 May 2018

%\cite{Crivellin:2017zlb}
\bibitem{Crivellin:2017zlb}
  A.~Crivellin, D.~M\"uller and T.~Ota,
  %``Simultaneous explanation of R(D$^{(∗)}$) and b→sμ$^{+}$ μ$^{−}$: the last scalar leptoquarks standing,''
  {\it JHEP} {\bf 1709} (2017) 040.
 % doi:10.1007/JHEP09(2017)040
 % [arXiv:1703.09226 [hep-ph]].
  %%CITATION = doi:10.1007/JHEP09(2017)040;%%
  %64 citations counted in INSPIRE as of 06 May 2018

%\cite{Capdevila:2017iqn}
\bibitem{Capdevila:2017iqn} B.~Capdevila {\it et al.},
  %B.~Capdevila, A.~Crivellin, S.~Descotes-Genon, L.~Hofer and J.~Matias,
  %``Searching for New Physics with $b\to s\tau^+\tau^-$ processes,''
  {\it Phys. Rev. Lett.}  {\bf 120} (2018) 181802.
 % doi:10.1103/PhysRevLett.120.181802
 % [arXiv:1712.01919 [hep-ph]].
  %%CITATION = doi:10.1103/PhysRevLett.120.181802;%%
  %5 citations counted in INSPIRE as of 06 May 2018

%\cite{Bobeth:2011st}
\bibitem{Bobeth:2011st}
  C.~Bobeth and U.~Haisch,
  %``New Physics in $\Gamma_{12}^s$: ($\bar{s} b$)$(\bar{\tau} \tau)$ Operators,''
 {\it Acta Phys. Polon.} {\bf B44} (2013) 127.
 % doi:10.5506/APhysPolB.44.127
 % [arXiv:1109.1826 [hep-ph]].
  %%CITATION = doi:10.5506/APhysPolB.44.127;%%
  %78 citations counted in INSPIRE as of 06 May 2018


%\cite{Aaij:2014pli}
\bibitem{Aaij:2014pli} LHCb Collaboration,
  % R.~Aaij {\it et al.} [LHCb Collaboration],
  %``Differential branching fractions and isospin asymmetries of $B \to K^{(*)} \mu^+ \mu^-$ decays,''
  {\it JHEP} {\bf 1406} (2014) 133.
 % doi:10.1007/JHEP06(2014)133
 % [arXiv:1403.8044 [hep-ex]].
  %%CITATION = doi:10.1007/JHEP06(2014)133;%%
  %201 citations counted in INSPIRE as of 06 May 2018

%\cite{Aaij:2012vr}
\bibitem{Aaij:2012vr} LHCb Collaboration,
 % R.~Aaij {\it et al.} [LHCb Collaboration],
  %``Differential branching fraction and angular analysis of the $B^{+} \rightarrow K^{+}\mu^{+}\mu^{-}$ decay,''
  {\it JHEP} {\bf 1302} (2013) 105.
%  doi:10.1007/JHEP02(2013)105
%  [arXiv:1209.4284 [hep-ex]].
  %%CITATION = doi:10.1007/JHEP02(2013)105;%%
  %90 citations counted in INSPIRE as of 06 May 2018
%
%%\cite{Aaij:2016cbx}
%%\bibitem{Aaij:2016cbx}
%%  R.~Aaij {\it et al.} [LHCb Collaboration],
%%  %``Measurement of the phase difference between short- and long-distance amplitudes in the $B^{+}\to K^{+}\mu^{+}\mu^{-}$ decay,''
% {\it Eur. Phys. J.} {\bf C77} (2017) 161.
%%  doi:10.1140/epjc/s10052-017-4703-2
%%  [arXiv:1612.06764 [hep-ex]].
%  %%CITATION = doi:10.1140/epjc/s10052-017-4703-2;%%
%  %24 citations counted in INSPIRE as of 08 May 2018


%\cite{Aaij:2016flj}
\bibitem{Aaij:2016flj} LHCb Collaboration,
 % R.~Aaij {\it et al.} [LHCb Collaboration],
  %``Measurements of the S-wave fraction in $B^{0}\rightarrow K^{+}\pi^{-}\mu^{+}\mu^{-}$ decays and the $B^{0}\rightarrow K^{\ast}(892)^{0}\mu^{+}\mu^{-}$ differential branching fraction,''
  {\it JHEP} {\bf 1611} (2016) 047
   [Err: {\bf 1704} (2017) 142];
 % doi:10.1007/JHEP11(2016)047, 10.1007/JHEP04(2017)142
 % [arXiv:1606.04731 [hep-ex]].
  %%CITATION = doi:10.1007/JHEP11(2016)047, 10.1007/JHEP04(2017)142;%%
  %57 citations counted in INSPIRE as of 06 May 2018
%
%\cite{Aaij:2015oid}
%\bibitem{Aaij:2015oid}
%  R.~Aaij {\it et al.} [LHCb Collaboration],
  %``Angular analysis of the $B^{0} \to K^{*0} \mu^{+} \mu^{-}$ decay using 3 fb$^{-1}$ of integrated luminosity,''
%  JHEP 
{\bf 1602} (2016) 104;
%  doi:10.1007/JHEP02(2016)104
%  [arXiv:1512.04442 [hep-ex]].
  %%CITATION = doi:10.1007/JHEP02(2016)104;%%
  %302 citations counted in INSPIRE as of 14 May 2018
%
%\cite{Aaij:2013iag}
%\bibitem{Aaij:2013iag}
%  R.~Aaij {\it et al.} [LHCb Collaboration],
%  %``Differential branching fraction and angular analysis of the decay $B^{0} \to K^{*0} \mu^{+}\mu^{-}$,''
%  JHEP 
{\bf 1308} (2013) 131;
%  doi:10.1007/JHEP08(2013)131
%  [arXiv:1304.6325 [hep-ex]].
%  %%CITATION = doi:10.1007/JHEP08(2013)131;%%
%  %211 citations counted in INSPIRE as of 14 May 2018
%
%\cite{Aaij:2013qta}
%\bibitem{Aaij:2013qta}
%  R.~Aaij {\it et al.} [LHCb Collaboration],
  %``Measurement of Form-Factor-Independent Observables in the Decay $B^{0} \to K^{*0} \mu^+ \mu^-$,''
  {\it Phys. Rev. Lett.}  {\bf 111} (2013) 191801.
%  doi:10.1103/PhysRevLett.111.191801
%  [arXiv:1308.1707 [hep-ex]].
  %%CITATION = doi:10.1103/PhysRevLett.111.191801;%%
  %422 citations counted in INSPIRE as of 14 May 2018

%\cite{Aaij:2015esa}
\bibitem{Aaij:2015esa} LHCb Collaboration,
 % R.~Aaij {\it et al.} [LHCb Collaboration],
  %``Angular analysis and differential branching fraction of the decay $B^0_s\to\phi\mu^+\mu^-$,''
  {\it JHEP} {\bf 1509} (2015) 179;
%  doi:10.1007/JHEP09(2015)179
%  [arXiv:1506.08777 [hep-ex]].
  %%CITATION = doi:10.1007/JHEP09(2015)179;%%
  %164 citations counted in INSPIRE as of 06 May 2018
%
%\cite{Aaij:2013aln}
%\bibitem{Aaij:2013aln}
 % R.~Aaij {\it et al.} [LHCb Collaboration],
  %``Differential branching fraction and angular analysis of the decay $B_s^0\to\phi\mu^{+}\mu^{-}$,''
 % JHEP 
 {\bf 1307} (2013) 084.
%  doi:10.1007/JHEP07(2013)084
%  [arXiv:1305.2168 [hep-ex]].
  %%CITATION = doi:10.1007/JHEP07(2013)084;%%
  %119 citations counted in INSPIRE as of 06 May 2018



%\cite{Aaij:2015xza}
\bibitem{Aaij:2015xza} LHCb Collaboration,
 % R.~Aaij {\it et al.} [LHCb Collaboration],
  %``Differential branching fraction and angular analysis of $\Lambda^{0}_{b} \rightarrow \Lambda \mu^+\mu^-$ decays,''
  {\it JHEP} {\bf 1506} (2015) 115;
 % doi:10.1007/JHEP06(2015)115
 % [arXiv:1503.07138 [hep-ex]].
  %%CITATION = doi:10.1007/JHEP06(2015)115;%%
  %60 citations counted in INSPIRE as of 06 May 2018
%
%\cite{Aaij:2013mna}
%\bibitem{Aaij:2013mna}
%  R.~Aaij {\it et al.} [LHCb Collaboration],
%  %``Measurement of the differential branching fraction of the decay $\Lambda_b^0\rightarrow\Lambda\mu^+\mu^-$,''
  {\it Phys. Lett.} {\bf B725} (2013) 25.
%  doi:10.1016/j.physletb.2013.06.060
%  [arXiv:1306.2577 [hep-ex]].
  %%CITATION = doi:10.1016/j.physletb.2013.06.060;%%
  %37 citations counted in INSPIRE as of 06 May 2018

%\cite{Aaboud:2018krd}
\bibitem{Aaboud:2018krd} ATLAS Collaboration,
%M.~Aaboud \textit{et al.} [ATLAS],
%``Angular analysis of $B^0_d \rightarrow K^{*}\mu^+\mu^-$ decays in $pp$ collisions at $\sqrt{s}= 8$ TeV with the ATLAS detector,''
{\it JHEP} \textbf{10} (2018) 047.
%doi:10.1007/JHEP10(2018)047
%[arXiv:1805.04000 [hep-ex]].
%70 citations counted in INSPIRE as of 19 Apr 2020


%\cite{Lees:2015ymt}
\bibitem{Lees:2015ymt} BaBar Collaboration,
 % J.~P.~Lees {\it et al.} [BaBar Collaboration],
  %``Measurement of angular asymmetries in the decays $B \to K^*ℓ^+ℓ^-$,''
  {\it Phys. Rev.} {\bf D93} (2016) 052015.
 % doi:10.1103/PhysRevD.93.052015
 % [arXiv:1508.07960 [hep-ex]].
  %%CITATION = doi:10.1103/PhysRevD.93.052015;%%
  %28 citations counted in INSPIRE as of 14 May 2018

%\cite{Wehle:2016yoi}
\bibitem{Wehle:2016yoi} Belle Collaboration,
 % S.~Wehle {\it et al.} [Belle Collaboration],
  %``Lepton-Flavor-Dependent Angular Analysis of $B\to K^\ast \ell^+\ell^-$,''
  {\it Phys. Rev. Lett.}  {\bf 118} (2017) 111801;
%  doi:10.1103/PhysRevLett.118.111801
%  [arXiv:1612.05014 [hep-ex]].
  %%CITATION = doi:10.1103/PhysRevLett.118.111801;%%
  %104 citations counted in INSPIRE as of 07 May 2018
%
%\cite{Wei:2009zv}
%\bibitem{Wei:2009zv}
%  J.-T.~Wei {\it et al.} [Belle Collaboration],
%  %``Measurement of the Differential Branching Fraction and Forward-Backword Asymmetry for $B \to K^{(*)}\ell^+\ell^-$,''
%  Phys.\ Rev.\ Lett.\  
{\bf 103} (2009) 171801;
%  doi:10.1103/PhysRevLett.103.171801
%  [arXiv:0904.0770 [hep-ex]].
%  %%CITATION = doi:10.1103/PhysRevLett.103.171801;%%
%  %402 citations counted in INSPIRE as of 14 May 2018
%
%\cite{Abdesselam:2016llu}
%\bibitem{Abdesselam:2016llu} Belle Collaboration,
 % A.~Abdesselam {\it et al.} [Belle Collaboration],
  %``Angular analysis of $B^0 \to K^\ast(892)^0 \ell^+ \ell^-$,''
  arXiv:1604.04042 [hep-ex].
  %%CITATION = ARXIV:1604.04042;%%
  %99 citations counted in INSPIRE as of 07 May 2018

%\cite{Aaltonen:2011ja}
\bibitem{Aaltonen:2011ja} CDF Collaboration,
 % T.~Aaltonen {\it et al.} [CDF Collaboration],
  %``Measurements of the Angular Distributions in the Decays $B \to K^{(*)} \mu^+ \mu^-$ at CDF,''
  {\it Phys. Rev. Lett.}  {\bf 108} (2012) 081807.
 % doi:10.1103/PhysRevLett.108.081807
 % [arXiv:1108.0695 [hep-ex]].
  %%CITATION = doi:10.1103/PhysRevLett.108.081807;%%
  %171 citations counted in INSPIRE as of 14 May 2018

%\cite{Sirunyan:2017dhj}
\bibitem{Sirunyan:2017dhj} CMS Collaboration,
  % A.~M.~Sirunyan {\it et al.} [CMS Collaboration],
  %``Measurement of angular parameters from the decay $\mathrm{B}^0 \to \mathrm{K}^{*0} \mu^+ \mu^-$ in proton-proton collisions at $\sqrt{s} = $ 8 TeV,''
  {\it Phys. Lett.} {\bf B781} (2018) 517;
%  doi:10.1016/j.physletb.2018.04.030
%  [arXiv:1710.02846 [hep-ex]].
  %%CITATION = doi:10.1016/j.physletb.2018.04.030;%%
  %11 citations counted in INSPIRE as of 14 May 2018
%  
%\cite{Khachatryan:2015isa}
%\bibitem{Khachatryan:2015isa} CMS Collaboration,
  %V.~Khachatryan {\it et al.} [CMS Collaboration],
  %``Angular analysis of the decay $B^0 \to K^{*0} \mu^+ \mu^-$ from pp collisions at $\sqrt  s = 8$ TeV,''
  {\it Phys. Lett.} {\bf B753} (2016) 424.
  %doi:10.1016/j.physletb.2015.12.020
 % [arXiv:1507.08126 [hep-ex]].
  %%CITATION = doi:10.1016/j.physletb.2015.12.020;%%
  %62 citations counted in INSPIRE as of 14 May 2018

%\cite{DescotesGenon:2012zf}
\bibitem{DescotesGenon:2012zf}
  S.~Descotes-Genon, J.~Matias, M.~Ramon and J.~Virto,
  %``Implications from clean observables for the binned analysis of $B -> K*\mu^+\mu^-$ at large recoil,''
  {\it JHEP} {\bf 1301} (2013) 048.
 % doi:10.1007/JHEP01(2013)048
 % [arXiv:1207.2753 [hep-ph]].
  %%CITATION = doi:10.1007/JHEP01(2013)048;%%
  %182 citations counted in INSPIRE as of 07 May 2018

%\cite{Descotes-Genon:2014uoa}
\bibitem{Descotes-Genon:2014uoa}
  S.~Descotes-Genon, L.~Hofer, J.~Matias and J.~Virto,
  %``On the impact of power corrections in the prediction of $B \to K^*\mu^+\mu^-$ observables,''
  {\it JHEP} {\bf 1412} (2014) 125.
 % doi:10.1007/JHEP12(2014)125
 % [arXiv:1407.8526 [hep-ph]].
  %%CITATION = doi:10.1007/JHEP12(2014)125;%%
  %171 citations counted in INSPIRE as of 07 May 2018

%\cite{Altmannshofer:2014rta}
\bibitem{Altmannshofer:2014rta}
  W.~Altmannshofer and D.~M.~Straub,
  %``New physics in $b\rightarrow s$ transitions after LHC run 1,''
  {\it Eur. Phys. J.} {\bf C75} (2015) 382.
%  doi:10.1140/epjc/s10052-015-3602-7
%  [arXiv:1411.3161 [hep-ph]].
  %%CITATION = doi:10.1140/epjc/s10052-015-3602-7;%%
  %251 citations counted in INSPIRE as of 16 May 2018

%\cite{Straub:2015ica}
\bibitem{Straub:2015ica}
  A.~Bharucha, D.~M.~Straub and R.~Zwicky,
  %``$B\to V\ell^+\ell^-$ in the Standard Model from light-cone sum rules,''
  {\it JHEP} {\bf 1608} (2016) 098.
%  doi:10.1007/JHEP08(2016)098
%  [arXiv:1503.05534 [hep-ph]].
  %%CITATION = doi:10.1007/JHEP08(2016)098;%%
  %181 citations counted in INSPIRE as of 16 May 2018

%\cite{Dettori:2018igw}
\bibitem{Dettori:2018igw}
  F.~Dettori [LHCb Collaboration],
  %``Search for new physics in $b \to q \ell \ell$ decays,''
  arXiv:1805.05073 [hep-ex].
  %%CITATION = ARXIV:1805.05073;%%
  
%%\cite{Capdevila:2017ert}
%\bibitem{Capdevila:2017ert}
%  B.~Capdevila, S.~Descotes-Genon, L.~Hofer and J.~Matias,
%  %``Hadronic uncertainties in $B \to K^* \mu^+ \mu^-$: a state-of-the-art analysis,''
%  {\it JHEP} {\bf 1704} (2017) 016.
% % doi:10.1007/JHEP04(2017)016
% % [arXiv:1701.08672 [hep-ph]].
%  %%CITATION = doi:10.1007/JHEP04(2017)016;%%
%  %39 citations counted in INSPIRE as of 08 May 2018
%
%%\cite{Jager:2014rwa}
%\bibitem{Jager:2014rwa}
%  S.~J\"ager and J.~Martin Camalich,
%  %``Reassessing the discovery potential of the $B \to K^{*} \ell^+\ell^-$ decays in the large-recoil region: SM challenges and BSM opportunities,''
%  {\it Phys. Rev.} {\bf D93} (2016) 014028.
%%  doi:10.1103/PhysRevD.93.014028
%%  [arXiv:1412.3183 [hep-ph]].
%  %%CITATION = doi:10.1103/PhysRevD.93.014028;%%
%  %133 citations counted in INSPIRE as of 08 May 2018
%
%%\cite{Ciuchini:2015qxb}
%\bibitem{Ciuchini:2015qxb} M.~Ciuchini \etal,
% % M.~Ciuchini, M.~Fedele, E.~Franco, S.~Mishima, A.~Paul, L.~Silvestrini and M.~Valli,
%  %``$B\to K^* \ell^+ \ell^-$ decays at large recoil in the Standard Model: a theoretical reappraisal,''
%  {\it JHEP} {\bf 1606} (2016) 116;
% % doi:10.1007/JHEP06(2016)116
% % [arXiv:1512.07157 [hep-ph]].
%  %%CITATION = doi:10.1007/JHEP06(2016)116;%%
%  %114 citations counted in INSPIRE as of 08 May 2018
%%
%%\cite{Ciuchini:2016weo}
%%\bibitem{Ciuchini:2016weo}
%%  M.~Ciuchini, M.~Fedele, E.~Franco, S.~Mishima, A.~Paul, L.~Silvestrini and M.~Valli,
%%  %``$B\to K^*\ell^+\ell^-$ in the Standard Model: Elaborations and Interpretations,''
%  PoS ICHEP {\bf 2016} (2016) 584.
% % [arXiv:1611.04338 [hep-ph]].
%  %%CITATION = ARXIV:1611.04338;%%
%  %12 citations counted in INSPIRE as of 08 May 2018
%
%%\cite{Chobanova:2017ghn}
%\bibitem{Chobanova:2017ghn}  V.~G.~Chobanova   \etal,
%%  V.~G.~Chobanova, T.~Hurth, F.~Mahmoudi, D.~Martinez Santos and S.~Neshatpour,
%  %``Large hadronic power corrections or new physics in the rare decay $B→K^{∗}\mu^{+}\mu^{−}$?,''
%  {\it JHEP} {\bf 1707} (2017) 025.
%%  doi:10.1007/JHEP07(2017)025
%%  [arXiv:1702.02234 [hep-ph]].
%  %%CITATION = doi:10.1007/JHEP07(2017)025;%%
%  %26 citations counted in INSPIRE as of 06 May 2018  
%
%%\cite{Bobeth:2017vxj}
%\bibitem{Bobeth:2017vxj}
%  C.~Bobeth, M.~Chrzaszcz, D.~van Dyk and J.~Virto,
%  %``Long-distance effects in $B\to K^*\ell\ell$ from Analyticity,''
%  arXiv:1707.07305 [hep-ph].
%  %%CITATION = ARXIV:1707.07305;%%
%  %12 citations counted in INSPIRE as of 08 May 2018
%
%
%%\cite{Blake:2017fyh}
%\bibitem{Blake:2017fyh}
%  T.~Blake, U.~Egede, P.~Owen, G.~Pomery and K.~A.~Petridis,
%  %``An empirical model of the long-distance contributions to $\bar{B}^{0} \rightarrow \bar{K}^{*0}\mu^{+}\mu^{-}$ transitions,''
%  arXiv:1709.03921 [hep-ph].
%  %%CITATION = ARXIV:1709.03921;%%
%  %4 citations counted in INSPIRE as of 08 May 2018

%\cite{Aaij:2017vbb}
\bibitem{Aaij:2017vbb} LHCb Collaboration,
 % R.~Aaij {\it et al.} [LHCb Collaboration],
  %``Test of lepton universality with $B^{0} \rightarrow K^{*0}\ell^{+}\ell^{-}$ decays,''
  {\it JHEP} {\bf 1708} (2017) 055.
 % doi:10.1007/JHEP08(2017)055
 % [arXiv:1705.05802 [hep-ex]].
  %%CITATION = doi:10.1007/JHEP08(2017)055;%%
  %168 citations counted in INSPIRE as of 07 May 2018

%\cite{Aaij:2019wad}
\bibitem{Aaij:2019wad} LHCb Collaboration,
% R.~Aaij \textit{et al.} [LHCb],
%``Search for lepton-universality violation in $B^+\to K^+\ell^+\ell^-$ decays,''
{\it Phys. Rev. Lett.} \textbf{122} (2019) 191801.
%doi:10.1103/PhysRevLett.122.191801
%[arXiv:1903.09252 [hep-ex]].
%151 citations counted in INSPIRE as of 19 Apr 2020


%%\cite{Aaij:2014ora}
%\bibitem{Aaij:2014ora} LHCb Collaboration,
%  %R.~Aaij {\it et al.} [LHCb Collaboration],
%  %``Test of lepton universality using $B^{+}\rightarrow K^{+}\ell^{+}\ell^{-}$ decays,''
%  {\it Phys. Rev. Lett.} {\bf 113} (2014) 151601.
% % doi:10.1103/PhysRevLett.113.151601
% % [arXiv:1406.6482 [hep-ex]].
%  %%CITATION = doi:10.1103/PhysRevLett.113.151601;%%
%  %559 citations counted in INSPIRE as of 07 May 2018




%\cite{Bobeth:2011nj}
\bibitem{Bobeth:2011nj}
  C.~Bobeth, G.~Hiller, D.~van Dyk and C.~Wacker,
  %``The Decay $B \to K \ell^+ \ell^-$ at Low Hadronic Recoil and Model-Independent $\Delta B = 1$ Constraints,''
  {\it JHEP} {\bf 1201} (2012) 107.
%  doi:10.1007/JHEP01(2012)107
%  [arXiv:1111.2558 [hep-ph]].
  %%CITATION = doi:10.1007/JHEP01(2012)107;%%
  %137 citations counted in INSPIRE as of 08 May 2018

%\cite{Hiller:2003js}
\bibitem{Hiller:2003js}
  G.~Hiller and F.~Kruger,
  %``More model-independent analysis of $b \to s$ processes,''
  {\it Phys. Rev.} {\bf D69} (2004) 074020.
%  doi:10.1103/PhysRevD.69.074020
%  [hep-ph/0310219].
  %%CITATION = doi:10.1103/PhysRevD.69.074020;%%
  %287 citations counted in INSPIRE as of 08 May 2018

%\cite{Bobeth:2007dw}
\bibitem{Bobeth:2007dw}
  C.~Bobeth, G.~Hiller and G.~Piranishvili,
  %``Angular distributions of $\bar{B} \to \bar{K} \ell^+\ell^-$ decays,''
  {\it JHEP} {\bf 0712} (2007) 040.
 % doi:10.1088/1126-6708/2007/12/040
 % [arXiv:0709.4174 [hep-ph]].
  %%CITATION = doi:10.1088/1126-6708/2007/12/040;%%
  %230 citations counted in INSPIRE as of 08 May 2018
  
%\cite{Bouchard:2013mia}
\bibitem{Bouchard:2013mia} HPQCD Collaboration,
 % C.~Bouchard {\it et al.} [HPQCD Collaboration],
  %``Standard Model Predictions for $B \to K \ell^+ \ell^-$ with Form Factors from Lattice QCD,''
  {\it Phys. Rev. Lett.} {\bf 111} (2013) 162002
   [Err: {\bf 112} (2014) 149902].
%  doi:10.1103/PhysRevLett.112.149902, 10.1103/PhysRevLett.111.162002
%  [arXiv:1306.0434 [hep-ph]].
  %%CITATION = doi:10.1103/PhysRevLett.112.149902, 10.1103/PhysRevLett.111.162002;%%
  %95 citations counted in INSPIRE as of 08 May 2018
  


%\cite{Bordone:2016gaq}
\bibitem{Bordone:2016gaq}
  M.~Bordone, G.~Isidori and A.~Pattori,
  %``On the Standard Model predictions for $R_K$ and $R_{K^*}$,''
 {\it Eur. Phys. J.} {\bf C76} (2016) 440.
 % doi:10.1140/epjc/s10052-016-4274-7
 % [arXiv:1605.07633 [hep-ph]].
  %%CITATION = doi:10.1140/epjc/s10052-016-4274-7;%%
  %115 citations counted in INSPIRE as of 08 May 2018

%\cite{Abdesselam:2019wac}
\bibitem{Abdesselam:2019wac} Belle Collaboration,
%A.~Abdesselam \textit{et al.} [Belle],
%``Test of lepton flavor universality in ${B\to K^\ast\ell^+\ell^-}$ decays at Belle,''
arXiv:1904.02440 [hep-ex].
%74 citations counted in INSPIRE as of 19 Apr 2020

%\cite{Abdesselam:2019lab}
\bibitem{Abdesselam:2019lab} Belle Collaboration,
%A.~Abdesselam \textit{et al.} [Belle],
%``Test of lepton flavor universality in $B \to K \ell^{+}\ell^{-}$ decays,''
arXiv:1908.01848 [hep-ex].
%19 citations counted in INSPIRE as of 19 Apr 2020

%%\cite{Capdevila:2017bsm}
%\bibitem{Capdevila:2017bsm}
%  B.~Capdevila, A.~Crivellin, S.~Descotes-Genon, J.~Matias and J.~Virto,
%  %``Patterns of New Physics in $b\to s\ell^+\ell^-$ transitions in the light of recent data,''
%  {\it JHEP} {\bf 1801} (2018) 093.
% % doi:10.1007/JHEP01(2018)093
% % [arXiv:1704.05340 [hep-ph]].
%  %%CITATION = doi:10.1007/JHEP01(2018)093;%%
%  %117 citations counted in INSPIRE as of 06 May 2018
%
%%\cite{Descotes-Genon:2015uva}
%\bibitem{Descotes-Genon:2015uva}
%  S.~Descotes-Genon, L.~Hofer, J.~Matias and J.~Virto,
%  %``Global analysis of $b\to s\ell\ell$ anomalies,''
%  {\it JHEP} {\bf 1606} (2016) 092.
%%  doi:10.1007/JHEP06(2016)092
%%  [arXiv:1510.04239 [hep-ph]].
%  %%CITATION = doi:10.1007/JHEP06(2016)092;%%
%  %248 citations counted in INSPIRE as of 16 May 2018
%
%%\cite{Altmannshofer:2017yso}
%\bibitem{Altmannshofer:2017yso}
%  W.~Altmannshofer, P.~Stangl and D.~M.~Straub,
%  %``Interpreting Hints for Lepton Flavor Universality Violation,''
%  {\it Phys. Rev.} {\bf D96} (2017) 055008.
% % doi:10.1103/PhysRevD.96.055008
% % [arXiv:1704.05435 [hep-ph]].
%  %%CITATION = doi:10.1103/PhysRevD.96.055008;%%
%  %97 citations counted in INSPIRE as of 08 May 2018
%
%%\cite{Geng:2017svp}
%\bibitem{Geng:2017svp} L.~S.~Geng   \etal,
%%  L.~S.~Geng, B.~Grinstein, S.~J\"ager, J.~Martin Camalich, X.~L.~Ren and R.~X.~Shi,
%  %``Towards the discovery of new physics with lepton-universality ratios of $b\to s\ell\ell$ decays,''
%  {\it Phys. Rev.} {\bf D96} (2017) 093006.
%%  doi:10.1103/PhysRevD.96.093006
%%  [arXiv:1704.05446 [hep-ph]].
%  %%CITATION = doi:10.1103/PhysRevD.96.093006;%%
%  %88 citations counted in INSPIRE as of 08 May 2018 
%
%%\cite{Hurth:2017hxg}
%\bibitem{Hurth:2017hxg}
%  T.~Hurth, F.~Mahmoudi, D.~Martinez Santos and S.~Neshatpour,
%  %``Lepton nonuniversality in exclusive $b{\rightarrow}s{\ell}{\ell}$ decays,''
%  {\it Phys. Rev.} {\bf D96} (2017) 095034.
%%  doi:10.1103/PhysRevD.96.095034
%%  [arXiv:1705.06274 [hep-ph]].
%  %%CITATION = doi:10.1103/PhysRevD.96.095034;%%
%  %22 citations counted in INSPIRE as of 06 May 2018\\

%\cite{Aebischer:2019mlg}
\bibitem{Aebischer:2019mlg} J.~Aebischer {\it et al.},
%J.~Aebischer, W.~Altmannshofer, D.~Guadagnoli, M.~Reboud, P.~Stangl and D.~M.~Straub,
%``B-decay discrepancies after Moriond 2019,''
{\it Eur. Phys. J.}  \textbf{C80} (2020) 252.
%doi:10.1140/epjc/s10052-020-7817-x
%[arXiv:1903.10434 [hep-ph]].
%99 citations counted in INSPIRE as of 19 Apr 2020

%\cite{Alguero:2019ptt}
\bibitem{Alguero:2019ptt} M.~Algueró {\it et al.},
%M.~Algueró, B.~Capdevila, A.~Crivellin, S.~Descotes-Genon, P.~Masjuan, J.~Matias, M.~Novoa and J.~Virto,
%``Emerging patterns of New Physics with and without Lepton Flavour Universal contributions,''
{\it Eur. Phys. J.}  \textbf{C79} (2019) 714.
%doi:10.1140/epjc/s10052-019-7216-3
%[arXiv:1903.09578 [hep-ph]].
%82 citations counted in INSPIRE as of 19 Apr 2020

%\cite{Ciuchini:2019usw}
\bibitem{Ciuchini:2019usw} M.~Ciuchini {\it et al.},
%M.~Ciuchini, A.~M.~Coutinho, M.~Fedele, E.~Franco, A.~Paul, L.~Silvestrini and M.~Valli,
%``New Physics in $b \to s \ell^+ \ell^-$ confronts new data on Lepton Universality,''
{\it Eur. Phys. J.}  \textbf{C79} (2019) 719.
%doi:10.1140/epjc/s10052-019-7210-9
%[arXiv:1903.09632 [hep-ph]].
%72 citations counted in INSPIRE as of 19 Apr 2020

%\cite{Datta:2019zca}
\bibitem{Datta:2019zca}
A.~Datta, J.~Kumar and D.~London,
%``The $B$ anomalies and new physics in $b \to s e^+ e^-$,''
{\it Phys. Lett.}  \textbf{B797} (2019) 134858.
%doi:10.1016/j.physletb.2019.134858
%[arXiv:1903.10086 [hep-ph]].
%34 citations counted in INSPIRE as of 19 Apr 2020

%\cite{Kowalska:2019ley}
\bibitem{Kowalska:2019ley}
K.~Kowalska, D.~Kumar and E.~M.~Sessolo,
%``Implications for new physics in $b\rightarrow s \mu \mu $ transitions after recent measurements by Belle and LHCb,''
{\it Eur. Phys. J.}  \textbf{C79} (2019) 840.
%doi:10.1140/epjc/s10052-019-7330-2
%[arXiv:1903.10932 [hep-ph]].
%41 citations counted in INSPIRE as of 19 Apr 2020

%\cite{Arbey:2019duh}
\bibitem{Arbey:2019duh} A.~Arbey {et al.},
%A.~Arbey, T.~Hurth, F.~Mahmoudi, D.~M.~Santos and S.~Neshatpour,
%``Update on the b→s anomalies,''
{\it Phys. Rev.}  \textbf{D100} (2019) 015045.
%doi:10.1103/PhysRevD.100.015045
%[arXiv:1904.08399 [hep-ph]].
%47 citations counted in INSPIRE as of 19 Apr 2020

%\cite{Alok:2019ufo}
\bibitem{Alok:2019ufo}
A.~K.~Alok, A.~Dighe, S.~Gangal and D.~Kumar,
%``Continuing search for new physics in $b \to s \mu \mu$ decays: two operators at a time,''
{\it JHEP} \textbf{06} (2019) 089.
%doi:10.1007/JHEP06(2019)089
%[arXiv:1903.09617 [hep-ph]].
%56 citations counted in INSPIRE as of 19 Apr 2020


%%\cite{Fajfer:2018bfj}
%\bibitem{Fajfer:2018bfj}
%  S.~Fajfer, N.~Ko\v{s}nik and L.~Vale Silva,
%  %``Footprints of leptoquarks: from $ R_{K^{(*)}} $ to $ K \rightarrow \pi \nu \bar{\nu }$,''
%  {\it Eur. Phys. J.}  {\bf C78} (2018) 275.
%%  doi:10.1140/epjc/s10052-018-5757-5
%%  [arXiv:1802.00786 [hep-ph]].
%  %%CITATION = doi:10.1140/epjc/s10052-018-5757-5;%%
%  %3 citations counted in INSPIRE as of 08 May 2018
%  
%%\cite{Marzocca:2018wcf}
%\bibitem{Marzocca:2018wcf}
%  D.~Marzocca,
%  %``Addressing the B-physics anomalies in a fundamental Composite Higgs Model,''
%  arXiv:1803.10972 [hep-ph].
%  %%CITATION = ARXIV:1803.10972;%%
%  %1 citations counted in INSPIRE as of 08 May 2018
%
% %\cite{Camargo-Molina:2018cwu}
%\bibitem{Camargo-Molina:2018cwu}
%  J.~E.~Camargo-Molina, A.~Celis and D.~A.~Faroughy,
%  %``Anomalies in Bottom from new physics in Top,''
%  arXiv:1805.04917 [hep-ph].
%  %%CITATION = ARXIV:1805.04917;%% 
%  
 
%\cite{Angelescu:2018tyl}
\bibitem{Angelescu:2018tyl}
A.~Angelescu, D.~Be\v{c}irevi\'c, D.~Faroughy and O.~Sumensari,
%``Closing the window on single leptoquark solutions to the $B$-physics anomalies,''
{\it JHEP} \textbf{10} (2018) 183.
%doi:10.1007/JHEP10(2018)183
%[arXiv:1808.08179 [hep-ph]].
%87 citations counted in INSPIRE as of 19 Apr 2020 
 
%\cite{Bifani:2018zmi}
\bibitem{Bifani:2018zmi}
S.~Bifani, S.~Descotes-Genon, A.~Romero Vidal and M.~Schune,
%``Review of Lepton Universality tests in $B$ decays,''
{\it J. Phys.} \textbf{G46} (2019) 023001.
%doi:10.1088/1361-6471/aaf5de
%[arXiv:1809.06229 [hep-ex]].
%65 citations counted in INSPIRE as of 19 Apr 2020 
 
 
 
  
%%%%%%%%%%%%%%%%%%%%%%%%%%%%%%%%%%%%%%%%%%%%%%%%%%%%%%%%%%
%
%%\bibitem{LHC} The ATLAS and CMS Collaborations, ATLAS-CONF-2011-157,
%%CMS PAS HIG-11-023 (2011).
%
%\bibitem{ATLAS} ATLAS Collaboration, ATLAS-CONF-2011-163.
%
%\bibitem{CMS} CMS Collaboration, CMS PAS HIG-11-032 (2011).
%
%\bibitem{PI:11b} A. Pich, arXiv:1010.5217.
%% Nucl.Phys.Proc.Suppl. 209 (2010) 182-187
%
%\bibitem{GW:77} S.L. Glashow and S. Weinberg, {\it Phys. Rev.} {\bf D15} (1977) 1958.


\end{thebibliography}
\end{document}